\documentclass[twocolumn]{aastex631}

\usepackage{ragged2e}
\usepackage{lineno}
\usepackage{color}
\usepackage{xspace}
\usepackage[version=4]{mhchem} % Molecular species! e.g. \ce{H2O}
\usepackage[nameinlink]{cleveref}
\usepackage{longtable}
\usepackage{hyperref}
\usepackage{amssymb}

\newcommand\jwst{\em JWST}

\newcommand\eureka{\texttt{Eureka!}\xspace}

\newcommand{\smarter}{\texttt{smarter}\xspace}

\newcommand\tiberius{\texttt{Tiberius}\xspace}
\newcommand\tswift{\texttt{Tswift}\xspace}
\newcommand{\planetname}{GJ\,341b\xspace}
\usepackage[symbol]{footmisc}

\DeclareSymbolFont{UPM}{U}{eur}{m}{n}
\DeclareMathSymbol{\umu}{0}{UPM}{"16}
\newcommand\micro{$\umu$}
\newcommand\microns{\micro m\xspace}

\newcommand\rearth{R$_{\oplus}$}
\newcommand\mearth{M$_{\oplus}$}
\usepackage[version=4]{mhchem} % Molecular species! e.g. \ce{H2O}

\defcitealias{Lustig-YaegerFu2023}{Lustig-Yaeger \& Fu et al.}
\defcitealias{MoranStevenson2023}{Moran \& Stevenson et al.}
\defcitealias{MayMacDonald2023}{May \& MacDonald et al.}

\makeatletter
\newcommand*{\linktocite}[2]{%
  \hyper@natlinkstart{#1}#2\hyper@natlinkend}
\makeatother

\submitjournal{AJ}

\shorttitle{JWST/NIRCam transmission spectroscopy of GJ\,341b}
\shortauthors{Kirk et al.}
\graphicspath{{./}{figs/}}
\begin{document}

\title{JWST/NIRCam Transmission Spectroscopy of the Nearby Sub-Earth GJ\,341b}

\author[0000-0002-4207-6615]{James Kirk}
\affil{Department of Physics, Imperial College London, Prince Consort Road, London, SW7 2AZ, UK}

\correspondingauthor{James Kirk}
\email{j.kirk22@imperial.ac.uk}

\author[0000-0002-7352-7941]{Kevin B. Stevenson}%\email{kevin.stevenson@jhuapl.edu}
\affiliation{Johns Hopkins APL, Laurel, MD, 20723, USA}

\author[0000-0002-3263-2251]{Guangwei Fu}%\email{guangweifu@gmail.com}
\affil{Department of Physics \& Astronomy, Johns Hopkins University, Baltimore, MD, USA}

\author[0000-0002-0746-1980]{Jacob Lustig-Yaeger}%\email{jacob.lustig-yaeger@jhuapl.edu}
\affil{Johns Hopkins APL, Laurel, MD, 20723, USA}

\author[0000-0002-6721-3284]{Sarah E. Moran}%\email{semoran@lpl.arizona.edu}
\affiliation{Department of Planetary Sciences and Lunar and Planetary Laboratory, University of Arizona, Tuscon, AZ, USA}

\author[0000-0002-1046-025X]{Sarah Peacock}
\affil{University of Maryland, Baltimore County, MD 21250, USA}
\affil{NASA Goddard Space Flight Center, Greenbelt, MD 20771, USA}

\author[0000-0003-4157-832X]{Munazza K. Alam}
\affil{Carnegie Earth \& Planets Laboratory, Washington, DC, 20015, USA}
\affil{Space Telescope Science Institute, Baltimore, MD 21218, USA}

\author[0000-0003-1240-6844]{Natasha E. Batalha}
\affil{NASA Ames Research Center, Moffett Field, CA, USA}

\author[0000-0002-9030-0132]{Katherine A. Bennett}
\affiliation{Department of Earth \& Planetary Sciences, Johns Hopkins University, Baltimore, MD, USA}

\author[0000-0002-9032-8530]{Junellie Gonzalez-Quiles}
\affiliation{Department of Earth \& Planetary Sciences, Johns Hopkins University, Baltimore, MD, USA}

 \author[0000-0003-3204-8183]{Mercedes L\'opez-Morales}
 \affil{Center for Astrophysics ${\rm \mid}$ Harvard {\rm \&} Smithsonian, 60 Garden St, Cambridge, MA 02138, USA}

\author[0000-0003-3667-8633]{Joshua D. Lothringer}
\affil{Department of Physics, Utah Valley University, Orem, UT, 84058 USA}

\author[0000-0003-4816-3469]{Ryan J. MacDonald}
\affil{Department of Astronomy, University of Michigan, 1085 S. University Ave., Ann Arbor, MI 48109, USA}
\affil{NHFP Sagan Fellow}

\author[0000-0002-2739-1465]{E. M. May}
\affiliation{Johns Hopkins APL, Laurel, MD, 20723, USA}

\author[0000-0002-4321-4581]{L. C. Mayorga}
\affil{Johns Hopkins APL, Laurel, MD, 20723, USA}

 \author[0000-0003-4408-0463]{Zafar Rustamkulov}
\affiliation{Department of Earth \& Planetary Sciences, Johns Hopkins University, Baltimore, MD, USA}

\author[0000-0001-6050-7645]{David K. Sing}
\affil{Department of Earth \& Planetary Sciences, Johns Hopkins University, Baltimore, MD, USA}
\affil{Department of Physics \& Astronomy, Johns Hopkins University, Baltimore, MD, USA}

\author[0000-0001-7393-2368]{Kristin S. Sotzen}
 \affiliation{Johns Hopkins APL, Laurel, MD, 20723, USA}

\author[0000-0003-3305-6281]{Jeff A. Valenti}
\affil{Space Telescope Science Institute, Baltimore, MD 21218, USA}

\author[0000-0003-4328-3867]{Hannah R. Wakeford}
\affil{School of Physics, HH Wills Physics Laboratory, University of Bristol, Bristol, UK}

% \author[0009-0009-3217-0403]{Alicia N. Highland}
% \affil{Department of Astronomy, University of Michigan, 1085 S. University Ave., Ann Arbor, MI 48109, USA}

% \author[0000-0002-0493-1342]{Ethan Kruse}
% \affil{NASA Goddard Space Flight Center, Greenbelt, MD 20771, USA}
% \affil{Department of Astronomy, University of Maryland, College Park, MD 20742.}
% \affil{Center for Research and Exploration in Space Science and Technology, NASA/GSFC, Greenbelt, MD 20771}




\begin{abstract}

We present a JWST/NIRCam transmission spectrum from 3.9--5.0\,\microns of the recently-validated sub-Earth GJ\,341b ($\mathrm{R_P} = 0.92$\,\rearth, $\mathrm{T_{eq}} = 540$\,K) orbiting a nearby bright M1 star ($\mathrm{d} = 10.4$\,pc, $\mathrm{K_{mag}}=5.6$). We use three independent pipelines to reduce the data from the three JWST visits and perform several tests to check for the significance of an atmosphere. Overall, our analysis does not uncover evidence of an atmosphere. Our null hypothesis tests find that none of our pipelines' transmission spectra can rule out a flat line, although there is weak evidence for a Gaussian feature in two spectra from different pipelines (at 2.3 and 2.9\,$\sigma$). However, the candidate features are seen at different wavelengths (4.3 \microns vs 4.7 \microns), and our retrieval analysis finds that different gas species can explain these features in the two reductions (CO$_2$ at $3.1\sigma$ compared to O$_3$ at $2.9\sigma$), suggesting that they are not real astrophysical signals. Our forward model analysis rules out a low mean molecular weight atmosphere ($< 350\times$ solar metallicity) to at least $3\sigma$, and disfavors CH$_4$-dominated atmospheres at $1-3\sigma$, depending on the reduction. Instead, the forward models find our transmission spectra are consistent with no atmosphere, a hazy atmosphere, or an atmosphere containing a species that does not have prominent molecular bands across the NIRCam/F444W bandpass, such as a water-dominated atmosphere. Our results demonstrate the unequivocal need for two or more transit observations analyzed with multiple reduction pipelines, alongside rigorous statistical tests, to determine the robustness of molecular detections for small exoplanet atmospheres.

\end{abstract}



\section{Introduction} 
\label{sec:introduction}

The M-dwarf opportunity is one of the most promising routes to measure the atmospheres of terrestrial, and potentially habitable, exoplanets \citep[e.g.][]{Charbonneau2007}. This opportunity arises from the favorable planet-to-star radius ratios, close-in habitable zones, and frequent transits of rocky planets transiting M dwarfs. However, the M-dwarf opportunity must contend with the high and prolonged X-ray and EUV emission from M-dwarf host stars \citep[e.g.,][]{Pizzolato2003,Preibisch2005,Shkolnik2014,Peacock2019b}, which could cause significant exoplanet atmospheric loss \citep[e.g.,][]{Lammer2003,Owen2012,Owen2016}. 

Indeed, atmosphere observations of terrestrial planets (R$_\mathrm{P} \leq 1.6$\,\rearth) around M dwarfs have so far revealed no conclusive detections of atmospheres \citep[e.g.,][]{deWit2016,deWit2018,Wakeford2019,DiamondLowe2018,DiamondLowe2022,Kreidberg2019}, with recent JWST studies of TRAPPIST-1b and -c being consistent with no, or minimal, atmospheres \citep{Greene2023,Lim2023,Zieba2023}. %Instead, the general conclusion has been that these terrestrial exoplanets have not retained hydrogen/helium atmospheres; however, in most cases, high mean molecular weight atmospheres cannot be ruled out. 

In order to assess the M-dwarf opportunity, it is necessary to determine which, if any, terrestrial planets around M dwarfs are capable of retaining an atmosphere. One way to address this question is to determine whether M dwarf exoplanets are separated by a `cosmic shoreline', a division in insolation--escape velocity space (I--v$_\mathrm{esc}$) separating bodies with atmospheres from those without \citep{Zahnle2017}. In the Solar System, \cite{Zahnle2017} demonstrated that this dividing line follows an $\mathrm{I} \propto \mathrm{v_{esc}}^4$ relation. 

Following the successful launch and commissioning of JWST \citep{Gardner2023,McElwain2023,Rigby2023}, we are now able to expand the I--v$_\mathrm{esc}$ parameter space over which we can address the existence of a cosmic shoreline to terrestrial planets around M-dwarf stars. This is the focus of JWST program 1981 (PIs: K.\ Stevenson and J.\ Lustig-Yaeger), which is obtaining JWST transmission spectra of five exoplanets around M dwarfs to determine whether a cosmic shoreline exists for M-dwarf systems.  

Our first study (\citetalias{Lustig-YaegerFu2023} \citeyear{Lustig-YaegerFu2023}) ruled out Earth-like, hydrogen/helium, water- and methane-dominated clear atmospheres for the Earth-sized exoplanet LHS\,475b. This is consistent with predictions from the cosmic shoreline. This study also confirmed LHS\,475b as a bona fide exoplanet, promoting it from a planet candidate in parallel with the efforts of \cite{Ment2023}.

Our second study (\citetalias{MoranStevenson2023} \citeyear{MoranStevenson2023}) was of the super-Earth GJ\,486b which sits close to the shoreline predicted by \cite{Zahnle2017} and thus provides an important test of its existence. We tentatively detected water in GJ\,486b's transmission spectrum, however, our NIRSpec/G395H observations could not determine the origin of the water, be it in the planet's atmosphere or the stellar atmosphere.

% In our second study, we tentatively detected water in the transmission spectrum of the super-Earth GJ\,486b (\citetalias{MoranStevenson2023} \citeyear{MoranStevenson2023}). However, our NIRSpec/G395H observations could not determine the origin of the water, be it in the planet's atmosphere or the stellar atmosphere. While GJ\,486b sits on the `dry' (no atmosphere) side of the cosmic shoreline, its close proximity to the shoreline means that a water-rich atmosphere would not be in strong tension with predictions from \cite{Zahnle2017} or that the Solar System cosmic shoreline requires adjustments for M-dwarf systems.

Most recently, in \citetalias{MayMacDonald2023} \citeyear{MayMacDonald2023}, we observed two transits of the super-Earth GJ\,1132b, another planet near the shoreline. Our study revealed the importance of multi-visit observations for small transiting exoplanets, since one visit was suggestive of spectral features while the other produced a flat transmission spectrum. Given the inconsistent transmission spectra between our two visits, we were unable to make a conclusive determination of the planet's likely atmospheric composition, although neither visit was consistent with a previous report of HCN \citep{Swain2021}.  

%Most recently, we observed two transits of GJ\,1132b, which sits on the dry side of the cosmic shoreline and has had conflicting results regarding claims of atmospheric detections \citep{Southworth2017,DiamondLowe2018,Swain2021,Mugnai2021,LibbyRoberts2022}. Our JWST observations of GJ\,1132b revealed the importance of multi-visit observations for small transiting exoplanets (\citetalias{MayMacDonald2023} \citeyear{MayMacDonald2023}), since one visit was suggestive of spectral features while the other produced a flat transmission spectrum. Given the inconsistent transmission spectra between our two visits, we were unable to make a conclusive determination of the planet's likely atmospheric composition, although we rule out previous claims about this planet's atmosphere \citep{Swain2021}. 

Here, we extend our program to GJ\,341b (TOI 741.01), a planet that was recently validated by \cite{DiTomasso2023} and which orbits a nearby bright M1 star ($K = 5.6$, 10.4\,pc, \citealt{gaia2016,gaia_dr3}). The TESS-derived planetary radius is $0.92 \pm 0.05$\,R$_{\oplus}$, making this a sub-Earth, with a $3\sigma$ upper limit on the planet's mass of 4.5\,M$_{\oplus}$ \citep{DiTomasso2023}. The mass--radius relation of \cite{Chen2017} gives a planet mass of $0.72 \pm 0.14$\,\mearth. Assuming a Bond albedo of zero and uniform heat redistribution, the planet's equilibrium temperature is 540\,K, while its irradiation temperature is 760\,K. Given the planet's escape velocity of 10\,km\,s$^{-1}$, the cosmic shoreline relation predicts that this planet has not retained a substantial atmosphere. Our study seeks to test this prediction.

The paper is laid out as follows: in Section \ref{sec:observations} we describe the observational setup. In Section \ref{sec:reduction} we explain our data reduction and light curve fitting, using three independent pipelines. We present our transmission spectra, flat line rejection tests, atmospheric forward models and atmospheric retrievals in Section \ref{sec:transmission_spectrum}. We discuss our results and present our conclusions in Section \ref{sec:discussion}.

\section{Observations}
\label{sec:observations}

We observed three transits of \planetname across three visits with JWST's Near Infrared Camera (NIRCam, \citealt{Rieke2023}) as part of GO program 1981 (PIs: K.\ Stevenson and J. Lustig-Yaeger). Unlike the other GO 1981 targets that were observed with NIRSpec, NIRCam was selected for GJ\,341 because this bright star ($\mathrm{K_{mag}} = 5.6$) saturates NIRSpec in one group. The three visits occurred on March 2, March 10, and April 17, 2023 UT, with each visit lasting 5.25 hours. We used the F444W filter, covering a wavelength range from 3.8--5.1\,\microns at a native spectral resolution of $R \approx 1650$, and the SUBGRISM128 subarray, giving us 2048 pixel columns and 128 pixel rows. Given the brightness of our target, we used 3 groups per integration and acquired 4652 integrations during each of our three visits. The effective integration time was 3.38\,s. 

\section{Data reduction}
\label{sec:reduction}

As with the previous studies in our program (\citetalias{Lustig-YaegerFu2023} \citeyear{Lustig-YaegerFu2023}; \citetalias{MoranStevenson2023} \citeyear{MoranStevenson2023}; \citetalias{MayMacDonald2023} \citeyear{MayMacDonald2023}), we used three different pipelines to reduce our data independently: \eureka, \tiberius, and \tswift. We describe the approach taken with each pipeline in more detail below and include a table in the appendix that summarizes each pipeline's approach (Table \ref{tab:reduction_choices}).

\subsection{\eureka}
\label{sec:r_eureka}

The {\eureka} data reduction pipeline \citep{Eureka2022} is well established, with published JWST analyses demonstrating reliable results from numerous observations (\citealt{ERSFirstLook, Ahrer2023, Alderson2023, Rustamkulov2023}; \citetalias{Lustig-YaegerFu2023} \citeyear{Lustig-YaegerFu2023}; \citetalias{MoranStevenson2023} \citeyear{MoranStevenson2023}).  As described below, we adopt similar methods for the analysis of \planetname.

Starting from the \texttt{uncal.fits} files, we masked a region within 30 pixels of the measured trace from pixel column 640 to 2048.  We then performed a double-iteration, group-level background subtraction by first fitting and removing the mean flux on a column-by-column basis and, second, fitting and removing the mean flux on a row-by-row basis.  Unlike NIRSpec and NIRISS, NIRCam detectors exhibit 1/f noise along pixel rows, thus necessitating this additional row-by-row step. This orientation of the 1/f noise is because NIRCam's rapid readout direction aligns with the wavelength dispersion direction \citep{schlawin_jwst_2020}. In Stage 1, we skipped the jump detection step because our observations only had 3 groups per integration, which makes the jump step unreliable.  We processed the \texttt{rateints.fits} files through the regular \texttt{jwst} Stage 2 pipeline.

In Stage 3, {\eureka} performed a full-frame outlier search with a 4$\sigma$ rejection threshold.  We then computed a clean median frame and corrected the spectrum's curvature by rolling pixel columns to bring the trace to the center of the detector.  We skipped any additional background subtraction and applied optimal spectral extraction \citep{Horne1986} using an aperture that was within 3 pixels of the measured trace.  We tested a range of aperture sizes and found that a 3-pixel half-width minimized the white light curve residuals. To calculate our wavelength solution, we used a polynomial from Flight Program 1076 (E.\ Schlawin, private communication). %By comparing to Phoenix model spectra \citep{Husser2013}, we found this polynomial gives a more accurate wavelength solution than the wavelength solution from the \texttt{jwst} pipeline, which was offset by approximately 0.01\,\microns (10 pixels).

We extracted the flux from 3.9--5.0 {\microns}, splitting the light into 55 spectroscopic channels, each 0.02\,\microns in width.  For each light curve, we applied a box-car filter of length 50 integrations and flagged any 3.5$\sigma$ outliers.  This procedure removed $<2$\% of all integrations for a given light curve.

We fitted all three white light curves simultaneously.  We used \texttt{batman} \citep{batman2015} to fit the system parameters and fixed the quadratic limb-darkening coefficients to those provided by \texttt{ExoTiC-LD} \citep{david_grant_2022_7437681} from 3D stellar models \citep{Magic2015}. For our stellar parameters, we used $\mathrm{T_{eff}} = 3745$\,K and $\log g = 4.71$\,g\,cm$^{-2}$ and assumed a solar metallicity ([Fe/H] = 0) (ExoMAST). Across the three visits, we fitted for a single planet-star radius ratio ($R_P/R_*$), orbital inclination ($i$), semi-major axis ($a/R_*$), and transit time ($T_0$). The period (P) was held fixed to 7.5768707 days (ExoMAST). For each visit, we fitted an independent exponential times linear ramp model to account for light curve systematics.  We also renormalized the error bars from each visit in order to achieve a reduced chi-squared of unity.  In total, there were 19 free parameters.  Table \ref{tab:system_params} lists our best-fit system parameters.  

\begin{figure*}
    \centering
    \includegraphics[width=1.0\textwidth]{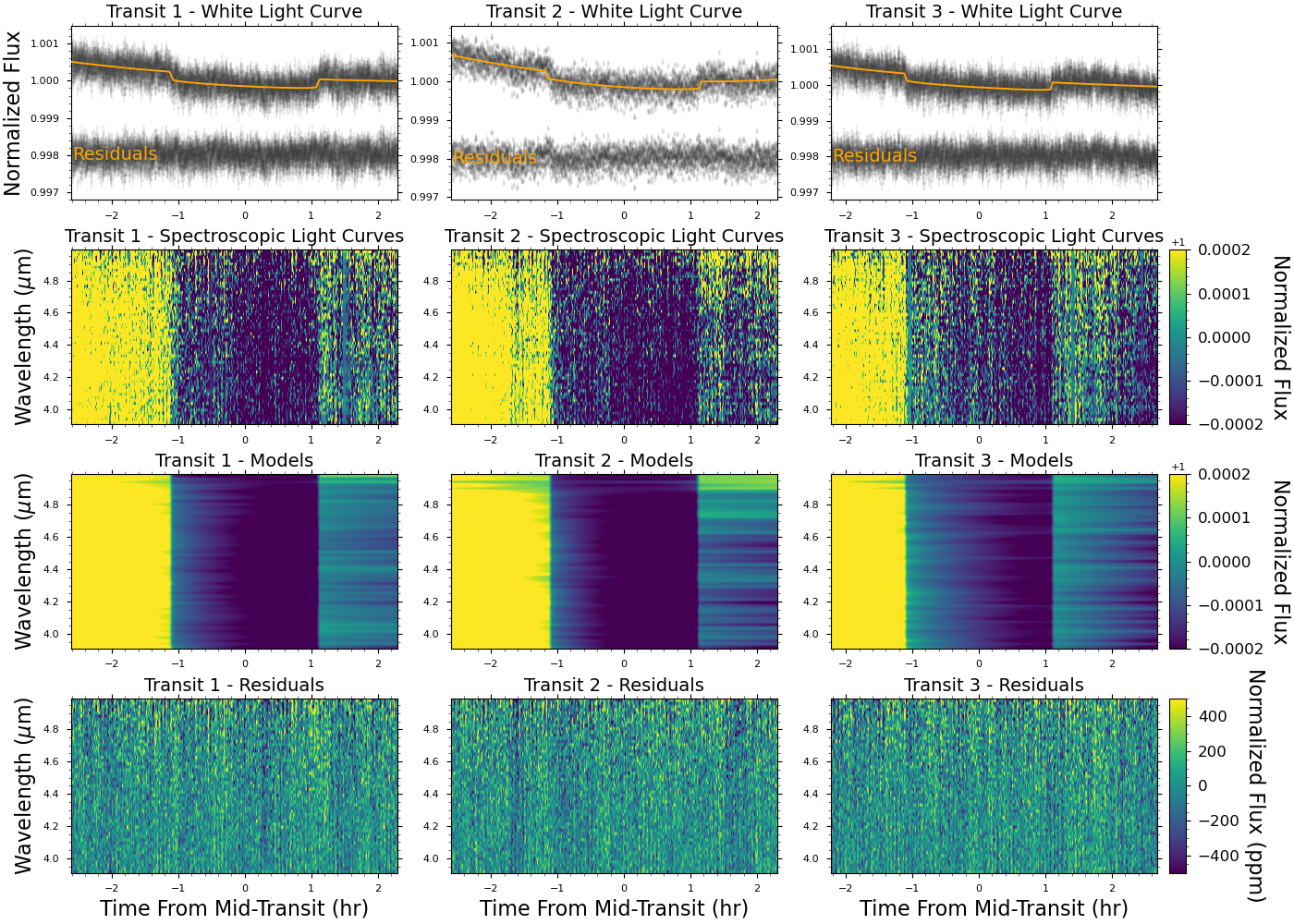}
    \caption{The light curves and best fit models resulting from the \eureka pipeline. The columns show, from left to right: transit 1, transit 2, transit 3. The rows show, from top to bottom: 1) the white light curve and residuals, 2) the spectroscopic light curves, 3) the best-fit models to the spectroscopic light curves, 4) the residuals from the spectroscopic light curve fitting.}
    \label{fig:lc_fits_eureka} 
\end{figure*}

\begin{figure}
    \centering
    \includegraphics[width=0.475\textwidth]{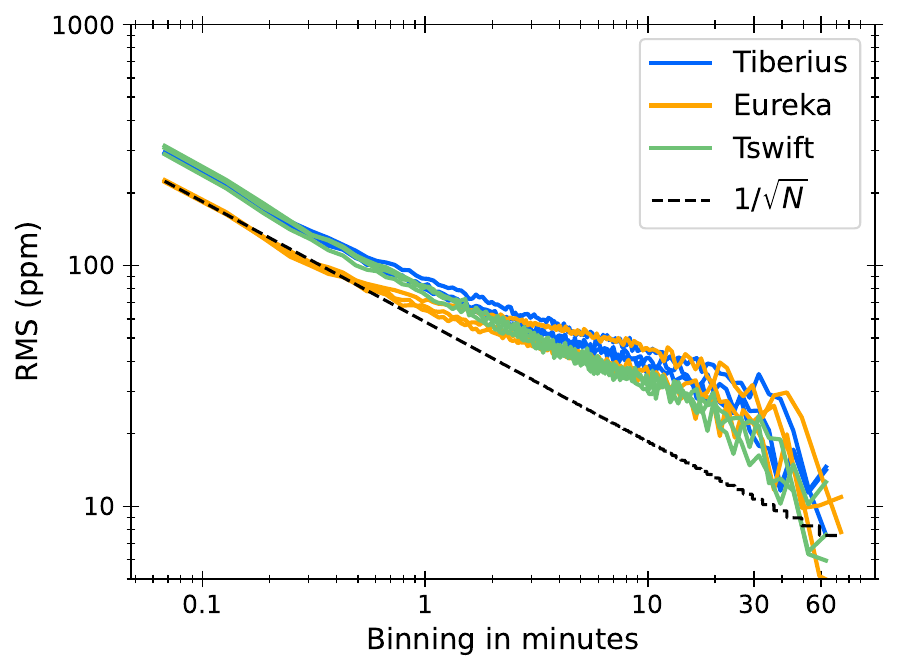}
    \caption{The Allan variance plot from the white light curve fits to each pipeline's reduction and for all three visits. The amplitude of the residuals from \tiberius (blue), \eureka (orange) and \tswift (green) are shown as a function of bin size in minutes. In the absence of red noise, the residuals would follow the dashed black line.}
    \label{fig:av_all} 
\end{figure}

When fitting the spectroscopic light curves, we fixed the orbital inclination, semi-major axis, and transit time to the white light curve best-fit values.  Again, we fixed the quadratic limb-darkening parameters to the values provided by \texttt{ExoTiC-LD} for each spectroscopic channel.  The planet-to-star radius ratio was the only physical free parameter.  For each spectroscopic channel, we fitted an independent exponential times linear ramp model to account for light curve systematics.  Including the error bar renormalization term, each spectroscopic fit had six free parameters.  
For both the white and spectroscopic light curves, we clipped the first 300 integrations from each visit to avoid fitting the exponential decrease in flux. Figure \ref{fig:lc_fits_eureka} shows our fits to the \eureka white and spectroscopic light curves, with the Allan variance plot of the residuals shown in Figure \ref{fig:av_all}.

\begin{table*}[]
    \centering
    \begin{tabular}{c|c|c|c|c} 
        Reduction & $T_0 (\mathrm{BJD_{TDB}})$ & $i (^{\circ})$ & $a/R_*$ & $R_P/R_*$ \\ \hline
        \eureka & $2460006.420147^{+0.000152}_{-0.000141}$ & $ 88.17^{+0.64}_{-0.51} $ & $ 20.12^{+2.93}_{-2.19} $ & $ 0.016079^{+0.000271}_{-0.000254} $ \\
        \tiberius & $2460006.420152^{+0.000193}_{-0.000196}$ & $ 88.77^{+0.88}_{-0.97} $ & $ 22.88^{+2.97}_{-4.37} $ & $ 0.015315^{+0.000316}_{-0.000297} $ \\
        \tswift & $2460006.420143^{+0.000241}_{-0.000245}$ & $ 88.72^{+0.41}_{-0.46} $ & $ 23.48^{+1.44}_{-1.95} $ & $0.016130^{+0.000295}_{-0.000337}$\\ \hline
        DiTomasso et al.\ & $2459301.771 \pm 0.002$ & $ 89.22^{+0.54}_{-0.96} $ & $ 24.50^{+2.07}_{-4.01} $ & $ 0.0161^{+0.0007}_{-0.0008} $ \\ \hline
    \end{tabular}
    \caption{GJ\,341b's system parameters from the three independent reductions' white light curve fits. We also include the parameters from the independent TESS and radial velocity analysis \protect\citep{DiTomasso2023}.}
    \label{tab:system_params}
\end{table*}

\subsection{\tiberius}
\label{sec:r_tiberius}

The \tiberius reduction began with the uncalibrated fits files (\texttt{uncal.fits}). We ran these through the stage 1 detector processing stages of the \texttt{jwst} pipeline. However, we did not perform the \texttt{jump} step as this has been found to increase the noise in time series spectrophotometry \citep[e.g.,][]{Rustamkulov2023}. Furthermore, unlike in the application of \tiberius to NIRSpec data (e.g., \citealt{Rustamkulov2023}; \citetalias{Lustig-YaegerFu2023} \citeyear{Lustig-YaegerFu2023}; \citetalias{MoranStevenson2023} \citeyear{MoranStevenson2023}) and in our \eureka reductions of the data presented here, we did not apply a 1/f correction to the data. This is because the 1/f noise is parallel to the dispersion dimension for NIRCam data, complicating its removal.% Despite not performing this correction, the noise in our \tiberius spectroscopic light curves is the lowest of the three reductions, likely resulting from our use of a common mode correction, as described later in this subsection.

Following the stage 1 detector processing, we worked with the integration level images (\texttt{gainscalestep.fits} files). While these were processed through the \texttt{gain\_scale} step, the \texttt{jwst} pipeline does not apply a gain correction if the default gain setting is used. For this reason, we worked in units of DN/s throughout the rest of our analysis. Since we work with normalized light curves, and rescale our photometric uncertainties, the gain correction is not necessary. 

Before extracting the stellar flux from each image, each pixel was compared to the median value for that pixel across the time series and $>5\sigma$ outliers were replaced by the median for that pixel. We then oversampled each pixel by a factor of 10 in the spatial dimension using a flux-conserving interpolation, which allows us to more precisely place the aperture boundaries. We found this step to reduce white noise by 10\,\% in previous tests on JWST data \citep{Rustamkulov2023}. Next, we traced the stellar spectrum using a fourth order polynomial between columns 700 and 2042 (inclusive, zero-indexed) and performed standard aperture extraction. We subtracted the background at every column using the median value across the full column, after masking a total of 45 pixels centered on the stellar trace, and summed up the stellar flux in an aperture with a full width of 5 pixels. We experimented with wider apertures and found these led to consistent transmission spectra and light curves but with greater noise.    

% To calculate our wavelength solution, we used a polynomial from Flight Program 1076 (E.\ Schlawin, private communication). By comparing to Phoenix model spectra \citep{Husser2013}, we found this polynomial gives a more accurate wavelength solution than the wavelength solution from the \texttt{jwst} pipeline, which was offset by approximately 0.01\,\microns (10 pixels). We then created our 57 spectroscopic light curves using 20-pixel-wide wavelength bins between 3.906 and 5.004\,\microns

We fitted our white light curves from the three visits simultaneously after subtracting one orbital period (7.576863 days, \citealt{DiTomasso2023}) from transit 2's time array and six orbital periods from transit 3's time array so that all time arrays could be fitted with a single $T_0$, which corresponded to transit 1. The 16 free parameters in our white light curve fits were $T_0$, $i$, $a/R_*$, $R_P/R_*$ and the 12 parameters defining our systematics model. We defined this model as a single component exponential ramp multiplied by a linear-in-time polynomial for each visit. The period was held fixed to 7.576863 days \citep{DiTomasso2023}. We used \texttt{batman} \citep{batman2015} to generate our analytic transit light curves. 

For our limb darkening calculations, we adopted $\mathrm{T_{eff}} = 3770 \pm 40$\,K and $\log g = 4.72 \pm 0.02$ g\,cm$^{-2}$ \citep{DiTomasso2023} and assumed a solar metallicity ([Fe/H] = 0). We used a quadratic limb darkening law with coefficients held fixed to values calculated from 3D stellar atmosphere models \citep{Magic2015} using ExoTiC-LD \citep{david_grant_2022_7437681} and our own custom instrument throughput. This was constructed by dividing our observed stellar spectrum by the appropriate Phoenix model spectrum \citep{Husser2013} and smoothing the result with a running median, following the approach of \cite{Alderson2023}.

We explored the parameter space with a Markov chain Monte Carlo (MCMC) using the \texttt{emcee} Python package \citep{emcee2013}. We ran three sets of chains each with 50,000 steps and 200 walkers. After the first run, we discarded the first 25,000 steps and calculated the median values from the second 25,000 steps of our chains which were used as starting points for the second run. After the second run, we again discarded the first 25,000 steps and used the best-fitting model to rescale our photometric uncertainties to give $\chi^2_\nu = 1$. Our third and final run used the rescaled photometric uncertainties. The medians, 16th and 84th percentiles of our posteriors from this third run are reported in Table \ref{tab:system_params}. These posteriors were constructed after discarding the first 100 steps and thinning by a factor of 10. 

Our spectroscopic light curves were created using 57 20-pixel-wide wavelength bins between 3.906 and 5.004\,\microns and we adopted the same wavelength solution as used in the \eureka reduction. For our spectroscopic light curve fits, we fixed the system parameters ($T_0$, $i$ and $a/R_*$) to the \tiberius values in Table \ref{tab:system_params} and fitted each visit independently, not jointly. Prior to fitting, we divided each spectroscopic light curve by a visit-dependent common noise model. This noise model was defined as WLC$_\mathrm{v}$/tm$_\mathrm{v}$ where WLC$_\mathrm{v}$ is the visit-dependent white light curve and tm$_\mathrm{v}$ is the visit-dependent best-fitting transit (\texttt{batman}) model. For our spectroscopic light curves, the exponential coefficients did not converge for every wavelength bin across each visit, despite experimenting with different priors. For this reason, at the spectroscopic light curve stage, we clipped the first 600 integrations (40 minutes) from each visit, after removing the common noise, and fitted the light curves with a linear polynomial and no exponential. This gave comparable residual red noise to the fits where the exponential coefficients did converge, indicating that the ramp was adequately removed by clipping the first 600 integrations. Figure \ref{fig:lc_fits_tiberius} in the Appendix shows the fits to our white and spectroscopic light curves, with the Allan variance plot of the residual red noise from our white light curve fits shown in Figure \ref{fig:av_all}.

\subsection{\tswift}
\label{sec:r_tswift}

Transit swift (\tswift) is a JWST data reduction routine developed to swiftly analyze time-series observation of transiting exoplanet atmospheres. Starting from the \texttt{uncal.fits} images, we used the default \texttt{jwst} pipeline with \texttt{steps.jump} turned off to produce stage one \texttt{rateints.fits} images. Next, we performed row-by-row and column-by-column background subtraction \citep{schlawin_jwst_2020} for each integration by subtracting out the median of the non-illuminated regions from each row. Next, we cross-correlated the spectral line spread function with each column to determine the vertical position of the spectral trace for that column. We then used a third-order polynomial to fit the vertical position versus the column number, determining the spectral trace location for each integration. The spectrum was extracted using a full-width 6-pixel wide aperture centered on this polynomial fit as it minimizes the scatter in the white light curve. Given the brief exposure time for each integration and the high number of integrations per visit, cosmic rays or hot pixels have a minimal impact on the transit light curve. Furthermore, we identified and discarded any outliers in time exceeding 5$\sigma$ based on the transit light curve baseline scatter for each column. Finally, we combined all wavelength channels to produce the white light curve of each transit.

The fitting of each individual white light curve employs a combination of \texttt{batman} \citep{batman2015} and \texttt{emcee} \citep{emcee2013}. This approach incorporates eight parameters: $T_0$, $R_P/R_*$, $a/R_*$, $i$, a linear time slope, a constant scaling factor, and two coefficients for the exponential ramp. The averaged best-fit parameters from the three white light curves are listed in Table \ref{tab:system_params}. To determine the quadratic limb darkening parameters, we used the 3D stellar models \citep{Magic2015} from the ExoTiC-LD \citep{david_grant_2022_7437681} package and subsequently fixed both quadratic terms, using the same stellar parameters as in the \eureka and \tiberius analyses. We tested fitting for the second quadratic term while fixing the first one, and it had no effect on the final transmission spectrum. %So both quadratic limb darkening parameters were fixed to the 3D stellar model values.

The NIRCam detector has four amplifiers and the spectrum is dispersed across the second to fourth amplifier in the F444W observation mode \citep{schlawin_jwst_2020}. As already discussed, there is 1/f noise along the dispersion direction in our NIRCam data. %The rapid readout direction aligns with the wavelength dispersion direction. This configuration of the instrument leads to amplifier-specific correlated 1/f readout noise along the dispersion direction. 
In an ideal scenario, this noise could be subtracted if every row in each amplifier had unilluminated pixels. However, due to the continuous dispersion of spectra across various amplifiers, it is not feasible to correct the entire spectrum. As a result, we observed lingering 1/f noise residuals that showed correlation across wavelengths in the light curve fits. We identified that these residual patterns are predominantly consistent across all amplifiers, with a few minor patterns unique to individual amplifiers. To better remove this correlated noise, we performed a white light fit within each amplifier, and the residuals were then used as a common-mode correction during the spectroscopic light curve fits. 

For our spectroscopic light curve fits, we used \texttt{batman} and \texttt{scipy.optimize.curvefit}. The model consists of five parameters: $R_P/R_*$, a constant scaling factor, two coefficients for the exponential ramp, and a scaling factor for the common-mode correction. We held $T_0$, $a/R_*$, and $i$ fixed to the corresponding averaged best-fit values from the white light curves (Table \ref{tab:system_params}). Figure \ref{fig:lc_fits_tswift} shows our fits to the \tswift white and spectroscopic light curves, with the Allan variance plot of the residuals shown in Figure \ref{fig:av_all}.

\subsection{Engineering parameters}

As can be seen in the Allan variance plots from our reductions (Figure \ref{fig:av_all}) there is residual red noise which becomes apparent at time scales longer than 1 minute. In order to remove this red noise, we explored detrending our light curves with engineering telemetry data timeseries from the JWST Engineering Database. The telemetry data are identified by mnemonics, which follow the naming convention X\_Y\_Z, where X corresponds to the system (I for Integrated Science Instrument Module, S for spacecraft), Y corresponds to the subsystem, instrument, controller, or component (1--4 letters), and Z corresponds to the telemetry name (2--20 characters). Throughout the rest of this section, we refer to these timeseries telemetry data sets as `mnemonics'.

We began by whittling down the 6176 NIRCam-related mnemonics to the 596 mnemonics of time-varying floats (as opposed to booleans and non-varying constants). We then resampled each mnemonic onto the same time-sampling as each of our three visits and proceeded to sequentially incorporate each of the 596 mnemnonics into our white light curve systematics model, for the \tiberius reduction, to test whether the addition of a mnemnonics model decreased the residual red noise.

For these fits we used a Levenberg-Marquadt alogorithm, with the systematics model being the summation of a linear polynomial describing the mnemonic and a linear-in-time polynomial, after clipping the first 600 integrations. We then compared the residual red noise in each case to our fiducial model: a linear-in-time polynomial. We defined the residual red noise as being the RMS of the residuals in bins of 200 integrations divided by the white noise expectation (equivalent to dividing the color lines in our Allan variance plots by the dashed black lines at $x = 14.5$\,mins, Figure \ref{fig:av_all}). Of the 596 possible mnemonics, we found that 112 mnemonics led to a decrease in the residual red noise compared to the fiducial model for all three visits. The ten mnemonics that led to the largest decrease in the RMS are given in Table \ref{tab:top10_mnemonics}. As shown in Table \ref{tab:top10_mnemonics}, these mnemonics mostly correspond to temperature.

\begin{table*}[]
    \centering
    \begin{tabular}{c|c|c}
         Top 10 mnemonics (all visits) & Short description & Residual red noise \\ \hline
         INRC\_FA\_PSC\_LVCBRD\_TEMP & LVC\_BOARD\_TEMP & 4.79 \\
         INRC\_FA\_PSC\_SCRBRD\_TEMP & SCR\_BOARD\_TEMP & 4.90 \\
         INRC\_FB\_PSC\_SCRBRD\_TEMP & SCR\_BOARD\_TEMP & 5.16 \\
         INRC\_FA\_ACE2\_SC\_TEMP\_HI & SC\_ASIC temperature high gain & 5.20 \\
         IGDP\_NRC\_FA\_ACE2\_SCTEMP & ASIC Temperature read by ACE card & 5.20 \\
         INRC\_FB\_PSC\_LVCBRD\_TEMP & LVC\_BOARD\_TEMP & 5.23 \\
         INRC\_FA\_ACE1\_RR\_RD & Voltage of ACE reference reference & 5.37 \\
         IGDP\_NRC\_FB\_ACE1\_SCTEMP & ASIC Temperature read by ACE card & 5.39 \\
         INRC\_FB\_ACE1\_SC\_TEMP\_HI & SC\_ASIC temperature high gain & 5.39 \\
         INRC\_FB\_ACE1\_D2P5\_V & Voltage of ACE Digital 2.5V voltage & 5.39 \\ \hline
         Fiducial & Linear-in-time polynomial & 5.50 \\ \hline
    \end{tabular}
    \caption{The ten engineering mnemonics that led to the largest improvement in the cumulative residual red noise from all three white light curve fits. These are ordered from largest improvement (top) to least improvement (bottom). With regards to the mnemonics, `INRC' is NIRCam telemetry reported by ISIM, the Integrated Science Instrument Module, `IGDP' are `ground data points' computed from downlinked engineering telemetry, and `TEMP' corresponds to temperature. We understand that `A' and `B' correspond to module A and module B, `ACE' corresponds to Application Specific Integrated Circuit (ASIC) control electronics, and `LV' corresponds to low voltage. We define the residual red noise in the text.}
    \label{tab:top10_mnemonics}
\end{table*}

\begin{figure}
    \centering
    \includegraphics[width=0.45\textwidth]{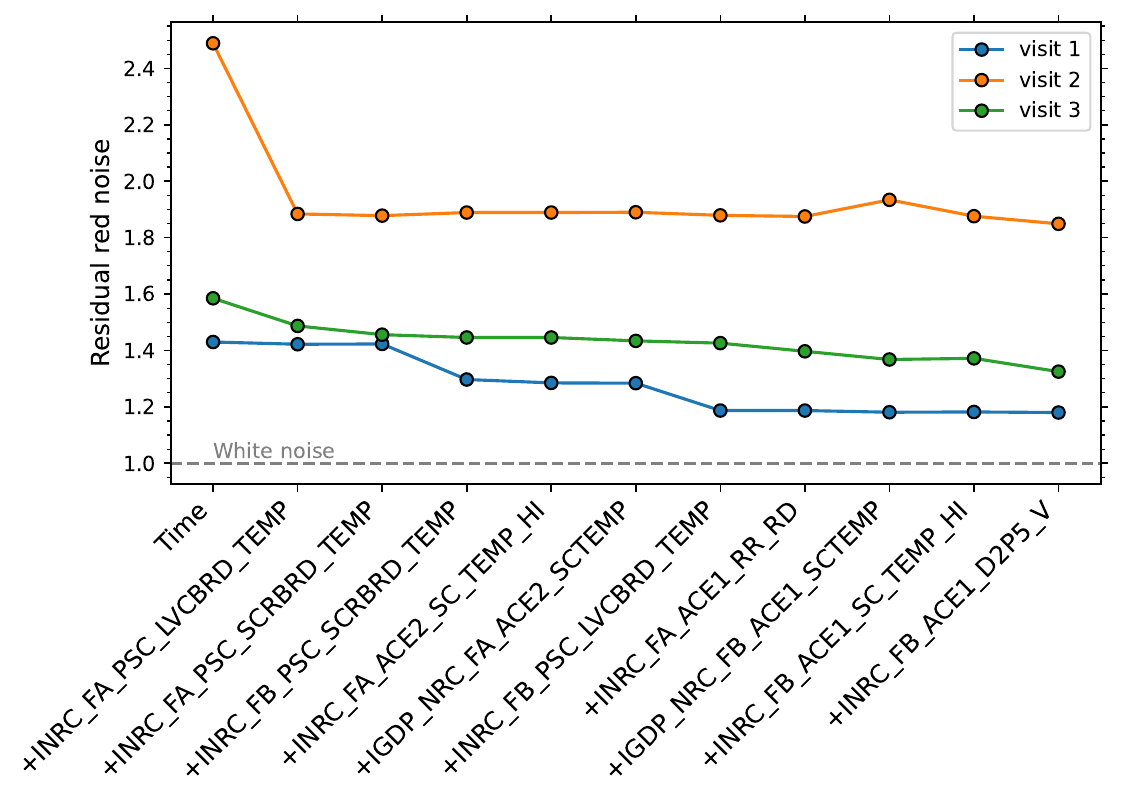}
    \caption{The improvement in residual red noise in the white light curves by adding an additional mnemonic / telemetry timeseries to the systematics model. The blue, orange and green lines indicate the residuals from visit 1, 2 and 3's white light curves. The fiducial model is a linear-in-time polynomial. Each x-tick label indicates an additional mnemonic is added. The dashed gray line indicates the white noise level that the residual noise would achieve if no red noise is present.}
    \label{fig:top10_mnemonics}
\end{figure}

Following this test, we ran a second test where we increased the number of mnemonics used to fit the white light curves, going from one to ten mnemonics in the order presented in Table \ref{tab:top10_mnemonics}. As can be seen in Figure \ref{fig:top10_mnemonics}, including additional mnemonics does not necessarily lead to an improvement in the residual red noise. This is likely due to some mnemonics representing similar telemetry data (e.g., INRC\_FA\_PSC\_LVCBRD\_TEMP and INRC\_FA\_PSC\_SCRBRD\_TEMP). This figure also demonstrates that the use of a mnemonic significantly improves the residual red noise in visit 2 while the changes in visits 1 and 3 are more subtle. Additionally, a combination of 10 mnemonics does not eradicate the residual red noise. As a further test, we fitted all 596 mnemonics jointly with a linear-in-time polynomial. In this case, the residual red noise was below the white noise expectations. We interpreted this as evidence that including a large number of mnemonics can lead to over-fitting. 

While independently fitting each visit's light curve with mnemonics demonstrated the residual red noise could be improved upon, we ended up not using the mnemonic data in our final white light curve fits. This was because we could not obtain acceptable fits when we tried to fit the three visits simultaneously with mnemonics, likely due to the fact that the amplitude of the systematics in our data is comparable to the transit depth. We advocate for a further exploration of mnemonics for exoplanets with deeper transits where the transit and systematics models are less degenerate.

\subsection{Short wavelength photometry}

Since our observations were conducted with NIRCam, we additionally have access to the short wavelength (SW) channel between 0.6--2.3\,\microns. This channel acquired photometry simultaneously with the long wavelength spectroscopy described above. We extracted the short wavelength photometry using \tiberius  to determine whether we could use these additional white light curves to improve the precision of our system parameters.

\begin{figure}
    \centering
    \includegraphics[scale=0.525]{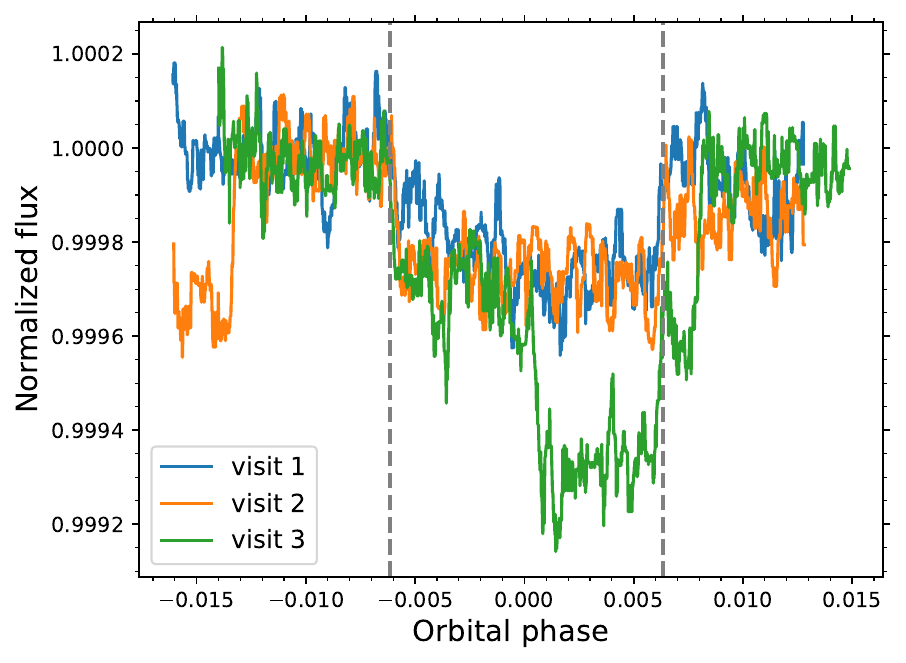}
    \caption{The short wavelength photometric light curves for each visit, revealing significant pre-transit noise (visit 2) and in-transit noise (visit 3). These have been smoothed with a running median calculated over a sliding box with a width of 51 integrations. The dashed vertical lines indicate the first and fourth contact points of the transit.}
    \label{fig:SW_photometry}
\end{figure}

As with the long wavelength data, we began by processing the \texttt{uncal.fits} files through stage 1 of the \texttt{jwst} pipeline. We then extracted the time-series photometry using \tiberius with an aperture 100 pixels wide and a center fixed at a pixel column of 1795. We also subtracted the background at every pixel row, which was calculated as the median of two 50-pixel-wide regions separated by 100 pixels from each side of the target aperture. The resulting SW light curves for each visit are shown in Figure \ref{fig:SW_photometry}. Given the significant noise in these light curves, we did not use these in our parameter estimation. 

\subsection{Measuring flux-calibrated spectra}
\label{sec:flux_cal}

Following our study of GJ\,486b (\citetalias{MoranStevenson2023} \citeyear{MoranStevenson2023}), we extracted the flux calibrated stellar spectrum of GJ\,341. We did this in an attempt to constrain the covering fractions of star spots and faculae and hence determine the impact of stellar contamination on our transmission spectra. To do this, we processed our stellar spectra through stage 2 of the \texttt{jwst} pipeline. Specifically, taking the \texttt{gainscalestep.fits} files, we processed these through the \texttt{assign\_wcs}, \texttt{flat\_field}, \texttt{extract\_2d}, \texttt{srctype}, \texttt{pathloss} and \texttt{photom} steps. We did not perform the \texttt{background} step, and instead used an alternate version of our stage 1 reduction where we did perform a group-level background subtraction. 

We then used \tiberius to perform standard aperture photometry on our \texttt{photomstep.fits} files using an aperture with a full width of 20 pixels. Despite running the \texttt{srctype} command with the \texttt{--source\_type POINT} argument defined, the resulting units of our spectra were MJy/sr. We therefore had to multiply our spectra by the solid angle of a pixel, listed as \texttt{PIXAR\_SR} in the \texttt{photomstep.fits} header, to convert to MJy. We defined the flux uncertainties in the flux calibrated spectra as being the standard deviation of each pixel's flux across the time series. We found no significant variability in the flux-calibrated spectra from visit-to-visit ($\leq 0.1$\,\%) and within each visit ($0.24$\,\%). 

We then computed multi-component forward models using the \cite{Allard2012} \texttt{PHOENIX} stellar models to quantify star spot and faculae covering fractions of the stellar surface for each of the three visits. We employed a weighted linear combination of three \texttt{PHOENIX} models to represent the background photosphere, spots ($T_{\rm eff}$ $\leq$ $T_{\rm eff, photosphere}$ - 100 K), and faculae ($T_{\rm eff}$ $\geq$ $T_{\rm eff, photosphere}$ + 100 K). We assumed that all spots have the same $T_{\rm eff}$, log($g$), and metallicity, as do the faculae, and each feature is required to not exceed 45\% of the total surface. The grid of \texttt{PHOENIX} models used for our analysis covered $T_{\rm eff}$ = 2800 -- 4000 K, log($g$) = 4 -- 5.5 cm s$^{-2}$, [Fe/H] = -0.5 -- 0.5 providing extensive coverage of possible spot and faculae temperatures for GJ 341 assuming photospheric values near the literature quoted $T_{\rm eff}$= 3770 $\pm$ 40 K and log($g$) = 4.72 $\pm$ 0.02 cm s$^{-2}$ \citep{DiTomasso2023}. The models were scaled by R$_*^2$/dist$^2$ using literature values for GJ 341: R$_*$=0.506 R$_\odot$ \citep{DiTomasso2023} and $d$ = 10.45 pc \citep{Gaia2021}. We also smoothed and interpolated the models to be the same resolution as the observations before calculating a reduced-$\chi^2$. In our reduced-$\chi^2$ calculations, we considered 868 wavelength points for each visit and eight fitted parameters (the $T_{\rm eff}$, log($g$), and [Fe/H] of the photosphere, the $T_{\rm eff}$ and coverage fraction of both spots and faculae, and a scaling factor). The scaling factor was multiplied by the R$_*^2$/$d$$^2$ term to account for uncertainty in either measured quantity and varied from 0.9 to 1.3.

\begin{figure*}
    \centering
    \includegraphics[scale=0.7]{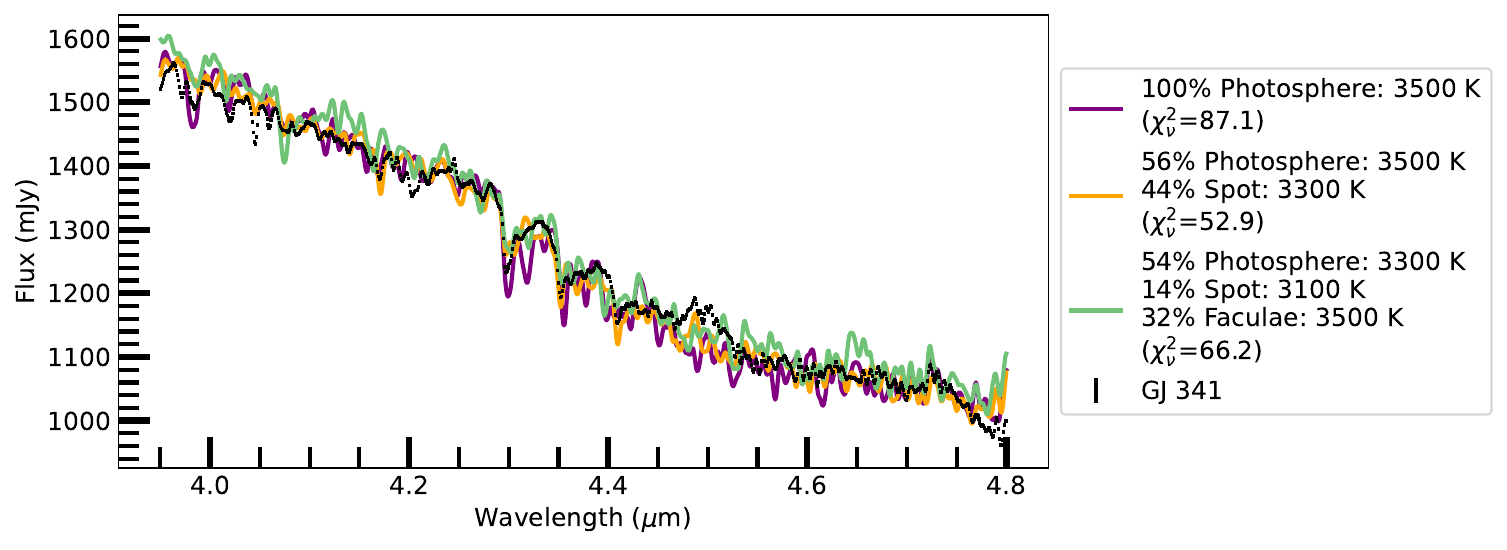}
    \caption{A comparison between our measured flux calibrated spectra (black) and superpositions of \texttt{PHOENIX} stellar atmosphere models: photosphere with no active region contribution (purple), photosphere with contribution from spots (orange), and photosphere with contributions from spots and faculae (green). Given the poor $\chi^2_\nu$, and lack of contamination evident in our transmission spectra, we do interpret these model results as evidence for stellar contamination in our data.}
    \label{fig:stellar_spectra}
\end{figure*}

Unlike in our application of this procedure to the stellar spectra of GJ\,486 (\citetalias{Lustig-YaegerFu2023} \citeyear{Lustig-YaegerFu2023}) and GJ\,1132 (\citetalias{MayMacDonald2023} \citeyear{MayMacDonald2023}), which used NIRSpec/G395H, we could not obtain an acceptable fit to the flux-calibrated NIRCam/F444W spectra ($\chi^2_\nu \sim 60$), as shown in Figure \ref{fig:stellar_spectra}. We do not know the reason for these poor fits, however, the shape of our extracted spectrum changes significantly when we apply the photometric calibration (\texttt{photom}) step in the \texttt{jwst} pipeline, suggesting that the NIRCam PHOTOM reference file may be responsible. Given the poor fits in Figure \ref{fig:stellar_spectra}, and lack of contamination evident in our transmission spectra (see section \ref{sec:transmission_spectrum}), we do interpret the results from our analysis of the flux calibrated spectrum as evidence for stellar contamination in our data.

\section{GJ\,341\lowercase{b's} transmission spectrum}
\label{sec:transmission_spectrum}

\begin{figure*}
    \centering
    \includegraphics[scale=0.5]{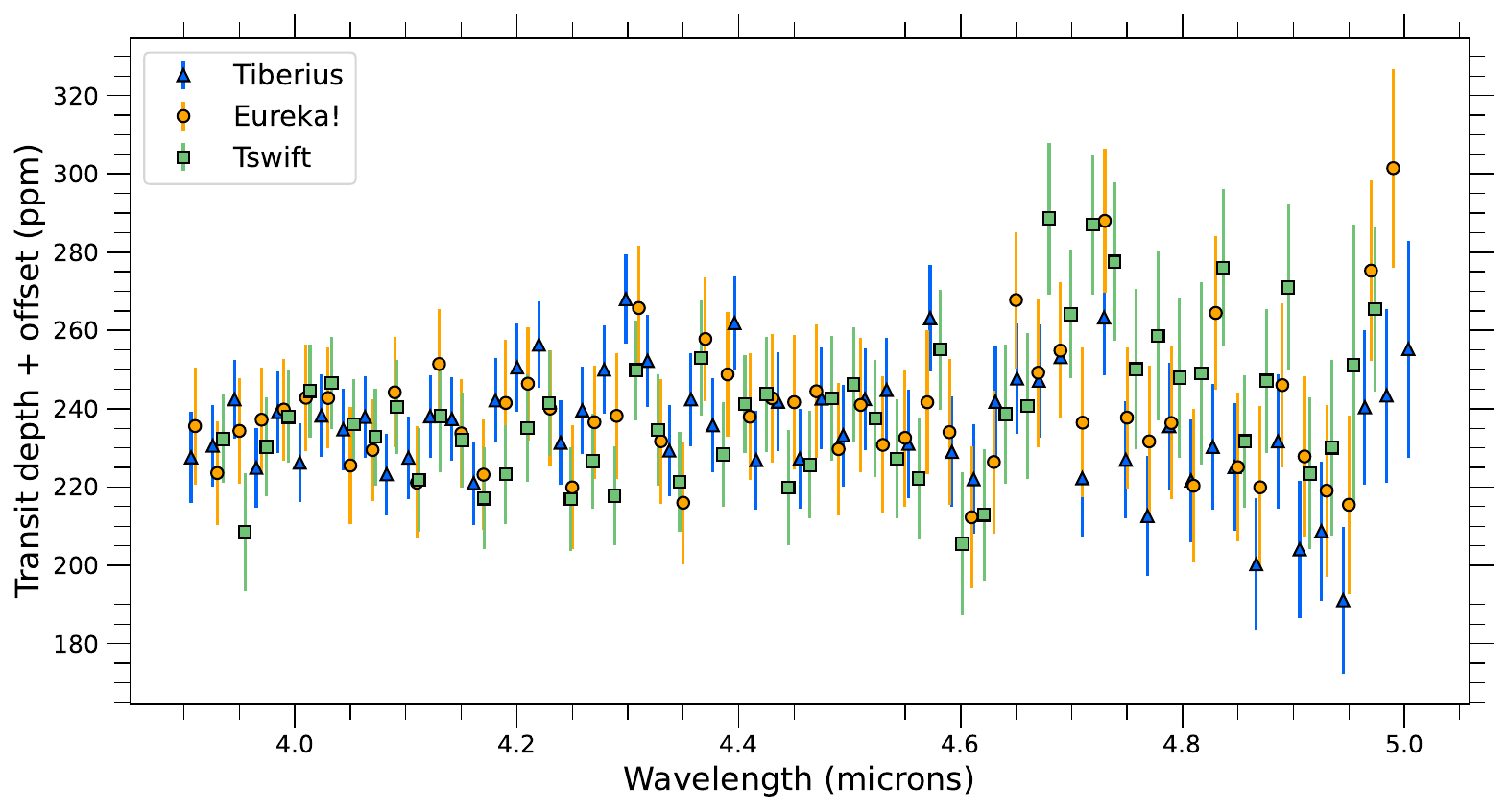}
    \caption{GJ\,341b's transmission spectrum as measured by \tiberius (blue), \eureka (orange) and \tswift (green). These spectra are the the weighted means of the three visits' spectra. The \eureka spectrum has been offset by -15\,ppm to make the median transit depths equal to \tiberius and \tswift between 4.0 and 4.8\,\microns.}
    \label{fig:transmission_spectrum}
\end{figure*}

The transmission spectra we measure for GJ\,341b from the three different reductions are shown in Figure \ref{fig:transmission_spectrum}. This figure demonstrates that the three independent reductions produce consistent transmission spectra. For this figure, we offset the \eureka spectrum by -15\,ppm so that its median transit depth matches that of \tiberius and \tswift between 4.0 and 4.8\,\microns. We include the transmission spectra measured for each visit by each reduction in the Appendix (Figure \ref{fig:visit_spectra}) which demonstrates that the spectra are also consistent between visits for each reduction. In the following subsections, we further investigate the transmission spectrum of GJ\,341b with flat line rejection tests, atmosphere forward models, and retrievals. 

\subsection{Gaussian flat line rejection tests}

% Intro to approach
We begin our assessment of the \planetname transmission spectrum by quantifying the statistical significance of any potential features in the spectrum relative to the null hypothesis. In this case, the null hypothesis is that the planet possesses a flat, featureless transmission spectrum that is devoid of atmospheric information, as would be the case if the planet does not possess an atmosphere or the atmospheric signal falls below our detection sensitivity (\citealp{Kreidberg2014}; \citetalias{Lustig-YaegerFu2023} \citeyear{Lustig-YaegerFu2023}).

% Description of method
Following the approaches of \citetalias{MoranStevenson2023} (\citeyear{MoranStevenson2023}) and \citetalias{MayMacDonald2023} (\citeyear{MayMacDonald2023}), we perform two statistical tests on the transmission spectra from each data reduction pipeline. The first test quantifies the (in)consistency of our measurements with the null hypothesis given the observational errors, while the second test quantifies the evidence favoring an agnostic Gaussian absorption feature in the spectrum over the flat null hypothesis. As both tests use slightly different approaches to quantify the significance of one or more features in the spectrum, they are both useful diagnostics. However, these tests are limited to statistical statements about the spectrum as opposed to physical explanations, which require the more detailed modeling presented in the forthcoming subsections. 

% Test 1 
To understand whether the observed spectrum is consistent with the null hypothesis, we generated 100,000 random spectra on the same wavelength grid subject to the assumption that the underlying true planet spectrum is the null hypothesis and the observational errors are unchanged from those observed. We calculated the reduced chi-squared ($\chi^{2}_{\nu}$) for each sample and build up the expected distribution of $\chi^{2}_{\nu}$ under the null hypothesis. That is, if the null hypothesis were true then we would expect that our measured $\chi^{2}_{\nu}$ would be a single sample from that distribution. Thus, we can quantify the probability that the measured $\chi^{2}_{\nu}$ value would be at least as large as the one we observed given the assumption of the null hypothesis (i.e., a p-value). 

% Test 2
To understand whether there is evidence of a spectral feature in the observed spectrum, we first fit a flat featureless spectrum to the data and then fit an agnostic Gaussian model to the spectrum. We then compare the Bayesian evidence of the two fits and calculate Bayes factors to estimate which model is preferred, as in \citetalias{MoranStevenson2023} (\citeyear{MoranStevenson2023}) and \citetalias{MayMacDonald2023} (\citeyear{MayMacDonald2023}). The Gaussian model contains four fitting parameters, which are all constrained by uniform priors. These include the wavelength independent transit depth upon which the Gaussian feature is added ${\sim}\mathcal{U}(0.0, 1.0)$, the wavelength of the Gaussian center ${\sim}\mathcal{U}(3.5, 5.5)$ \microns, the width of the feature ${\sim}\mathcal{U}(\Delta x_0, \Delta x_1)$ \microns, where $\Delta x_0$ is the minimum spectroscopic bin width and $\Delta x_1$ is the width of the entire wavelength range of the data, and the amplitude of the Gaussian feature ${\sim}\mathcal{U}(-10 \sigma_y, 10 \sigma_y)$, where $\sigma_y$ is the standard deviation of the transmission spectrum measurements (i.e., features should not greatly exceed the scatter in the data). 

The results of these tests are shown in Table \ref{tab:flat}. This demonstrates that the \eureka spectrum is the flattest of the three spectra, ruling out our null hypothesis at only $0.2\sigma$, while the Gaussian feature test favors a flat line over a feature at $2.2\sigma$. The \tiberius and \tswift spectra are less flat. However, in the case of the \tiberius spectrum we believe this is due to the smaller uncertainties in its transmission spectrum (see Section \ref{sec:errors}). For these reductions, the null hypothesis is in weak tension, at $1.3\sigma$ for \tiberius and $1.4\sigma$ for \tswift. The Gaussian feature test results in $2.3\sigma$ evidence for a feature in the \tiberius spectrum (broad and centered near 4.3 \microns) and $2.9\sigma$ in the \tswift spectrum (width of ${\sim}0.2$ \microns centered near 4.7 \microns). While these look promising at face value, as we see from these tests and  demonstrate with our subsequent retrieval analysis (section \ref{sec:retrievals}), the two candidate features occur at different wavelengths between the \tiberius and \tswift spectra and are therefore less likely to be astrophysical.  

% Test 2

% Results in Table 

% Table of flat tests
\begin{deluxetable}{l||l|r|c}
\tablewidth{0.98\textwidth}
\tabletypesize{\small}
\tablecaption{Is it Flat? \label{tab:flat}}
\tablehead{
%\colhead{} & \multicolumn{3}{c}{\bf Visit 1} & \multicolumn{3}{c}{\bf Visit 2} & \multicolumn{3}{c}{\bf Combined}  \\ 
\colhead{Reduction} & \colhead{$\chi^{2}_{\nu}$} & \colhead{p-value ($\sigma$)} & \colhead{Feature?}}
\startdata
\tiberius &          1.15 &      20\%  (1.3$\sigma$) &   2.3$\sigma$ \\
\eureka   &          0.82 &      83\%  (0.2$\sigma$) &  -2.2$\sigma$ \\
\tswift   &          1.18 &      17\%  (1.4$\sigma$) &   2.9$\sigma$ \\
\enddata
\tablecomments{$\chi^{2}_{\nu}$ is the reduced chi-squared resulting from the best fitting featureless fit (null hypothesis) to the observed spectrum, ``p-value'' refers to the probability that $\chi^{2}_{\nu}$ would be at least as extreme as the observed value under the assumption that the null hypothesis is correct and is displayed along with the corresponding ``sigma'' value, and the ``Feature?'' column shows the level of confidence in the detection of an agnostic Gaussian absorption feature in the spectrum over the null hypothesis.} 
\end{deluxetable}

\subsection{Sensitivity to Spectrum Uncertainties}
\label{sec:errors}

\begin{figure}
    \centering
    \includegraphics[width=0.47\textwidth]{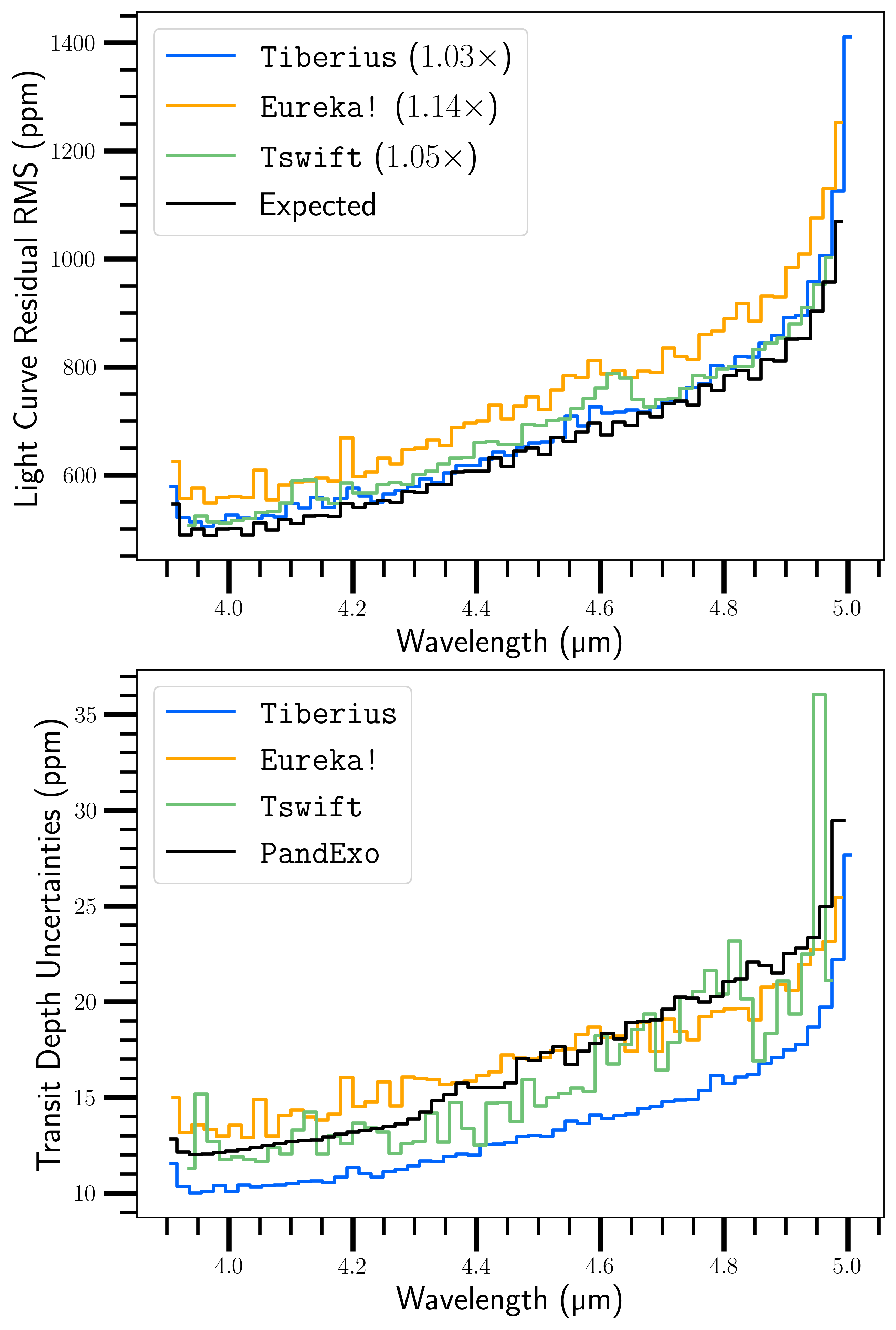}
    \caption{Light curve residual RMS (upper panel) and transit depth uncertainties (lower panel) for each reduction (color lines). The expected noise (photon + readnoise + dark current) calculated from the data is shown in the upper panel (black line) and the median RMS relative to the expected noise is listed in the legend for each reduction. The predicted noise from \texttt{PandExo} \citep{Batalha2017} is shown in the lower panel (black line). }
    \label{fig:error_comparison}
\end{figure}

The results of the aforementioned flat line rejection tests depend not just on potential features in the spectrum, but also on the size of the error bars, which are not equivalent for each reduction. The 1$\sigma$ spectroscopic transit depth uncertainties are shown in Figure \ref{fig:error_comparison} (lower panel). Since all reductions used approximately the same spectral bin widths, we can readily compare their 1$\sigma$ uncertainties and apply the errors from one reduction to the others to estimate the role of uncertainties in driving the statistical results of our null hypothesis tests. Table \ref{tab:errors} shows $\chi^{2}_{\nu}$ values for flat line fits to each reduction's transmission spectrum data points (rows) with errors interpolated from the other reductions (columns). 

% Table of error changes
\begin{deluxetable}{l|c|c|c}
\tablewidth{0.98\textwidth}
\tabletypesize{\small}
\tablecaption{Reduced-$\chi^2$ Values for a flat line fit when exchanging errors between reductions \label{tab:errors}}
\tablehead{
%\colhead{} & \multicolumn{3}{c}{\bf Visit 1} & \multicolumn{3}{c}{\bf Visit 2} & \multicolumn{3}{c}{\bf Combined}  \\ 
\colhead{} & \colhead{\tiberius Errors} & \colhead{\eureka Errors} & \colhead{\tswift Errors}}
\startdata
\tiberius & 1.15 & 0.68 & 0.79 \\
\eureka  &  1.28 & 0.82 & 0.75 \\
\tswift  &  1.81 & 1.10 & 1.18 \\
\enddata 
\end{deluxetable}

Figure \ref{fig:error_comparison} (lower panel) and Table \ref{tab:errors} show that the differences in uncertainties between reductions leads to statistically significant differences in $\chi^{2}_{\nu}$. The \tiberius reduction has the smallest uncertainties and when those are applied to the \eureka and \tswift spectra, those spectra deviate more from a flat line. On the other hand, the \tswift spectrum has $\chi^{2}_{\nu} > 1$ regardless of which errors are applied, consistent with the larger deviations seen in the \tswift spectrum. The \eureka spectrum has relatively large errors, but relatively  small spectroscopic deviations, which together explain our statistical result and subsequent physical interpretations.   

The upper panel of Figure \ref{fig:error_comparison} shows the light curve residual RMS and the expected noise for comparison, which includes the photon noise, read noise and dark current. The \tiberius, \tswift, and \eureka reductions achieve a median precision of 1.03$\times$, 1.05$\times$, and 1.14$\times$ the noise limit, respectively. This ordering is different to the white light curve residuals, where \eureka obtained the lowest residual RMS (Figure \ref{fig:av_all}). The lower RMS of \tiberius and \tswift at the spectroscopic light curve stage comes from those pipelines' use of a common mode correction. Although the \tiberius transit depth uncertainties are lower than the predicted \texttt{PandExo} uncertainties, this is likely due to improved performance relative to the model-based prediction, since the \tiberius reduction still sits above the noise limit (Figure \ref{fig:error_comparison}, upper panel).     

\subsection{Atmosphere forward models}
\label{sec:forward_models}

As with previous targets in program JWST-GO-1981, we compare our transmission spectra of \planetname to a set of atmospheric forward models to infer whether the data suggests that the planet possesses an appreciable atmosphere, and if so, its composition (\citetalias{Lustig-YaegerFu2023} \citeyear{Lustig-YaegerFu2023}; \citetalias{MoranStevenson2023} \citeyear{MoranStevenson2023}; \citetalias{MayMacDonald2023} \citeyear{MayMacDonald2023}). Since \planetname has no published mass constraints beyond an upper limit \citep[$\leq$ 3~M$_{\oplus}$ at 1$\sigma$;][]{DiTomasso2023}, we generate forward models using a planetary mass of 0.73 M$_{\oplus}$. This mass assumes that the planetary radius of 0.92 R$_{\oplus}$ is consistent with a rocky bulk composition, as \citet{Luque2022} showed that very small ($\leq$ 1 R$_{\oplus}$) planets around M-dwarfs, like \planetname, appear to have very tightly constrained densities similar to Earth-like iron-silicate bulk compositions.

\begin{figure*}
    \centering
    \includegraphics[scale=0.475]{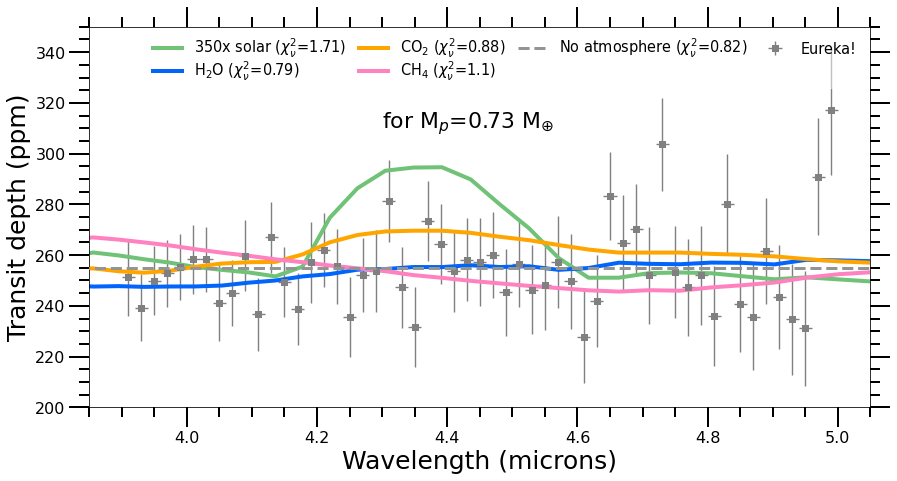}
    %{figs/gj341_twsift_350xsolar_v2.png}
    %{figs/gj341_Tiberius_350xsolar_v2.png}
    \caption{Atmospheric forward models compared to the \eureka reduction from {\jwst}/NIRCam. Because of the planet's small size and low equilibrium temperature, only hydrogen-dominated or methane atmospheres are strongly ruled out or disfavored by the data. Atmospheres of water, carbon dioxide, or an opaque haze layer -- or no atmosphere -- are all statistically consistent with the data to within $\sim$1$\sigma$ across all three reductions.}
    \label{fig:forward_models}
\end{figure*}

Our atmospheric forward models include two cases: 1) \texttt{CHIMERA} thermochemical equilibrium models \citep{Line2013a,Line2014-C/O} from 100-1000$\times$ solar metallicity, using a parameterized temperature-pressure profile \citep{Guillot2010} with an equilibrium temperature of 540 K and opacity contributions from H$_2$O, CH$_4$, CO, CO$_2$, NH$_3$,	HCN, H$_2$S,	H$_2$, and He; and 2) \texttt{PICASO} \citep{Batalha2019} models of single-gas atmospheres with an isothermal temperature-pressure profile at 540 K. The \texttt{PICASO} models include atmospheres of pure 1 bar H$_2$O, CO$_2$, or CH$_4$, as well as a flat line model representative of no atmosphere or a uniform aerosol layer. For all model configurations, we compute synthetic transmission spectra with \texttt{PICASO}'s radiative transfer routine using molecular opacities resampled from R = 60,000 \citep{batalha2020}, rebinned to the resolution of each data reduction. We compute best fits by calculating the $\chi^2_{\nu}$ between each reduction and each forward model, with 57, 54, or 55 degrees of freedom (dof) for the \tiberius, \tswift, and \eureka reductions, respectively. We include an offset in relative transit depth to the mean of each reduction in our fitting. Our results for \eureka are shown in Figure \ref{fig:forward_models}, while those for \tiberius and \tswift are shown in Figures \ref{fig:forward_models_tiberius} and \ref{fig:forward_models_tswift}. We include the complete set of statistical constraints for each reduction compared to each forward model case in Table \ref{tab:forward}.
%can add figure sets if wanted...

Across all three reductions, hydrogen-dominated or methane atmospheric models do not fit the data. For the \tswift reduction, hydrogen-dominated atmospheres as metal-rich as 1000$\times$ solar can be ruled out to 3$\sigma$, while this constraint is reduced to 350$\times$ solar for \tiberius and \eureka. Such hydrogen-rich atmospheres would be unexpected to persist against escape, so this result is unsurprising. Methane atmospheres of 1 bar are disfavored at 3$\sigma$, 2.5$\sigma$, or 1$\sigma$ between \tswift, \tiberius, and \eureka.

On the other hand, water atmospheres or a flat line -- which could stem either from an airless world or from a hazy world -- are consistent with each reduction, with $\chi^2_{\nu}$ $\lesssim$ 1. Though water is a ubiquitous molecule and common in warm planetary atmospheres, we are unable to rule it out here since water has no strong absorption bandheads and instead acts as a continuum-like absorber from 4-5 $\mu$m. Both \eureka and \tiberius also are consistent ($\chi^2_{\nu}$ $\lesssim$ 1) with a carbon dioxide atmosphere, while \tswift disfavors this scenario to 2$\sigma$. 

%this paragraph is Sarah trying to explain why our constraints are like this, feel free to cut, move, or spin for more optimism...
Despite the excellent agreement between each data reduction (Figure \ref{fig:transmission_spectrum}), there are obvious differences in goodness of fit between atmospheric models. This occurs due to 1) the planet's small size and temperature, which results in a very small atmospheric scale height, and 2) the ratio of the sub-Earth radius (R$_P$=0.92 R$_{\oplus}$) to the relatively large radius of the star for an M-dwarf (R$_*$=0.506 R$_\odot$), resulting in a very shallow transit. Given these physical parameters, our 350$\times$ solar, hydrogen-dominated atmosphere model has a peak feature size of 40 ppm while a carbon dioxide atmosphere model has a peak feature size of only 15 ppm in the NIRCam/F444W bandpass. Even with the extremely high precision obtained by our program (average uncertainties of 15\,ppm per 0.02\,$\mu$m bin), such small feature sizes mean that minute differences between reductions driven by random noise or data treatments drive divergent atmospheric interpretations. Overall, either no atmosphere, a hazy atmosphere, or an atmosphere containing a species that does not have prominent molecular bands across the NIRCam/F444W bandpass -- e.g., a water-dominated atmosphere -- provide the best explanation of the data across the three reductions compared to our forward models.

% Table of forward model results
\begin{deluxetable*}{l||r|r|r|c}
\tablewidth{0.9\textwidth}
\tabletypesize{\small}
\tablecaption{Atmospheric Constraints \label{tab:forward}}
\tablehead{
%\colhead{} & \multicolumn{3}{c}{\bf Visit 1} & \multicolumn{3}{c}{\bf Visit 2} & \multicolumn{3}{c}{\bf Combined}  \\ 
\colhead{Model}  & \colhead{\tiberius}  & \colhead{\eureka}  & \colhead{\tswift} & Interpretation}
\startdata
350$\times$ solar & 1.78  &  1.71 & 2.91 & rejected at $\geq$ 3$\sigma$ \\
H$_2$O   &          1.18 &   0.79 & 1.11 & $\sim$ consistent with data  \\
CO$_2$   &          1.02 &   0.88 & 1.41 & inconclusive \\
CH$_4$   &          1.48 &   1.10 & 1.70 & ruled out to $\geq$1$\sigma$ \\
Flat line   &       1.15 &   0.82 & 1.22 & $\sim$ consistent with data  \\
\enddata
\tablecomments{The $\chi^2_\nu$ of each model is reported in the columns for each reduction name. All atmospheric models are computed for 1 bar and an equilibrium temperature of 540 K. The ``flat line'' model can indicate no atmosphere or an opaque aerosol layer.} %could cut flat line row since it's redundant with previous table?} 
\end{deluxetable*}

% also talk about how we tried different masses to 2sigma of C&K MR relation and didn't strongly change results??

\subsection{Atmosphere retrievals}
\label{sec:retrievals}

% Retrievals intro
To examine a broader range of atmospheric parameter space that is (in)consistent with our measurements, we ran a set of retrievals using the Spectral Mapping Atmospheric Radiative Transfer Exoplanet Retrieval code (\citealp[\smarter;][]{Lustig-Yaeger2022, Lustig-Yaeger2023earth}; \citetalias{Lustig-YaegerFu2023} \citeyear{Lustig-YaegerFu2023}). We use the same \smarter modeling approach and assumptions as described in \citetalias{Lustig-YaegerFu2023} (\citeyear{Lustig-YaegerFu2023}), except slightly modified to account for the different planet parameters. We run separate retrievals on each of the final coadded spectra from the \eureka, \tiberius, and \tswift reductions. 

% Priors
Our nominal retrieval setup for \planetname uses 11 free parameters. We include the following six spectroscopically active gases that have (or partially have) absorption bands in the NIRCam F444W wavelength range and are plausible candidate gases for secondary atmospheres at temperatures at or near \planetname's equilibrium temperature of 540\,K: \ce{H2O}, \ce{CH4}, \ce{CO}, \ce{CO2}, \ce{N2O}, and \ce{O3}. Each gas is fitted in terms of the log$_{10}$ evenly-mixed volume mixing ratio (VMR) with a flat prior, VMR $\sim \mathcal{U}(-12,0)$. The bulk atmospheric constituent is left agnostic by fitting for the mean molecular weight of the primary constituent using a flat prior, $\mu \sim \mathcal{U}(2.5, 50)$ g/mol, which ranges from a low-mass mix of \ce{H2}+\ce{He} to heavier than \ce{CO2} (44 g/mol) and \ce{O3} (48 g/mol). Although aerosols are not explicitly included in our retrievals, we fit for the opaque planet radius using a uniform prior $R_p \sim \mathcal{U}(0.83, 1.01) R_{\oplus}$, which could be due to a cloud deck, the solid planetary surface \citep{Benneke2012}, additional continuum opacity, or the critical refraction limit \citep{Betremieux2013, Misra2014}. We fit for the pressure at the opaque surface using a uniform prior, $P_0 \sim \mathcal{U}(-1, 6)$ Pa. An isothermal temperature structure is assumed with a uniform prior, $T \sim \mathcal{U}(300, 700)$ K. Finally, we allow the planet mass to vary uniformly between $M_p \sim \mathcal{U}(0.3, 2.2) M_{\oplus}$. 

% Running retrievals
Our \smarter retrievals use the \texttt{dynesty} \citep{Speagle2020} nested sampling algorithm \citep{Skilling2004} with 600 live points. We run the model until the contribution to the total evidence from the remaining prior volume drops below \texttt{dlogz=0.075}. 

\begin{figure*}
    \centering
    \includegraphics[width=0.95\textwidth]{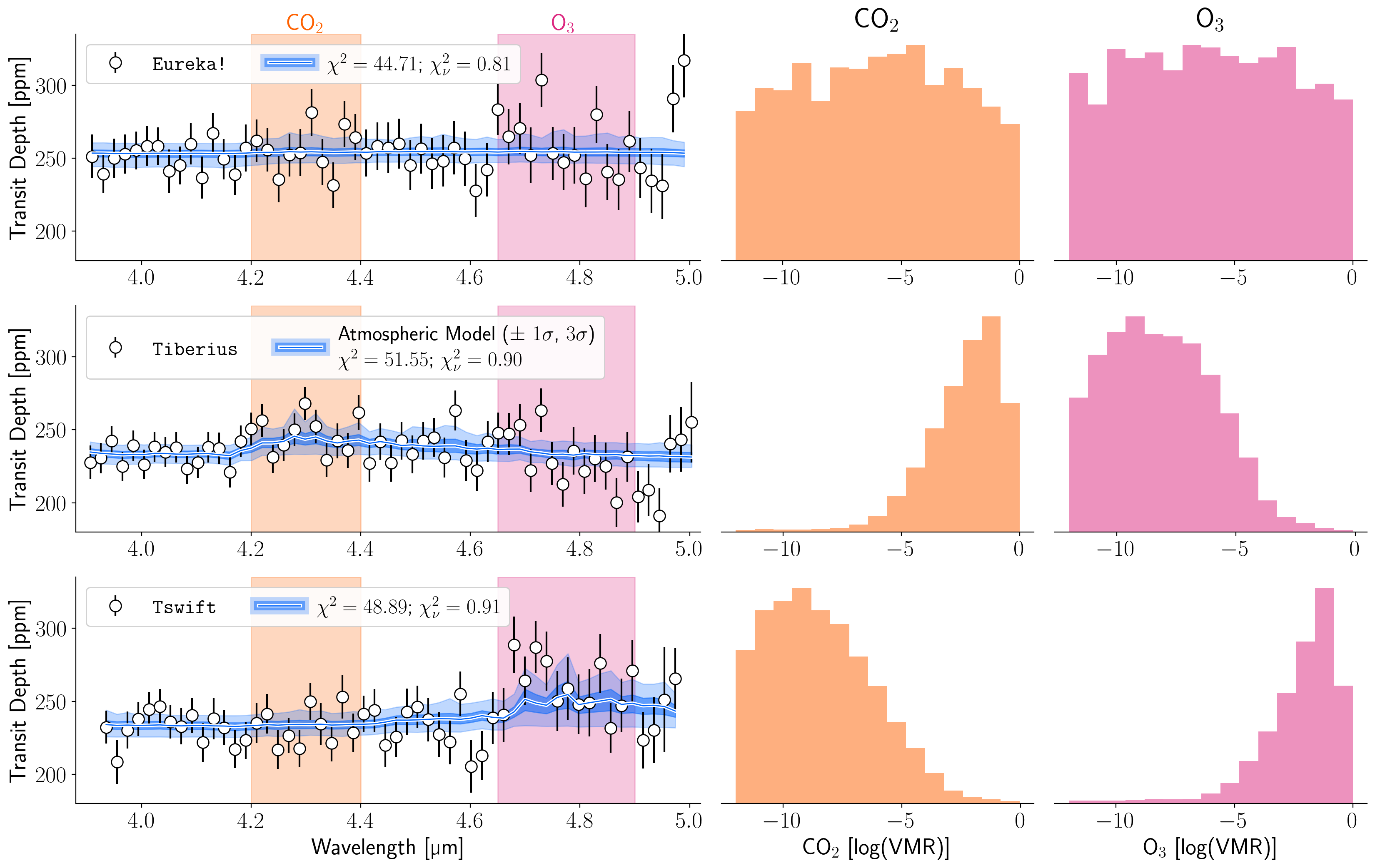}
    \caption{Comparison of retrieval results on data reduced by the \eureka, \tiberius, and \tswift pipelines, respectively (rows of subplots). For each reduction's retrieval, the median spectral fit is shown in the left panel, the 1D marginalized \ce{CO2} posterior is shown in the middle panel, and the 1D marginalized \ce{O3} posterior is shown in the right panel. Although either \ce{CO2} or \ce{O3} are weakly detected in retrievals of the \tiberius and \tswift spectra, respectively, (the \eureka result is unconstrained) these findings are incompatible with one another. Thus, relatively small differences in the reductions can cause comparatively large changes in the subsequent physical interpretation, including spurious molecular detections.}
    \label{fig:retrieval_compare}
\end{figure*}

% Retrieval results top-line, eureka
Figure \ref{fig:retrieval_compare} shows a summary of our retrieval results, comparing top-level findings for each data reduction pipeline. Figures \ref{fig:retrieval_eureka},  \ref{fig:retrieval_tiberius}, and \ref{fig:retrieval_tswift} (in Appendix \ref{sec:app:retrievals}) show results for the \eureka, \tiberius, and \tswift pipelines, respectively. Based on the $\chi^2_{\nu}$ values from the median retrieved spectrum, the retrievals are overfitting the spectrum, and therefore likely fitting noise. Nonetheless, they reveal important aspects of our observational sensitivity to the possible atmosphere of \planetname as well as the agreement/disagreement between the reduction pipelines. As previously discussed, the \eureka reduction appears the most flat and as a result the retrieval rules out a limited slice of parameter space that corresponds to only the thickest (reference pressures $P_0 > 10^{4}$ Pa) and most extended atmospheres, with high abundances of absorbing gases, that we explored. 

% Retrieval results, tiberius and tswift, the plot thickens 
On the other hand, the \tiberius and \tswift reductions tell a different story due to their low significance detection of a spectroscopic feature (Table \ref{tab:flat}), but they do not agree on the wavelength/molecule responsible for the feature, implying that these are not real astrophysical signals. The retrieval results for \tiberius favor high abundances of \ce{CO2} ($-2.1^{+1.2}_{-1.9}$ dex) with moderate abundances of \ce{N2O} ($-4.3^{+2.1}_{-3.9}$ dex) and low abundances of \ce{O3} ($-8.3^{+2.5}_{-2.7}$ dex; 2$\sigma$ upper limit of $-3.6$ dex). The retrieval results for \tswift show the opposite, favoring low abundances of \ce{CO2} ($-8.7^{+2.6}_{-2.1}$ dex; 2$\sigma$ upper limit of $-3.6$ dex) and \ce{N2O} ($-8.2^{+2.8}_{-2.4}$ dex; 2$\sigma$ upper limit of $-2.7$ dex) and high abundances of \ce{O3} ($-1.8^{+1.0}_{-1.8}$ dex). These \ce{O3} and \ce{CO2} constraints differ at ${\sim}1.5{\sigma}$. Our Bayesian model comparisons (using the Bayesian evidence ratios computed by \texttt{dynesty}) find that the \tiberius reduction implies a weak detection of \ce{CO2} (3.1$\sigma$), but no \ce{O3}, while the \tswift reduction instead leads to a weak detection of \ce{O3} (2.9$\sigma$), but no \ce{CO2}. Conversely, the \eureka reduction favors no detected gases at ${>}1\sigma$ significance. However, both \tiberius and \tswift retrievals levy similar constraints on the reference pressure, isothermal temperature, reference radius, planet mass, and mean molecular weight which are all in agreement at better than 1$\sigma$. These findings suggest that different noise instances with otherwise similar noise properties dominate the character of each reduced spectrum, as opposed to planetary characteristics. Furthermore, the statistical significance of results from any one pipeline must be interpreted as lower than the assessment of that spectrum in isolation would suggest.

\section{Discussion and Conclusions}
\label{sec:discussion}

We have measured the 3.9--5.0\,\microns transmission spectrum of the sub-Earth sized exoplanet GJ\,341b using JWST's NIRCam instrument. Our transmission spectrum is the result of co-adding the transmission spectra measured from three separate transit observations and represents the first observation of this planet's atmosphere. We reduced the data using three independent pipelines (\eureka, \tiberius and \tswift) and performed several tests to determine the significance of an atmosphere detection.

\subsection{The NIRCam/F444W data quality}

Our NIRCam/F444W data is dominated by an exponential ramp that varies between the three visits (Figure \ref{fig:lc_fits_eureka}). Despite fitting for this ramp, there is residual red noise in the white light curve data (Figure \ref{fig:av_all}). This red noise becomes apparent at timescales between $\sim 1-30$ minutes. We attempted to remove this red noise by using up to ten engineering telemetry timeseries (`mnemonics') in our systematics model (Figure \ref{fig:top10_mnemonics}). Despite encouraging signs, we were unable to significantly improve the residual red noise. We believe that the JWST engineering mnemonics do present an opportunity to detrend JWST spectrophotometry in general, however, we leave further attempts to do so to future work. Fortunately, this red noise is not significant in the spectroscopic light curves with our light curves reaching between 1.03 and 1.14$\times$ the noise limit (Fig. \ref{fig:error_comparison}). As a result, we were able to obtain a very precise transmission spectrum from our three coadded visits, with an average uncertainty of 15\,ppm in 0.02\,\micron\ bins. For these reasons, we recommend the use of NIRCam/F444W for studies of bright exoplanet systems. While NIRCam also provides us with short wavelength photometry (as used successfully in the ERS program, \citealp{Ahrer2023}), our extraction of this here shows that is not always beneficial.   

\subsection{GJ\,341b's transmission spectrum}

The main conclusion we draw from our high precision transmission spectrum of GJ\,341b is that the planet has not retained a low mean molecular weight atmosphere. This is consistent with expectations from the cosmic shoreline \citep{Zahnle2017} that small planets like GJ\,341b should be stripped of any primordial H/He-dominated atmosphere. Digging deeper, we find that the very high precision of our data, combined with the small expected feature amplitudes, leads our statistical tests, forward models and retrievals to slightly different conclusions.

In our null hypothesis tests, the \eureka spectrum was consistent with being flat to within $0.2\sigma$ while the \tiberius and \tswift spectra were less consistent at $1.3\sigma$ and $1.4\sigma$, respectively. However, we believe that the relative consistency with being flat is driven by the relative size of the uncertainties for each transmission spectrum. Repeating the null hypothesis tests but switching the uncertainties between the reductions, we find that \tiberius is inherently the flattest spectrum, followed by \eureka and then \tswift. 

Our Gaussian feature tests, using the default uncertainties for each reduction, find $2.2\sigma$ evidence \textit{against} a feature for \eureka and $2.3\sigma$ and $2.9\sigma$ evidence \textit{for} features in the \tiberius and \tswift spectra, respectively. However, our atmospheric retrievals find that different species are responsible for these candidate features: CO$_2$ in the \tiberius spectrum at $3.1\sigma$ and O$_3$ in the \tswift spectrum at $2.9\sigma$, with a relative difference between the CO$_2$ and O$_3$ abundance constraints of ${\sim}1.5\sigma$. These differences suggest that these `features' are not real astrophysical signals.

Our forward model analysis finds that each reduction's spectrum rejects a $< 285\times$ solar metallicity atmosphere to $\geq 3\sigma$, with the \tswift spectrum rejecting a $< 1000 \times$ solar metallicity atmosphere. Instead, the forward models find our transmission spectra are consistent with no atmosphere, a hazy atmosphere, or an atmosphere containing a species that does not have prominent molecular bands across the NIRCam/F444W bandpass, e.g. H$_2$O. Furthermore, CH$_4$-dominated atmospheres are ruled out to $1\sigma$--$3\sigma$ depending on the reduction, while CO$_2$ atmospheres are consistent with the \eureka and \tiberius spectra to $< 1\sigma$ and inconsistent with the \tswift spectrum to $> 2 \sigma$.

At the radius and equilibrium temperature of GJ\,341b, atmospheres dominated by CO$_2$, H$_2$O or O$_2$/O$_3$, resulting from post-runaway atmospheres, are all plausible and even predicted \citep{Luger2015,Gao2015,Schaefer2016, Lincowski2018,Kite2021,Krissansen-Totton2022}. Therefore, the single gas end member atmospheres that result from our forward modelling, and that are consistent with our retrievals, are in line with expectations. However, since a rocky exoplanet atmosphere has yet to be detected and characterized, the possibilities are largely restricted to theoretical predictions and the range of possible temperatures, pressures and gas abundance combinations that are consistent with the observations is vast. Furthermore the fact that each pipeline prefers a different gas demonstrates that the evidence for these gases is not due to real astrophysical signals.

Our study of this shallow transit (240\,ppm) demonstrates that small differences between pipelines' spectra can lead to differences in the conclusions drawn from statistical tests, atmospheric forward models and retrievals. Therefore, it is imperative that such studies use multiple independent pipelines, two or more visits, and several statistical tests to determine the robustness of future claims of atmospheric absorption for small exoplanets.  

\subsection{The Cosmic Shoreline for M-dwarfs}

\begin{figure}[t]
    \centering
    \includegraphics[scale=0.13]{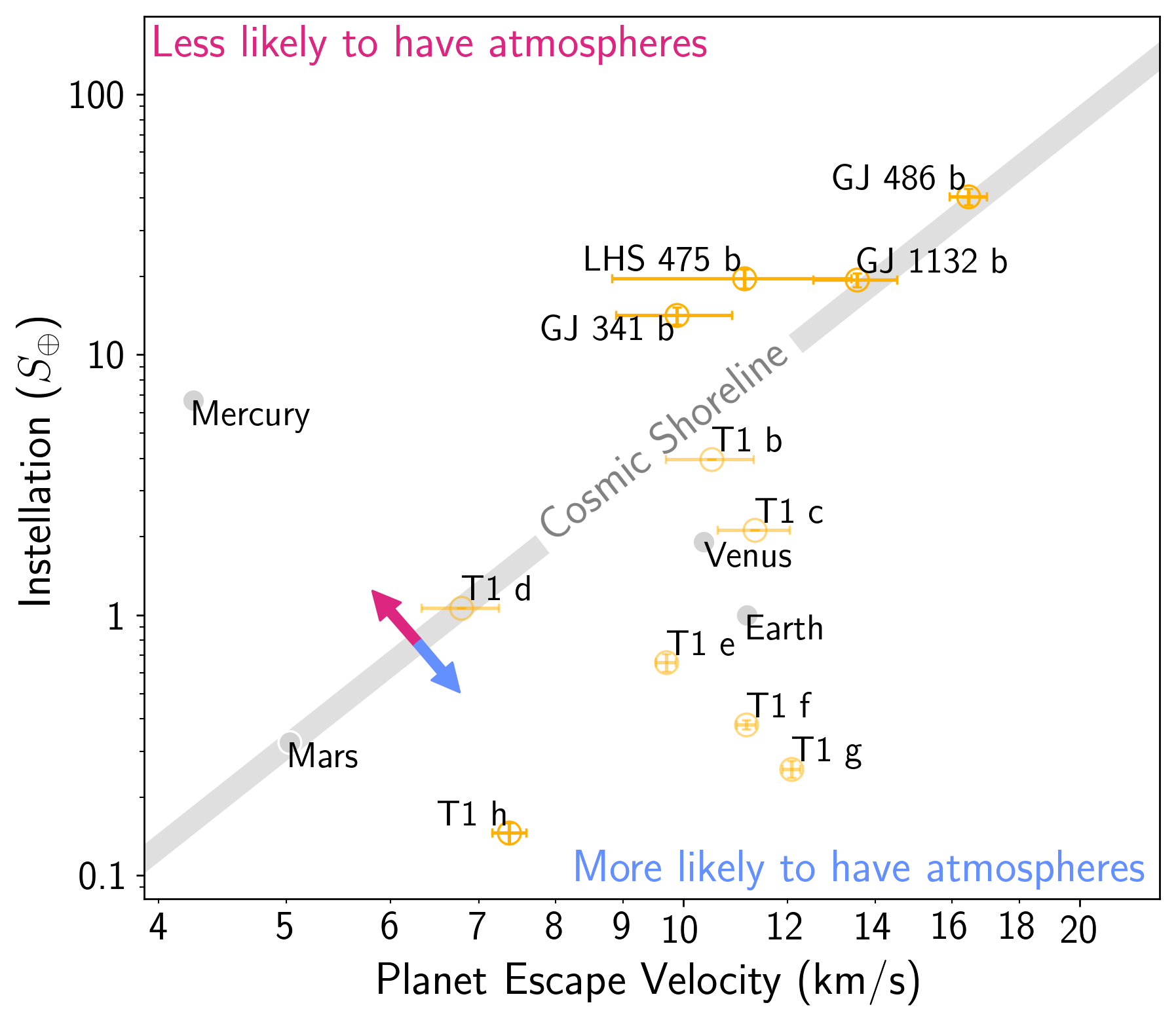}
    \caption{The M-dwarf Cosmic Shoreline \protect\citep{Zahnle2017} plotted in escape velocity--instellation space. Planets that sit below the shoreline (gray line) are more likely to have atmospheres and those above are less likely. The targets in our program are: GJ\,341b (this paper), LHS\,475b (\citetalias{Lustig-YaegerFu2023} \citeyear{Lustig-YaegerFu2023}), GJ\,486b (\citetalias{MoranStevenson2023} \citeyear{MoranStevenson2023}), GJ\,1132b (\citetalias{MayMacDonald2023} \citeyear{MayMacDonald2023}) and TRAPPIST-1h (Rustamkulov \& Peacock et al., in prep.). We include the other TRAPPIST-1 planets for reference.}
    \label{fig:cosmic_shoreline}
\end{figure}

GJ\,341b is the fourth planet to be studied as part of our reconnaissance program (PID:1981) to test the existence of an M-dwarf Cosmic Shoreline \citep{Zahnle2017}. Our results so far are consistent with expectations (Figure \ref{fig:cosmic_shoreline}). GJ\,341b (this work) and LHS\,475b (\citetalias{Lustig-YaegerFu2023} \citeyear{Lustig-YaegerFu2023}) sit on the dry/airless side of the shoreline, which is consistent with our transmission spectra. GJ\,486b (\citetalias{MoranStevenson2023} \citeyear{MoranStevenson2023}) and GJ\,1132b (\citetalias{MayMacDonald2023} \citeyear{MayMacDonald2023}) sit at the edge of the shoreline, where atmospheres and atmosphere-less are both possible outcomes. GJ\,486b shows water in its transmission spectrum, although this could be stellar contamination. GJ\,1132b paints an inconsistent picture between the two visits' transits, making inferences more challenging. Our remaining target (TRAPPIST-1h) is likely to have an atmosphere based on expectations from the shoreline. Our upcoming transit observations will shed light on the existence of its atmosphere. Taken together, our program has demonstrated the tight constraints that can be placed on Earth-sized exoplanet atmospheres.

Looking beyond our program to other JWST studies of Earth-sized exoplanets around M-dwarfs, TRAPPIST-1b and c are both consistent with having no atmosphere \citep{Greene2023,Lim2023,Zieba2023}. TRAPPIST-1b is close to the Cosmic Shoreline and thus the lack of an atmosphere is not in tension with expectations. TRAPPIST-1c, on the other hand, sits on the atmosphere side. However, the shoreline presented in Figure \ref{fig:cosmic_shoreline} is plotted in terms of the relative instellation. The XUV cosmic shoreline would sit below this \citep{Zahnle2017} and may be consistent with the TRAPPIST-1c result. If the results of no atmospheres for these planets hold, then this could be emerging evidence that XUV dominates the cosmic shoreline in these late M-dwarf systems, consistent with past escape predictions \citep{Wheatley2017,Dong2018}. Only by combining results from wider and deeper surveys, including ongoing JWST programs (e.g., PID: 2512), will we be able to confidently determine the boundaries of a cosmic shoreline for M-dwarfs, if indeed one exists.

% \section{Conclusions}
% \label{sec:conclusions}

%% Also note that the akcnowlodgment environment does not support long amounts of text. If you have a lot of people and institutions to acknowledge, do not use this command. Instead, create a new \section{Acknowledgments}.
\begin{acknowledgments}

We thank the anonymous reviewer for their constructive comments that improved the manuscript. We also thank Victoria DiTomasso for sharing the system parameters of GJ\,341b from her joint analysis of the radial velocity and TESS data. JK acknowledges financial support from Imperial College London through an Imperial College Research Fellowship grant.

\end{acknowledgments}


\vspace{5mm}
\facilities{JWST(NIRCam)}

All of the data presented in this paper were obtained from the Mikulski Archive for Space Telescopes (MAST) at the Space Telescope Science Institute. The specific observations analyzed can be accessed via \dataset[DOI]{https://doi.org/10.17909/kt9e-h986}.


\software{ Astropy \citep{astropy,astropy2}, \texttt{batman} \citep{batman2015}, \texttt{CHIMERA} \citep{Line2013b,Line2014-C/O}, \texttt{Dynesty} \citep{Speagle2020}, \texttt{emcee} \citep{emcee2013}, \eureka \citep{Eureka2022}, ExoCTK \citep{exoctk}, ExoTiC-LD \citep{david_grant_2022_7437681} \texttt{FIREFLy} \citep{Rustamkulov2022}, \texttt{Forecaster} \citep{Chen2017}, IPython \citep{ipython}, \texttt{jwst} \citep{jwstpipeline2022}, Matplotlib \citep{matplotlib}, NumPy \citep{numpy, numpynew}, \texttt{PHOENIX} \citep{Allard2012} \texttt{PICASO} \citep{Batalha2019}, PyMC3 \citep{Salvatier2016}, pysynphot\citep{STScIDevelopmentTeam2013}, SciPy \citep{scipy}, \texttt{smarter} \citep{Lustig-Yaeger2022, Lustig-Yaeger2023earth}, 
\texttt{Tiberius} \citep{Kirk2019,Kirk2021}}



\appendix

\section{The reduction and fitting approaches taken by \eureka, \tiberius and \tswift}

% \rotatebox{90}{% Rotate the minipage by the desired angle
%     \begin{minipage}{1.1\textwidth}
\begin{table*}
\centering
    \caption{Comparison of the different reduction and analysis choices used by the three pipelines, along with the resulting noise properties.}
    \label{tab:reduction_choices}
    \begin{tabular}{l|c|c|c} \hline
         & \eureka & \tiberius & \tswift \\ \hline \hline
         \texttt{jwst} pipeline version & 1.8.2 & 1.8.2 & 1.8.1 \\ \hline
        % Reference files used & & mask\_0063, saturation\_0097, superbias\_0142, & \\ 
        % & & linearity\_0052, dark\_0363, readnoise\_0199 & \\ \hline
        Group-level & col.-by-col. mean & None & None \\
        background subtraction?$^{a}$ & \& row-by-row mean & & \\ \hline
        Outlier detection & Full-frame: 4$\sigma$ & Full-frame: 5$\sigma$ & Full-frame: 5$\sigma$ \\
        / sigma clipping & Light curves: 3.5$\sigma$ & Light curves: 4$\sigma$ & Light curves: 5$\sigma$ \\ \hline
        Pixel supersampling? & No & Yes, $10\times$ & No \\ \hline
        Standard / optimal extraction & Optimal & Standard & Standard \\ \hline
        Aperture full-width & 6 & 5 & 6 \\
        (pixels) & & & \\ \hline
        Integration-level & None & col.-by-col. median & row-by-row median \\
        background subtraction? & & & \& col.-by-col. median \\ \hline
        Spectral range extracted & 3.90--5.00 & 3.89--5.02 & 3.92--4.99 \\
        (\microns) & & & \\ \hline
        Wavelength bin width & 0.02 & 0.02 & 0.02 \\ 
        (\microns) & & & \\ \hline
        Number of integrations trimmed & First 300 from WLC$^b$  & 0 from WLC & None \\
        & First 300 from SLC$^c$ & First 600 from SLC & \\ \hline
        WLC model & Exponential $\times$ linear ramp & Exponential $\times$ linear ramp & Exponential $\times$ linear ramp \\
        & & & (amplifier-dependent) \\ \hline
        SLC model & Exponential $\times$ linear ramp & Linear ramp & Exponential $\times$ linear ramp \\ \hline
        Common mode correction used? & No & Yes & Yes \\ 
        & & (visit dependent) & (visit \& amplifier dependent) \\ \hline
        WLC RMS & 223 & 292 & 303 \\ 
        (ppm) &  &  & \\ \hline
        SLC RMS & 716 & 645 & 661 \\
        (ppm) &  &  &  \\ \hline
        Transit depth precision & 17 & 13 & 15 \\
        (ppm) &  &  &  \\ \hline
        \multicolumn{4}{l}{$^{a}$The goal of group-level background subtraction is to remove 1/f noise.} \\
        \multicolumn{4}{l}{$^{b}$WLC corresponds to the white light curves.} \\
        \multicolumn{4}{l}{$^{c}$SLC corresponds to the spectroscopic light curves.} 
    \end{tabular}
    \end{table*}
    % \end{minipage}}

\section{The light curves and residuals from each pipeline}

\begin{figure*}
    \centering
    \includegraphics[width=1.0\textwidth]{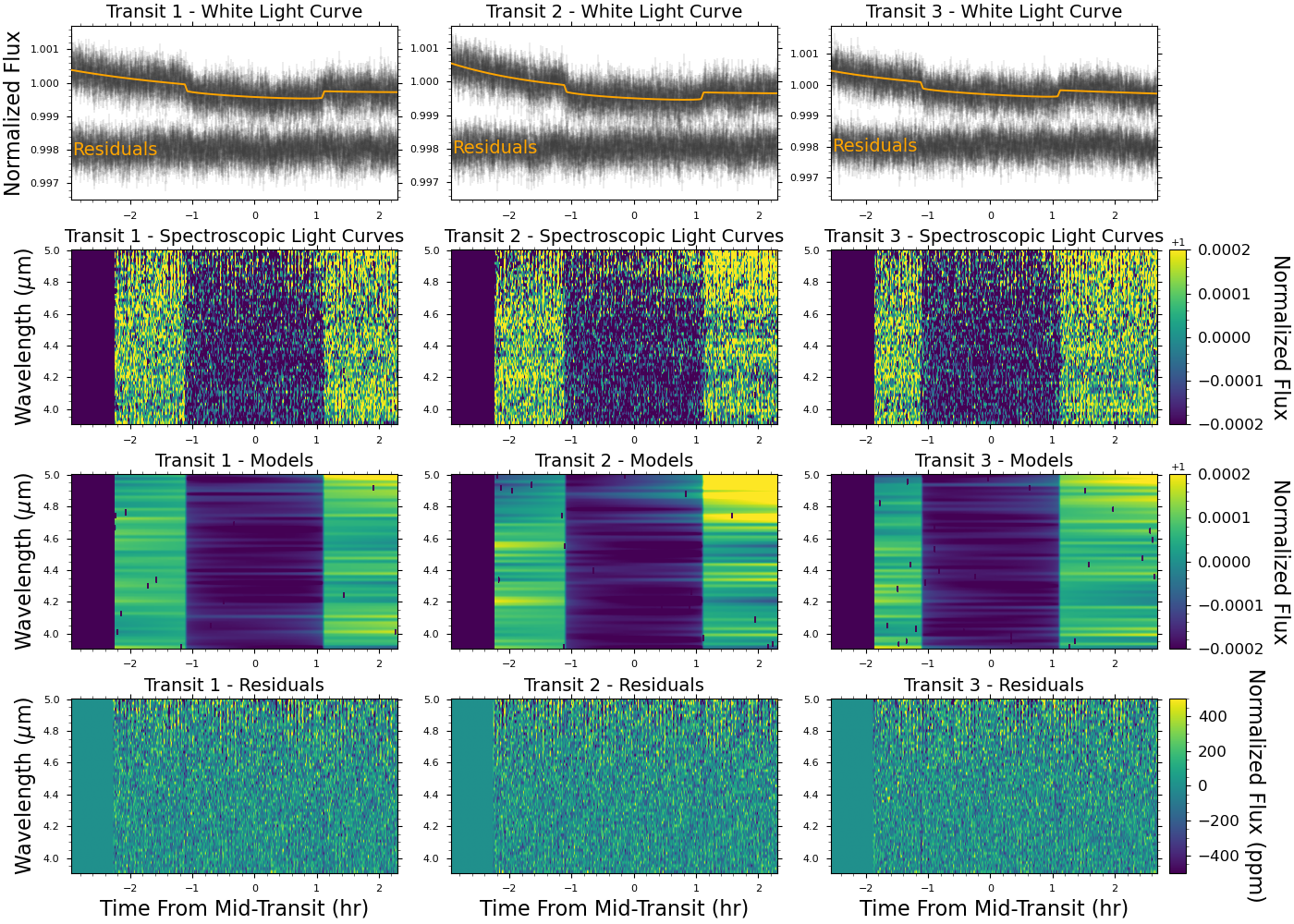}
    \caption{The light curves and best fit models resulting from the \tiberius pipeline. The columns show, from left to right: transit 1, transit 2, transit 3. The rows show, from top to bottom: 1) the white light curve and residuals, 2) the spectroscopic light curves, 3) the best-fit models to the spectroscopic light curves, 4) the residuals from the spectroscopic light curve fitting. The vertical band at the left hand edge of the panels in rows 2--4 correspond to the 600 integrations that were masked during the spectroscopic light curve fitting stage.}
    \label{fig:lc_fits_tiberius} 
\end{figure*}

\begin{figure*}
    \centering
    \includegraphics[width=1.0\textwidth]{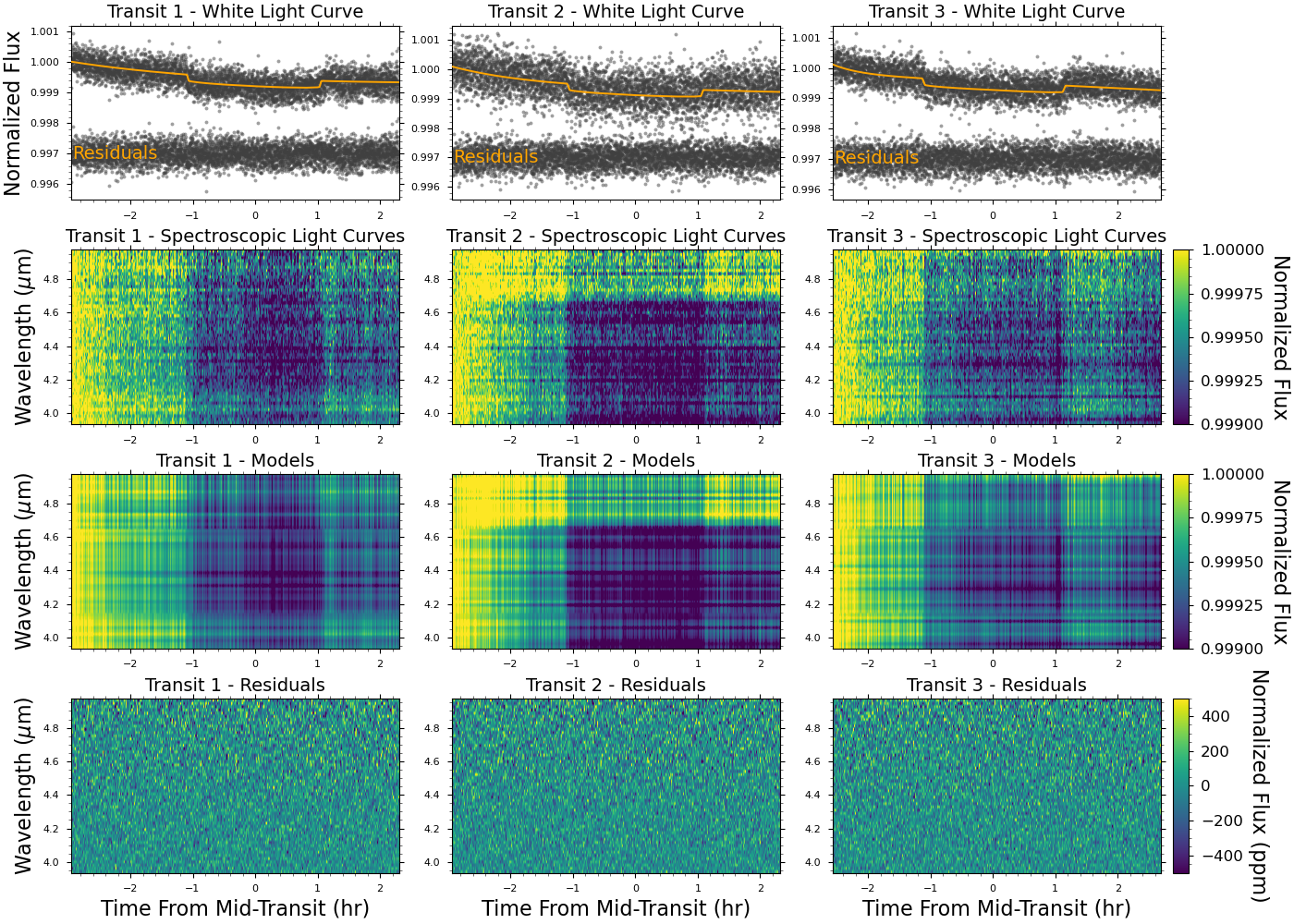}
    \caption{The light curves and best fit models resulting from the \tswift pipeline. The columns show, from left to right: transit 1, transit 2, transit 3. The rows show, from top to bottom: 1) the white light curve and residuals, 2) the spectroscopic light curves, 3) the best-fit models to the spectroscopic light curves, 4) the residuals from the spectroscopic light curve fitting.  }
    \label{fig:lc_fits_tswift} 
\end{figure*}

\section{Transmission spectra measured for each visit}

\begin{figure}[ht!]
    \centering
    \includegraphics[scale=0.5]{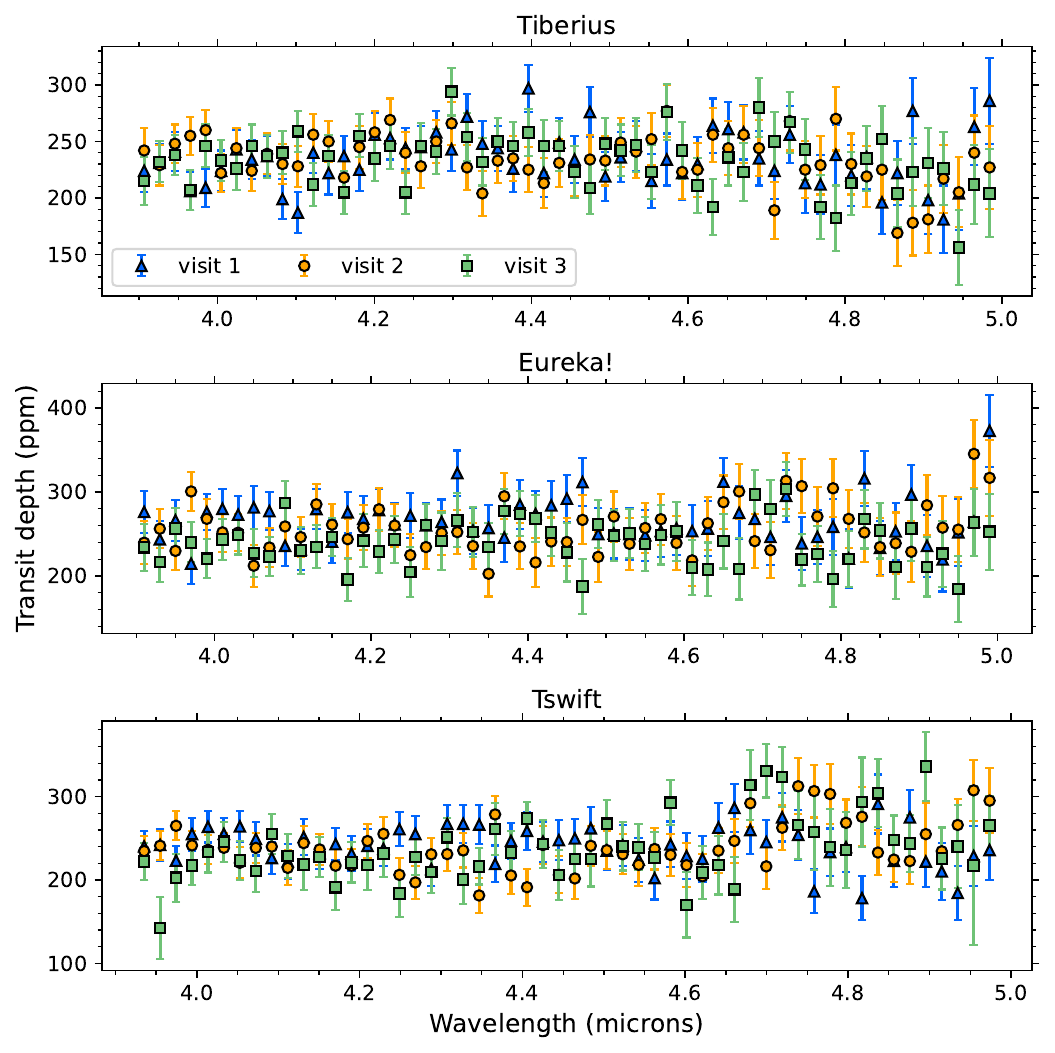}
    \caption{The transmission spectra for visit 1 (blue), visit 2 (orange) and visit 3 (green) as measured by \tiberius (top panel), \eureka (middle panel) and \tswift (bottom panel).}
    \label{fig:visit_spectra}
\end{figure}

The transmission spectra measured for each visit by each reduction are shown in Figure \ref{fig:visit_spectra}.

\section{Forward models for all reductions} 

\begin{figure*}
    \centering
    \includegraphics[scale=0.475]{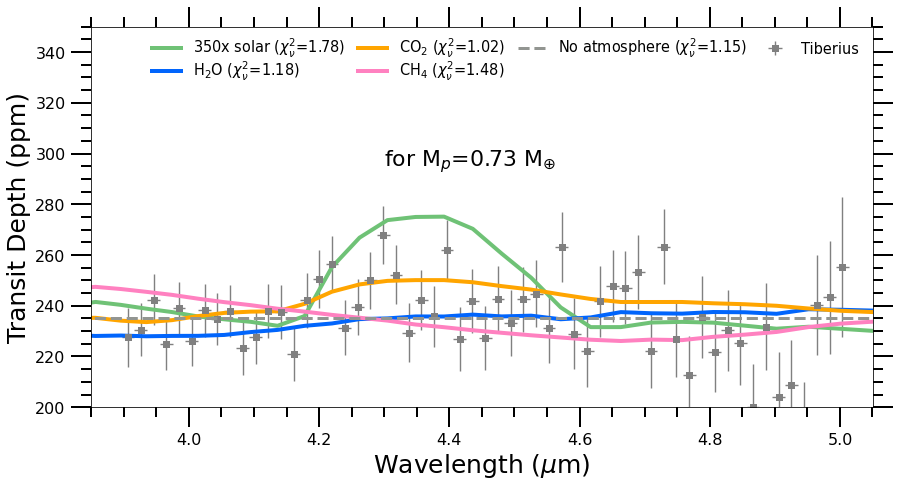}
    \caption{Atmospheric forward models compared to the \tiberius reduction from {\jwst}/NIRCam.}
    \label{fig:forward_models_tiberius}
\end{figure*}

\begin{figure*}
    \centering
    \includegraphics[scale=0.475]{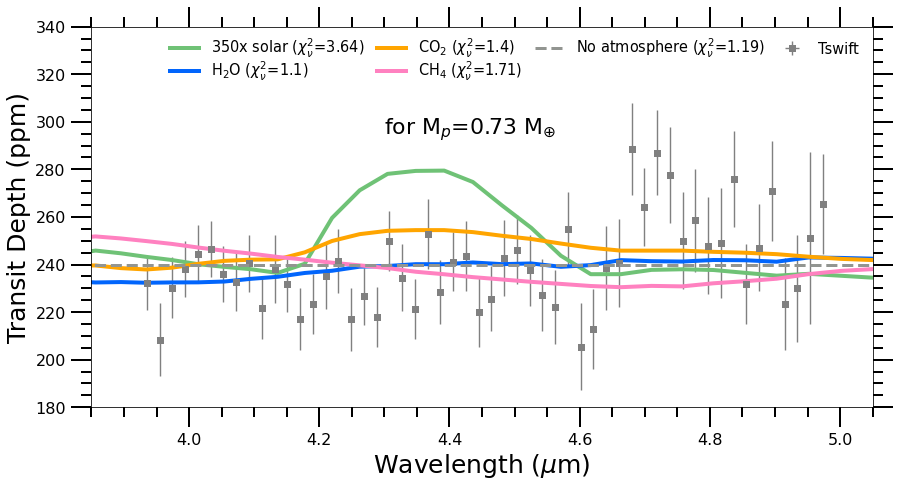}
    \caption{Atmospheric forward models compared to the \tswift reduction from {\jwst}/NIRCam.}
    \label{fig:forward_models_tswift}
\end{figure*}

\section{Retrieval Posteriors for all Reductions} \label{sec:app:retrievals}

\begin{figure*}
    \centering
    \includegraphics[width=0.95\textwidth]{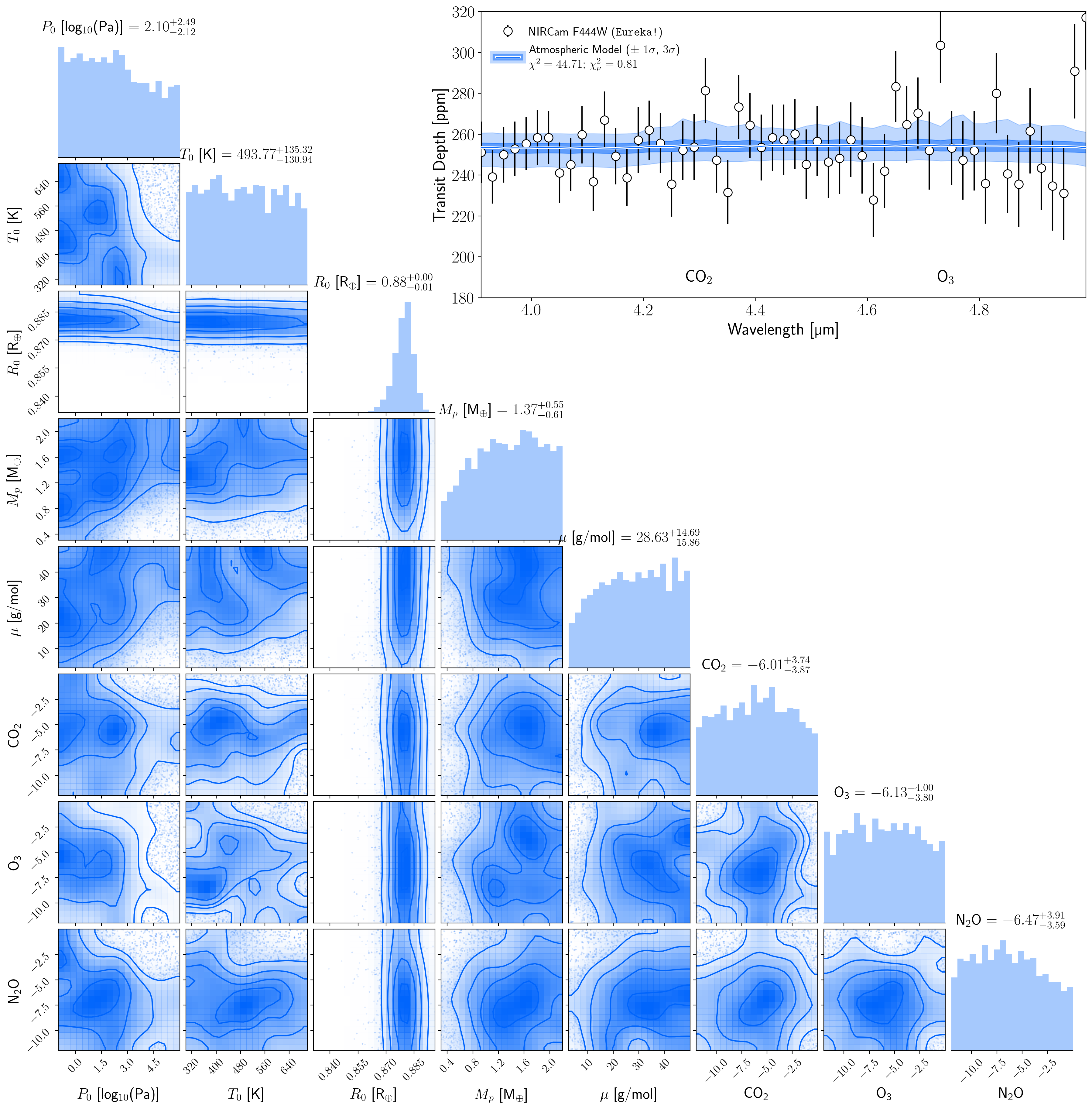}
    \caption{Atmospheric retrieval posteriors (lower left corner pairs) and range of spectral fits (upper right) from our coadded NIRCam observations of \planetname reduced with \eureka. A subset of model parameters and covariances are shown, with \ce{H2O}, \ce{CH4}, and \ce{CO} not displayed (nor constrained) for clarity. Analogous results are shown for the \tiberius and \tswift data reduction pipelines in Figures \ref{fig:retrieval_tiberius} and \ref{fig:retrieval_tswift}, respectively.}
    \label{fig:retrieval_eureka}
\end{figure*}

\begin{figure*}
    \centering
    \includegraphics[width=0.95\textwidth]{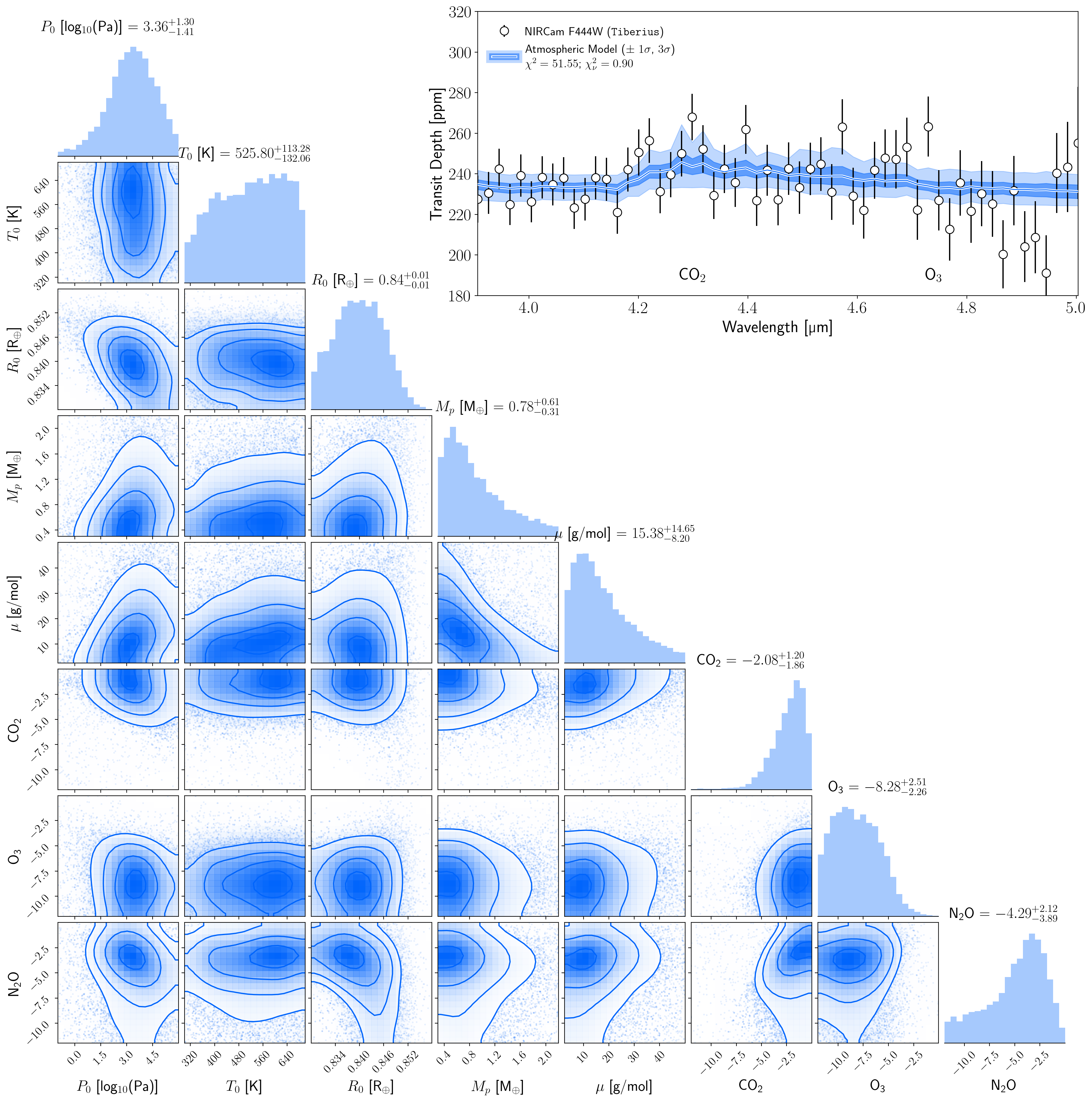}
    \caption{Atmospheric retrieval posteriors (lower left corner pairs) and range of spectral fits (upper right) from our coadded NIRCam observations of \planetname reduced with \tiberius. A subset of model parameters and covariances are shown, with \ce{H2O}, \ce{CH4}, and \ce{CO} not displayed (nor constrained) for clarity. }
    \label{fig:retrieval_tiberius}
\end{figure*}

\begin{figure*}
    \centering
    \includegraphics[width=0.95\textwidth]{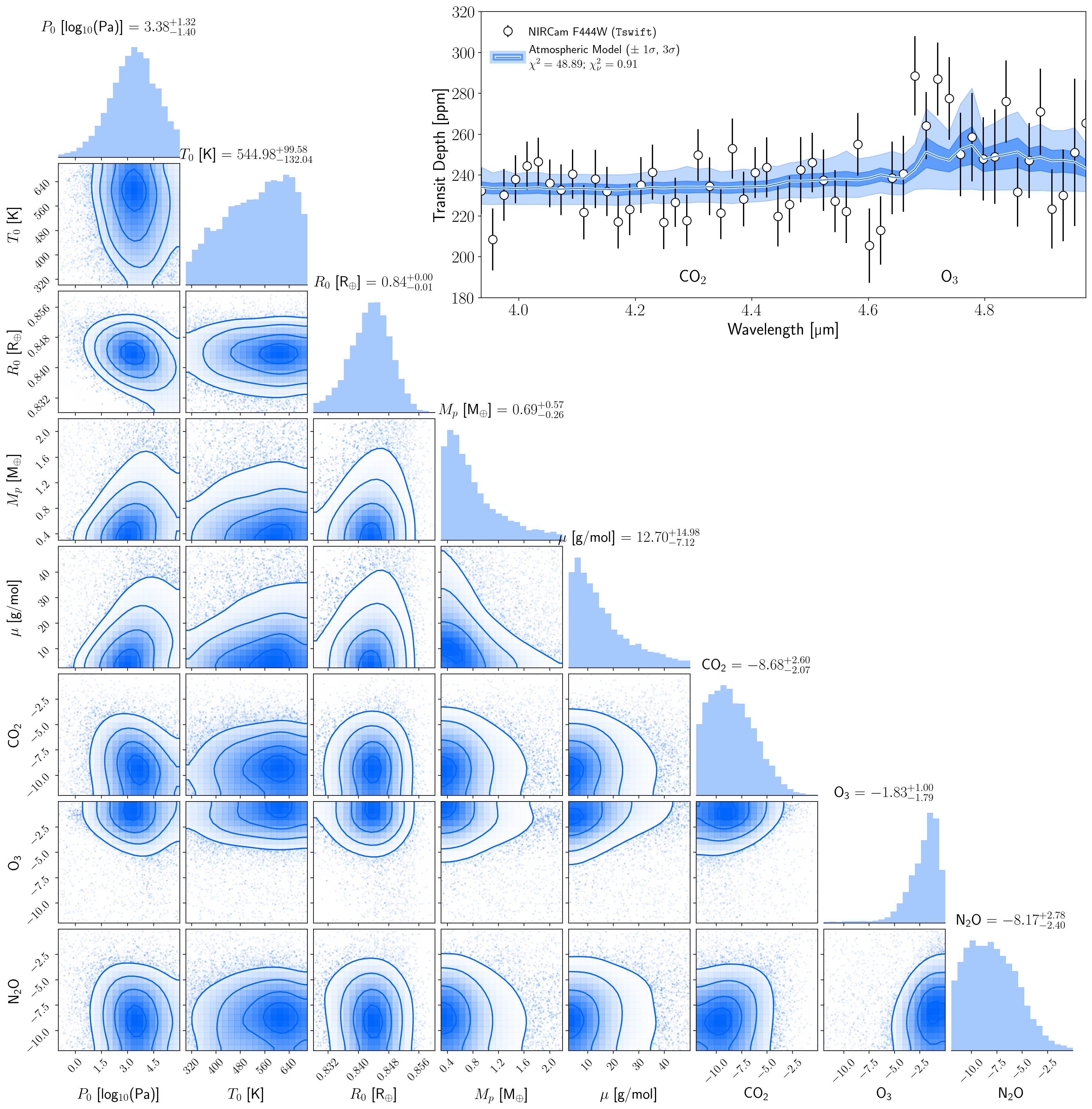}
    \caption{Atmospheric retrieval posteriors (lower left corner pairs) and range of spectral fits (upper right) from our coadded NIRCam observations of \planetname reduced with \tswift. A subset of model parameters and covariances are shown, with \ce{H2O}, \ce{CH4}, and \ce{CO} not displayed (nor constrained) for clarity. }
    \label{fig:retrieval_tswift}
\end{figure*}

\bibliography{NoAir_GJ341b_AJ}{}

\begin{thebibliography}{}
\expandafter\ifx\csname natexlab\endcsname\relax\def\natexlab#1{#1}\fi
\providecommand{\url}[1]{\href{#1}{#1}}
\providecommand{\dodoi}[1]{doi:~\href{http://doi.org/#1}{\nolinkurl{#1}}}
\providecommand{\doeprint}[1]{\href{http://ascl.net/#1}{\nolinkurl{http://ascl.net/#1}}}
\providecommand{\doarXiv}[1]{\href{https://arxiv.org/abs/#1}{\nolinkurl{https://arxiv.org/abs/#1}}}

\bibitem[{{Ahrer} {et~al.}(2023){Ahrer}, {Stevenson}, {Mansfield}, {Moran},
  {Brande}, {Morello}, {Murray}, {Nikolov}, {Petit dit de la Roche},
  {Schlawin}, {Wheatley}, {Zieba}, {Batalha}, {Damiano}, {Goyal}, {Lendl},
  {Lothringer}, {Mukherjee}, {Ohno}, {Batalha}, {Battley}, {Bean}, {Beatty},
  {Benneke}, {Berta-Thompson}, {Carter}, {Cubillos}, {Daylan}, {Espinoza},
  {Gao}, {Gibson}, {Gill}, {Harrington}, {Hu}, {Kreidberg}, {Lewis}, {Line},
  {L{\'o}pez-Morales}, {Parmentier}, {Powell}, {Sing}, {Tsai}, {Wakeford},
  {Welbanks}, {Alam}, {Alderson}, {Allen}, {Anderson}, {Barstow}, {Bayliss},
  {Bell}, {Blecic}, {Bryant}, {Burleigh}, {Carone}, {Casewell}, {Changeat},
  {Chubb}, {Crossfield}, {Crouzet}, {Decin}, {D{\'e}sert}, {Feinstein},
  {Flagg}, {Fortney}, {Gizis}, {Heng}, {Iro}, {Kempton}, {Kendrew}, {Kirk},
  {Knutson}, {Komacek}, {Lagage}, {Leconte}, {Lustig-Yaeger}, {MacDonald},
  {Mancini}, {May}, {Mayne}, {Miguel}, {Mikal-Evans}, {Molaverdikhani},
  {Palle}, {Piaulet}, {Rackham}, {Redfield}, {Rogers}, {Roy}, {Rustamkulov},
  {Shkolnik}, {Sotzen}, {Taylor}, {Tremblin}, {Tucker}, {Turner}, {de
  Val-Borro}, {Venot}, \& {Zhang}}]{Ahrer2023}
{Ahrer}, E.-M., {Stevenson}, K.~B., {Mansfield}, M., {et~al.} 2023, \nat, 614,
  653, \dodoi{10.1038/s41586-022-05590-4}

\bibitem[{{Alderson} {et~al.}(2023){Alderson}, {Wakeford}, {Alam}, {Batalha},
  {Lothringer}, {Adams Redai}, {Barat}, {Brande}, {Damiano}, {Daylan},
  {Espinoza}, {Flagg}, {Goyal}, {Grant}, {Hu}, {Inglis}, {Lee}, {Mikal-Evans},
  {Ramos-Rosado}, {Roy}, {Wallack}, {Batalha}, {Bean}, {Benneke},
  {Berta-Thompson}, {Carter}, {Changeat}, {Col{\'o}n}, {Crossfield},
  {D{\'e}sert}, {Foreman-Mackey}, {Gibson}, {Kreidberg}, {Line},
  {L{\'o}pez-Morales}, {Molaverdikhani}, {Moran}, {Morello}, {Moses},
  {Mukherjee}, {Schlawin}, {Sing}, {Stevenson}, {Taylor}, {Aggarwal}, {Ahrer},
  {Allen}, {Barstow}, {Bell}, {Blecic}, {Casewell}, {Chubb}, {Crouzet},
  {Cubillos}, {Decin}, {Feinstein}, {Fortney}, {Harrington}, {Heng}, {Iro},
  {Kempton}, {Kirk}, {Knutson}, {Krick}, {Leconte}, {Lendl}, {MacDonald},
  {Mancini}, {Mansfield}, {May}, {Mayne}, {Miguel}, {Nikolov}, {Ohno}, {Palle},
  {Parmentier}, {Petit dit de la Roche}, {Piaulet}, {Powell}, {Rackham},
  {Redfield}, {Rogers}, {Rustamkulov}, {Tan}, {Tremblin}, {Tsai}, {Turner}, {de
  Val-Borro}, {Venot}, {Welbanks}, {Wheatley}, \& {Zhang}}]{Alderson2023}
{Alderson}, L., {Wakeford}, H.~R., {Alam}, M.~K., {et~al.} 2023, \nat, 614,
  664, \dodoi{10.1038/s41586-022-05591-3}

\bibitem[{{Allard} {et~al.}(2012){Allard}, {Homeier}, \&
  {Freytag}}]{Allard2012}
{Allard}, F., {Homeier}, D., \& {Freytag}, B. 2012, Philosophical Transactions
  of the Royal Society of London Series A, 370, 2765,
  \dodoi{10.1098/rsta.2011.0269}

\bibitem[{{Astropy Collaboration} {et~al.}(2013){Astropy Collaboration},
  {Robitaille}, {Tollerud}, {Greenfield}, {Droettboom}, {Bray}, {Aldcroft},
  {Davis}, {Ginsburg}, {Price-Whelan}, {Kerzendorf}, {Conley}, {Crighton},
  {Barbary}, {Muna}, {Ferguson}, {Grollier}, {Parikh}, {Nair}, {Unther},
  {Deil}, {Woillez}, {Conseil}, {Kramer}, {Turner}, {Singer}, {Fox}, {Weaver},
  {Zabalza}, {Edwards}, {Azalee Bostroem}, {Burke}, {Casey}, {Crawford},
  {Dencheva}, {Ely}, {Jenness}, {Labrie}, {Lim}, {Pierfederici}, {Pontzen},
  {Ptak}, {Refsdal}, {Servillat}, \& {Streicher}}]{astropy}
{Astropy Collaboration}, {Robitaille}, T.~P., {Tollerud}, E.~J., {et~al.} 2013,
  \aap, 558, A33, \dodoi{10.1051/0004-6361/201322068}

\bibitem[{{Astropy Collaboration} {et~al.}(2018){Astropy Collaboration},
  {Price-Whelan}, {Sip{\H{o}}cz}, {G{\"u}nther}, {Lim}, {Crawford}, {Conseil},
  {Shupe}, {Craig}, {Dencheva}, {Ginsburg}, {Vand erPlas}, {Bradley},
  {P{\'e}rez-Su{\'a}rez}, {de Val-Borro}, {Aldcroft}, {Cruz}, {Robitaille},
  {Tollerud}, {Ardelean}, {Babej}, {Bach}, {Bachetti}, {Bakanov}, {Bamford},
  {Barentsen}, {Barmby}, {Baumbach}, {Berry}, {Biscani}, {Boquien}, {Bostroem},
  {Bouma}, {Brammer}, {Bray}, {Breytenbach}, {Buddelmeijer}, {Burke},
  {Calderone}, {Cano Rodr{\'\i}guez}, {Cara}, {Cardoso}, {Cheedella}, {Copin},
  {Corrales}, {Crichton}, {D'Avella}, {Deil}, {Depagne}, {Dietrich}, {Donath},
  {Droettboom}, {Earl}, {Erben}, {Fabbro}, {Ferreira}, {Finethy}, {Fox},
  {Garrison}, {Gibbons}, {Goldstein}, {Gommers}, {Greco}, {Greenfield},
  {Groener}, {Grollier}, {Hagen}, {Hirst}, {Homeier}, {Horton}, {Hosseinzadeh},
  {Hu}, {Hunkeler}, {Ivezi{\'c}}, {Jain}, {Jenness}, {Kanarek}, {Kendrew},
  {Kern}, {Kerzendorf}, {Khvalko}, {King}, {Kirkby}, {Kulkarni}, {Kumar},
  {Lee}, {Lenz}, {Littlefair}, {Ma}, {Macleod}, {Mastropietro}, {McCully},
  {Montagnac}, {Morris}, {Mueller}, {Mumford}, {Muna}, {Murphy}, {Nelson},
  {Nguyen}, {Ninan}, {N{\"o}the}, {Ogaz}, {Oh}, {Parejko}, {Parley}, {Pascual},
  {Patil}, {Patil}, {Plunkett}, {Prochaska}, {Rastogi}, {Reddy Janga},
  {Sabater}, {Sakurikar}, {Seifert}, {Sherbert}, {Sherwood-Taylor}, {Shih},
  {Sick}, {Silbiger}, {Singanamalla}, {Singer}, {Sladen}, {Sooley},
  {Sornarajah}, {Streicher}, {Teuben}, {Thomas}, {Tremblay}, {Turner},
  {Terr{\'o}n}, {van Kerkwijk}, {de la Vega}, {Watkins}, {Weaver}, {Whitmore},
  {Woillez}, {Zabalza}, \& {Astropy Contributors}}]{astropy2}
{Astropy Collaboration}, {Price-Whelan}, A.~M., {Sip{\H{o}}cz}, B.~M., {et~al.}
  2018, \aj, 156, 123, \dodoi{10.3847/1538-3881/aabc4f}

\bibitem[{Batalha {et~al.}(2020)Batalha, Freedman, Gharib-Nezhad, \&
  Lupu}]{batalha2020}
Batalha, N., Freedman, R., Gharib-Nezhad, E., \& Lupu, R. 2020, Resampled
  Opacity Database for PICASO, 2.0,  Zenodo, \dodoi{10.5281/zenodo.6928501}

\bibitem[{{Batalha} {et~al.}(2019){Batalha}, {Lewis}, {Fortney}, {Batalha},
  {Kempton}, {Lewis}, \& {Line}}]{Batalha2019}
{Batalha}, N.~E., {Lewis}, T., {Fortney}, J.~J., {et~al.} 2019, \apjl, 885,
  L25, \dodoi{10.3847/2041-8213/ab4909}

\bibitem[{{Batalha} {et~al.}(2017){Batalha}, {Mandell}, {Pontoppidan},
  {Stevenson}, {Lewis}, {Kalirai}, {Greene}, {Albert}, {Nielsen}, \&
  {Earl}}]{Batalha2017}
{Batalha}, N.~E., {Mandell}, A., {Pontoppidan}, K., {et~al.} 2017, ArXiv
  e-prints.
\newblock \doarXiv{1702.01820}

\bibitem[{{Bell} {et~al.}(2022){Bell}, {Ahrer}, {Brande}, {Carter},
  {Feinstein}, {Guzman Caloca}, {Mansfield}, {Zieba}, {Piaulet}, {Benneke},
  {Filippazzo}, {May}, {Roy}, {Kreidberg}, \& {Stevenson}}]{Eureka2022}
{Bell}, T.~J., {Ahrer}, E.-M., {Brande}, J., {et~al.} 2022, arXiv e-prints,
  arXiv:2207.03585.
\newblock \doarXiv{2207.03585}

\bibitem[{{Benneke} \& {Seager}(2012)}]{Benneke2012}
{Benneke}, B., \& {Seager}, S. 2012, \apj, 753, 100,
  \dodoi{10.1088/0004-637X/753/2/100}

\bibitem[{{B{\'e}tr{\'e}mieux} \& {Kaltenegger}(2013)}]{Betremieux2013}
{B{\'e}tr{\'e}mieux}, Y., \& {Kaltenegger}, L. 2013, \apjl, 772, L31,
  \dodoi{10.1088/2041-8205/772/2/L31}

\bibitem[{Bourque {et~al.}(2021)Bourque, Espinoza, Filippazzo, Fix, King,
  Martlin, Medina, Batalha, Fox, Fowler, Fraine, Hill, Lewis, Stevenson,
  Valenti, \& Wakeford}]{exoctk}
Bourque, M., Espinoza, N., Filippazzo, J., {et~al.} 2021, The Exoplanet
  Characterization Toolkit (ExoCTK), 1.0.0,  Zenodo,
  \dodoi{10.5281/zenodo.4556063}

\bibitem[{Bushouse {et~al.}(2022)Bushouse, Eisenhamer, Dencheva, Davies,
  Greenfield, Morrison, Hodge, Simon, Grumm, Droettboom, Slavich, Sosey, Pauly,
  Miller, Jedrzejewski, Hack, Davis, Crawford, Law, Gordon, Regan, Cara,
  MacDonald, Bradley, Shanahan, Jamieson, Teodoro, \&
  Williams}]{jwstpipeline2022}
Bushouse, H., Eisenhamer, J., Dencheva, N., {et~al.} 2022, JWST Calibration
  Pipeline, 1.8.2,  Zenodo, \dodoi{10.5281/zenodo.7325378}

\bibitem[{{Charbonneau} \& {Deming}(2007)}]{Charbonneau2007}
{Charbonneau}, D., \& {Deming}, D. 2007, arXiv e-prints, arXiv:0706.1047,
  \dodoi{10.48550/arXiv.0706.1047}

\bibitem[{{Chen} \& {Kipping}(2017)}]{Chen2017}
{Chen}, J., \& {Kipping}, D. 2017, \apj, 834, 17,
  \dodoi{10.3847/1538-4357/834/1/17}

\bibitem[{{de Wit} {et~al.}(2016){de Wit}, {Wakeford}, {Gillon}, {Lewis},
  {Valenti}, {Demory}, {Burgasser}, {Burdanov}, {Delrez}, {Jehin}, {Lederer},
  {Queloz}, {Triaud}, \& {Van Grootel}}]{deWit2016}
{de Wit}, J., {Wakeford}, H.~R., {Gillon}, M., {et~al.} 2016, \nat, 537, 69,
  \dodoi{10.1038/nature18641}

\bibitem[{{de Wit} {et~al.}(2018){de Wit}, {Wakeford}, {Lewis}, {Delrez},
  {Gillon}, {Selsis}, {Leconte}, {Demory}, {Bolmont}, {Bourrier}, {Burgasser},
  {Grimm}, {Jehin}, {Lederer}, {Owen}, {Stamenkovi{\'c}}, \&
  {Triaud}}]{deWit2018}
{de Wit}, J., {Wakeford}, H.~R., {Lewis}, N.~K., {et~al.} 2018, Nature
  Astronomy, 2, 214, \dodoi{10.1038/s41550-017-0374-z}

\bibitem[{{Diamond-Lowe} {et~al.}(2018){Diamond-Lowe}, {Berta-Thompson},
  {Charbonneau}, \& {Kempton}}]{DiamondLowe2018}
{Diamond-Lowe}, H., {Berta-Thompson}, Z., {Charbonneau}, D., \& {Kempton}, E.
  M.~R. 2018, \aj, 156, 42, \dodoi{10.3847/1538-3881/aac6dd}

\bibitem[{{Diamond-Lowe} {et~al.}(2022){Diamond-Lowe}, {Mendon{\c{c}}a},
  {Charbonneau}, \& {Buchhave}}]{DiamondLowe2022}
{Diamond-Lowe}, H., {Mendon{\c{c}}a}, J.~M., {Charbonneau}, D., \& {Buchhave},
  L.~A. 2022, arXiv e-prints, arXiv:2210.11809,
  \dodoi{10.48550/arXiv.2210.11809}

\bibitem[{{DiTomasso} {et~al.}(2023){DiTomasso}, {L\'{o}pez-Morales},
  {Peacock}, \& {Malavolta}}]{DiTomasso2023}
{DiTomasso}, V., {L\'{o}pez-Morales}, M., {Peacock}, S., \& {Malavolta}, L.
  2023, ApJL, submitted

\bibitem[{{Dong} {et~al.}(2018){Dong}, {Jin}, {Lingam}, {Airapetian}, {Ma}, \&
  {van der Holst}}]{Dong2018}
{Dong}, C., {Jin}, M., {Lingam}, M., {et~al.} 2018, Proceedings of the National
  Academy of Science, 115, 260, \dodoi{10.1073/pnas.1708010115}

\bibitem[{{Foreman-Mackey} {et~al.}(2013){Foreman-Mackey}, {Hogg}, {Lang}, \&
  {Goodman}}]{emcee2013}
{Foreman-Mackey}, D., {Hogg}, D.~W., {Lang}, D., \& {Goodman}, J. 2013, \pasp,
  125, 306, \dodoi{10.1086/670067}

\bibitem[{{Gaia Collaboration} {et~al.}(2016){Gaia Collaboration}, {Prusti},
  {de Bruijne}, {Brown}, {Vallenari}, {Babusiaux}, {Bailer-Jones}, {Bastian},
  {Biermann}, {Evans}, {Eyer}, {Jansen}, {Jordi}, {Klioner}, {Lammers},
  {Lindegren}, {Luri}, {Mignard}, {Milligan}, {Panem}, {Poinsignon},
  {Pourbaix}, {Randich}, {Sarri}, {Sartoretti}, {Siddiqui}, {Soubiran},
  {Valette}, {van Leeuwen}, {Walton}, {Aerts}, {Arenou}, {Cropper}, {Drimmel},
  {H{\o}g}, {Katz}, {Lattanzi}, {O'Mullane}, {Grebel}, {Holland}, {Huc},
  {Passot}, {Bramante}, {Cacciari}, {Casta{\~n}eda}, {Chaoul}, {Cheek}, {De
  Angeli}, {Fabricius}, {Guerra}, {Hern{\'a}ndez}, {Jean-Antoine-Piccolo},
  {Masana}, {Messineo}, {Mowlavi}, {Nienartowicz}, {Ord{\'o}{\~n}ez-Blanco},
  {Panuzzo}, {Portell}, {Richards}, {Riello}, {Seabroke}, {Tanga},
  {Th{\'e}venin}, {Torra}, {Els}, {Gracia-Abril}, {Comoretto},
  {Garcia-Reinaldos}, {Lock}, {Mercier}, {Altmann}, {Andrae}, {Astraatmadja},
  {Bellas-Velidis}, {Benson}, {Berthier}, {Blomme}, {Busso}, {Carry},
  {Cellino}, {Clementini}, {Cowell}, {Creevey}, {Cuypers}, {Davidson}, {De
  Ridder}, {de Torres}, {Delchambre}, {Dell'Oro}, {Ducourant}, {Fr{\'e}mat},
  {Garc{\'\i}a-Torres}, {Gosset}, {Halbwachs}, {Hambly}, {Harrison}, {Hauser},
  {Hestroffer}, {Hodgkin}, {Huckle}, {Hutton}, {Jasniewicz}, {Jordan},
  {Kontizas}, {Korn}, {Lanzafame}, {Manteiga}, {Moitinho}, {Muinonen},
  {Osinde}, {Pancino}, {Pauwels}, {Petit}, {Recio-Blanco}, {Robin}, {Sarro},
  {Siopis}, {Smith}, {Smith}, {Sozzetti}, {Thuillot}, {van Reeven}, {Viala},
  {Abbas}, {Abreu Aramburu}, {Accart}, {Aguado}, {Allan}, {Allasia},
  {Altavilla}, {{\'A}lvarez}, {Alves}, {Anderson}, {Andrei}, {Anglada Varela},
  {Antiche}, {Antoja}, {Ant{\'o}n}, {Arcay}, {Atzei}, {Ayache}, {Bach},
  {Baker}, {Balaguer-N{\'u}{\~n}ez}, {Barache}, {Barata}, {Barbier}, {Barblan},
  {Baroni}, {Barrado y Navascu{\'e}s}, {Barros}, {Barstow}, {Becciani},
  {Bellazzini}, {Bellei}, {Bello Garc{\'\i}a}, {Belokurov}, {Bendjoya},
  {Berihuete}, {Bianchi}, {Bienaym{\'e}}, {Billebaud}, {Blagorodnova},
  {Blanco-Cuaresma}, {Boch}, {Bombrun}, {Borrachero}, {Bouquillon}, {Bourda},
  {Bouy}, {Bragaglia}, {Breddels}, {Brouillet}, {Br{\"u}semeister},
  {Bucciarelli}, {Budnik}, {Burgess}, {Burgon}, {Burlacu}, {Busonero}, {Buzzi},
  {Caffau}, {Cambras}, {Campbell}, {Cancelliere}, {Cantat-Gaudin}, {Carlucci},
  {Carrasco}, {Castellani}, {Charlot}, {Charnas}, {Charvet}, {Chassat},
  {Chiavassa}, {Clotet}, {Cocozza}, {Collins}, {Collins}, {Costigan}, {Crifo},
  {Cross}, {Crosta}, {Crowley}, {Dafonte}, {Damerdji}, {Dapergolas}, {David},
  {David}, {De Cat}, {de Felice}, {de Laverny}, {De Luise}, {De March}, {de
  Martino}, {de Souza}, {Debosscher}, {del Pozo}, {Delbo}, {Delgado},
  {Delgado}, {di Marco}, {Di Matteo}, {Diakite}, {Distefano}, {Dolding}, {Dos
  Anjos}, {Drazinos}, {Dur{\'a}n}, {Dzigan}, {Ecale}, {Edvardsson}, {Enke},
  {Erdmann}, {Escolar}, {Espina}, {Evans}, {Eynard Bontemps}, {Fabre},
  {Fabrizio}, {Faigler}, {Falc{\~a}o}, {Farr{\`a}s Casas}, {Faye}, {Federici},
  {Fedorets}, {Fern{\'a}ndez-Hern{\'a}ndez}, {Fernique}, {Fienga}, {Figueras},
  {Filippi}, {Findeisen}, {Fonti}, {Fouesneau}, {Fraile}, {Fraser}, {Fuchs},
  {Furnell}, {Gai}, {Galleti}, {Galluccio}, {Garabato}, {Garc{\'\i}a-Sedano},
  {Gar{\'e}}, {Garofalo}, {Garralda}, {Gavras}, {Gerssen}, {Geyer}, {Gilmore},
  {Girona}, {Giuffrida}, {Gomes}, {Gonz{\'a}lez-Marcos},
  {Gonz{\'a}lez-N{\'u}{\~n}ez}, {Gonz{\'a}lez-Vidal}, {Granvik}, {Guerrier},
  {Guillout}, {Guiraud}, {G{\'u}rpide}, {Guti{\'e}rrez-S{\'a}nchez}, {Guy},
  {Haigron}, {Hatzidimitriou}, {Haywood}, {Heiter}, {Helmi}, {Hobbs},
  {Hofmann}, {Holl}, {Holland}, {Hunt}, {Hypki}, {Icardi}, {Irwin}, {Jevardat
  de Fombelle}, {Jofr{\'e}}, {Jonker}, {Jorissen}, {Julbe}, {Karampelas},
  {Kochoska}, {Kohley}, {Kolenberg}, {Kontizas}, {Koposov}, {Kordopatis},
  {Koubsky}, {Kowalczyk}, {Krone-Martins}, {Kudryashova}, {Kull}, {Bachchan},
  {Lacoste-Seris}, {Lanza}, {Lavigne}, {Le Poncin-Lafitte}, {Lebreton},
  {Lebzelter}, {Leccia}, {Leclerc}, {Lecoeur-Taibi}, {Lemaitre}, {Lenhardt},
  {Leroux}, {Liao}, {Licata}, {Lindstr{\o}m}, {Lister}, {Livanou}, {Lobel},
  {L{\"o}ffler}, {L{\'o}pez}, {Lopez-Lozano}, {Lorenz}, {Loureiro},
  {MacDonald}, {Magalh{\~a}es Fernandes}, {Managau}, {Mann}, {Mantelet},
  {Marchal}, {Marchant}, {Marconi}, {Marie}, {Marinoni}, {Marrese},
  {Marschalk{\'o}}, {Marshall}, {Mart{\'\i}n-Fleitas}, {Martino}, {Mary},
  {Matijevi{\v{c}}}, {Mazeh}, {McMillan}, {Messina}, {Mestre}, {Michalik},
  {Millar}, {Miranda}, {Molina}, {Molinaro}, {Molinaro}, {Moln{\'a}r},
  {Moniez}, {Montegriffo}, {Monteiro}, {Mor}, {Mora}, {Morbidelli}, {Morel},
  {Morgenthaler}, {Morley}, {Morris}, {Mulone}, {Muraveva}, {Musella},
  {Narbonne}, {Nelemans}, {Nicastro}, {Noval}, {Ord{\'e}novic},
  {Ordieres-Mer{\'e}}, {Osborne}, {Pagani}, {Pagano}, {Pailler}, {Palacin},
  {Palaversa}, {Parsons}, {Paulsen}, {Pecoraro}, {Pedrosa}, {Pentik{\"a}inen},
  {Pereira}, {Pichon}, {Piersimoni}, {Pineau}, {Plachy}, {Plum}, {Poujoulet},
  {Pr{\v{s}}a}, {Pulone}, {Ragaini}, {Rago}, {Rambaux}, {Ramos-Lerate},
  {Ranalli}, {Rauw}, {Read}, {Regibo}, {Renk}, {Reyl{\'e}}, {Ribeiro},
  {Rimoldini}, {Ripepi}, {Riva}, {Rixon}, {Roelens}, {Romero-G{\'o}mez},
  {Rowell}, {Royer}, {Rudolph}, {Ruiz-Dern}, {Sadowski}, {Sagrist{\`a}
  Sell{\'e}s}, {Sahlmann}, {Salgado}, {Salguero}, {Sarasso}, {Savietto},
  {Schnorhk}, {Schultheis}, {Sciacca}, {Segol}, {Segovia}, {Segransan},
  {Serpell}, {Shih}, {Smareglia}, {Smart}, {Smith}, {Solano}, {Solitro},
  {Sordo}, {Soria Nieto}, {Souchay}, {Spagna}, {Spoto}, {Stampa}, {Steele},
  {Steidelm{\"u}ller}, {Stephenson}, {Stoev}, {Suess}, {S{\"u}veges}, {Surdej},
  {Szabados}, {Szegedi-Elek}, {Tapiador}, {Taris}, {Tauran}, {Taylor},
  {Teixeira}, {Terrett}, {Tingley}, {Trager}, {Turon}, {Ulla}, {Utrilla},
  {Valentini}, {van Elteren}, {Van Hemelryck}, {van Leeuwen}, {Varadi},
  {Vecchiato}, {Veljanoski}, {Via}, {Vicente}, {Vogt}, {Voss}, {Votruba},
  {Voutsinas}, {Walmsley}, {Weiler}, {Weingrill}, {Werner}, {Wevers},
  {Whitehead}, {Wyrzykowski}, {Yoldas}, {{\v{Z}}erjal}, {Zucker}, {Zurbach},
  {Zwitter}, {Alecu}, {Allen}, {Allende Prieto}, {Amorim},
  {Anglada-Escud{\'e}}, {Arsenijevic}, {Azaz}, {Balm}, {Beck}, {Bernstein},
  {Bigot}, {Bijaoui}, {Blasco}, {Bonfigli}, {Bono}, {Boudreault}, {Bressan},
  {Brown}, {Brunet}, {Bunclark}, {Buonanno}, {Butkevich}, {Carret}, {Carrion},
  {Chemin}, {Ch{\'e}reau}, {Corcione}, {Darmigny}, {de Boer}, {de Teodoro}, {de
  Zeeuw}, {Delle Luche}, {Domingues}, {Dubath}, {Fodor}, {Fr{\'e}zouls},
  {Fries}, {Fustes}, {Fyfe}, {Gallardo}, {Gallegos}, {Gardiol}, {Gebran},
  {Gomboc}, {G{\'o}mez}, {Grux}, {Gueguen}, {Heyrovsky}, {Hoar}, {Iannicola},
  {Isasi Parache}, {Janotto}, {Joliet}, {Jonckheere}, {Keil}, {Kim},
  {Klagyivik}, {Klar}, {Knude}, {Kochukhov}, {Kolka}, {Kos}, {Kutka}, {Lainey},
  {LeBouquin}, {Liu}, {Loreggia}, {Makarov}, {Marseille}, {Martayan},
  {Martinez-Rubi}, {Massart}, {Meynadier}, {Mignot}, {Munari}, {Nguyen},
  {Nordlander}, {Ocvirk}, {O'Flaherty}, {Olias Sanz}, {Ortiz}, {Osorio},
  {Oszkiewicz}, {Ouzounis}, {Palmer}, {Park}, {Pasquato}, {Peltzer}, {Peralta},
  {P{\'e}turaud}, {Pieniluoma}, {Pigozzi}, {Poels}, {Prat}, {Prod'homme},
  {Raison}, {Rebordao}, {Risquez}, {Rocca-Volmerange}, {Rosen}, {Ruiz-Fuertes},
  {Russo}, {Sembay}, {Serraller Vizcaino}, {Short}, {Siebert}, {Silva},
  {Sinachopoulos}, {Slezak}, {Soffel}, {Sosnowska}, {Strai{\v{z}}ys}, {ter
  Linden}, {Terrell}, {Theil}, {Tiede}, {Troisi}, {Tsalmantza}, {Tur},
  {Vaccari}, {Vachier}, {Valles}, {Van Hamme}, {Veltz}, {Virtanen}, {Wallut},
  {Wichmann}, {Wilkinson}, {Ziaeepour}, \& {Zschocke}}]{gaia2016}
{Gaia Collaboration}, {Prusti}, T., {de Bruijne}, J.~H.~J., {et~al.} 2016,
  \aap, 595, A1, \dodoi{10.1051/0004-6361/201629272}

\bibitem[{{Gaia Collaboration} {et~al.}(2021){Gaia Collaboration}, {Brown},
  {Vallenari}, {Prusti}, {de Bruijne}, {Babusiaux}, {Biermann}, {Creevey},
  {Evans}, {Eyer}, {Hutton}, {Jansen}, {Jordi}, {Klioner}, {Lammers},
  {Lindegren}, {Luri}, {Mignard}, {Panem}, {Pourbaix}, {Randich}, {Sartoretti},
  {Soubiran}, {Walton}, {Arenou}, {Bailer-Jones}, {Bastian}, {Cropper},
  {Drimmel}, {Katz}, {Lattanzi}, {van Leeuwen}, {Bakker}, {Cacciari},
  {Casta{\~n}eda}, {De Angeli}, {Ducourant}, {Fabricius}, {Fouesneau},
  {Fr{\'e}mat}, {Guerra}, {Guerrier}, {Guiraud}, {Jean-Antoine Piccolo},
  {Masana}, {Messineo}, {Mowlavi}, {Nicolas}, {Nienartowicz}, {Pailler},
  {Panuzzo}, {Riclet}, {Roux}, {Seabroke}, {Sordo}, {Tanga}, {Th{\'e}venin},
  {Gracia-Abril}, {Portell}, {Teyssier}, {Altmann}, {Andrae}, {Bellas-Velidis},
  {Benson}, {Berthier}, {Blomme}, {Brugaletta}, {Burgess}, {Busso}, {Carry},
  {Cellino}, {Cheek}, {Clementini}, {Damerdji}, {Davidson}, {Delchambre},
  {Dell'Oro}, {Fern{\'a}ndez-Hern{\'a}ndez}, {Galluccio}, {Garc{\'\i}a-Lario},
  {Garcia-Reinaldos}, {Gonz{\'a}lez-N{\'u}{\~n}ez}, {Gosset}, {Haigron},
  {Halbwachs}, {Hambly}, {Harrison}, {Hatzidimitriou}, {Heiter},
  {Hern{\'a}ndez}, {Hestroffer}, {Hodgkin}, {Holl}, {Jan{\ss}en}, {Jevardat de
  Fombelle}, {Jordan}, {Krone-Martins}, {Lanzafame}, {L{\"o}ffler}, {Lorca},
  {Manteiga}, {Marchal}, {Marrese}, {Moitinho}, {Mora}, {Muinonen}, {Osborne},
  {Pancino}, {Pauwels}, {Petit}, {Recio-Blanco}, {Richards}, {Riello},
  {Rimoldini}, {Robin}, {Roegiers}, {Rybizki}, {Sarro}, {Siopis}, {Smith},
  {Sozzetti}, {Ulla}, {Utrilla}, {van Leeuwen}, {van Reeven}, {Abbas}, {Abreu
  Aramburu}, {Accart}, {Aerts}, {Aguado}, {Ajaj}, {Altavilla}, {{\'A}lvarez},
  {{\'A}lvarez Cid-Fuentes}, {Alves}, {Anderson}, {Anglada Varela}, {Antoja},
  {Audard}, {Baines}, {Baker}, {Balaguer-N{\'u}{\~n}ez}, {Balbinot}, {Balog},
  {Barache}, {Barbato}, {Barros}, {Barstow}, {Bartolom{\'e}}, {Bassilana},
  {Bauchet}, {Baudesson-Stella}, {Becciani}, {Bellazzini}, {Bernet}, {Bertone},
  {Bianchi}, {Blanco-Cuaresma}, {Boch}, {Bombrun}, {Bossini}, {Bouquillon},
  {Bragaglia}, {Bramante}, {Breedt}, {Bressan}, {Brouillet}, {Bucciarelli},
  {Burlacu}, {Busonero}, {Butkevich}, {Buzzi}, {Caffau}, {Cancelliere},
  {C{\'a}novas}, {Cantat-Gaudin}, {Carballo}, {Carlucci}, {Carnerero},
  {Carrasco}, {Casamiquela}, {Castellani}, {Castro-Ginard}, {Castro Sampol},
  {Chaoul}, {Charlot}, {Chemin}, {Chiavassa}, {Cioni}, {Comoretto}, {Cooper},
  {Cornez}, {Cowell}, {Crifo}, {Crosta}, {Crowley}, {Dafonte}, {Dapergolas},
  {David}, {David}, {de Laverny}, {De Luise}, {De March}, {De Ridder}, {de
  Souza}, {de Teodoro}, {de Torres}, {del Peloso}, {del Pozo}, {Delbo},
  {Delgado}, {Delgado}, {Delisle}, {Di Matteo}, {Diakite}, {Diener},
  {Distefano}, {Dolding}, {Eappachen}, {Edvardsson}, {Enke}, {Esquej}, {Fabre},
  {Fabrizio}, {Faigler}, {Fedorets}, {Fernique}, {Fienga}, {Figueras},
  {Fouron}, {Fragkoudi}, {Fraile}, {Franke}, {Gai}, {Garabato},
  {Garcia-Gutierrez}, {Garc{\'\i}a-Torres}, {Garofalo}, {Gavras}, {Gerlach},
  {Geyer}, {Giacobbe}, {Gilmore}, {Girona}, {Giuffrida}, {Gomel}, {Gomez},
  {Gonzalez-Santamaria}, {Gonz{\'a}lez-Vidal}, {Granvik},
  {Guti{\'e}rrez-S{\'a}nchez}, {Guy}, {Hauser}, {Haywood}, {Helmi}, {Hidalgo},
  {Hilger}, {H{\l}adczuk}, {Hobbs}, {Holland}, {Huckle}, {Jasniewicz},
  {Jonker}, {Juaristi Campillo}, {Julbe}, {Karbevska}, {Kervella}, {Khanna},
  {Kochoska}, {Kontizas}, {Kordopatis}, {Korn}, {Kostrzewa-Rutkowska},
  {Kruszy{\'n}ska}, {Lambert}, {Lanza}, {Lasne}, {Le Campion}, {Le Fustec},
  {Lebreton}, {Lebzelter}, {Leccia}, {Leclerc}, {Lecoeur-Taibi}, {Liao},
  {Licata}, {Lindstr{\o}m}, {Lister}, {Livanou}, {Lobel}, {Madrero Pardo},
  {Managau}, {Mann}, {Marchant}, {Marconi}, {Marcos Santos}, {Marinoni},
  {Marocco}, {Marshall}, {Martin Polo}, {Mart{\'\i}n-Fleitas}, {Masip},
  {Massari}, {Mastrobuono-Battisti}, {Mazeh}, {McMillan}, {Messina},
  {Michalik}, {Millar}, {Mints}, {Molina}, {Molinaro}, {Moln{\'a}r},
  {Montegriffo}, {Mor}, {Morbidelli}, {Morel}, {Morris}, {Mulone}, {Munoz},
  {Muraveva}, {Murphy}, {Musella}, {Noval}, {Ord{\'e}novic}, {Orr{\`u}},
  {Osinde}, {Pagani}, {Pagano}, {Palaversa}, {Palicio}, {Panahi}, {Pawlak},
  {Pe{\~n}alosa Esteller}, {Penttil{\"a}}, {Piersimoni}, {Pineau}, {Plachy},
  {Plum}, {Poggio}, {Poretti}, {Poujoulet}, {Pr{\v{s}}a}, {Pulone}, {Racero},
  {Ragaini}, {Rainer}, {Raiteri}, {Rambaux}, {Ramos}, {Ramos-Lerate}, {Re
  Fiorentin}, {Regibo}, {Reyl{\'e}}, {Ripepi}, {Riva}, {Rixon}, {Robichon},
  {Robin}, {Roelens}, {Rohrbasser}, {Romero-G{\'o}mez}, {Rowell}, {Royer},
  {Rybicki}, {Sadowski}, {Sagrist{\`a} Sell{\'e}s}, {Sahlmann}, {Salgado},
  {Salguero}, {Samaras}, {Sanchez Gimenez}, {Sanna}, {Santove{\~n}a},
  {Sarasso}, {Schultheis}, {Sciacca}, {Segol}, {Segovia}, {S{\'e}gransan},
  {Semeux}, {Shahaf}, {Siddiqui}, {Siebert}, {Siltala}, {Slezak}, {Smart},
  {Solano}, {Solitro}, {Souami}, {Souchay}, {Spagna}, {Spoto}, {Steele},
  {Steidelm{\"u}ller}, {Stephenson}, {S{\"u}veges}, {Szabados}, {Szegedi-Elek},
  {Taris}, {Tauran}, {Taylor}, {Teixeira}, {Thuillot}, {Tonello}, {Torra},
  {Torra}, {Turon}, {Unger}, {Vaillant}, {van Dillen}, {Vanel}, {Vecchiato},
  {Viala}, {Vicente}, {Voutsinas}, {Weiler}, {Wevers}, {Wyrzykowski}, {Yoldas},
  {Yvard}, {Zhao}, {Zorec}, {Zucker}, {Zurbach}, \& {Zwitter}}]{Gaia2021}
{Gaia Collaboration}, {Brown}, A.~G.~A., {Vallenari}, A., {et~al.} 2021, \aap,
  649, A1, \dodoi{10.1051/0004-6361/202039657}

\bibitem[{{Gaia Collaboration} {et~al.}(2023){Gaia Collaboration}, {Vallenari},
  {Brown}, {Prusti}, {de Bruijne}, {Arenou}, {Babusiaux}, {Biermann},
  {Creevey}, {Ducourant}, {Evans}, {Eyer}, {Guerra}, {Hutton}, {Jordi},
  {Klioner}, {Lammers}, {Lindegren}, {Luri}, {Mignard}, {Panem}, {Pourbaix},
  {Randich}, {Sartoretti}, {Soubiran}, {Tanga}, {Walton}, {Bailer-Jones},
  {Bastian}, {Drimmel}, {Jansen}, {Katz}, {Lattanzi}, {van Leeuwen}, {Bakker},
  {Cacciari}, {Casta{\~n}eda}, {De Angeli}, {Fabricius}, {Fouesneau},
  {Fr{\'e}mat}, {Galluccio}, {Guerrier}, {Heiter}, {Masana}, {Messineo},
  {Mowlavi}, {Nicolas}, {Nienartowicz}, {Pailler}, {Panuzzo}, {Riclet}, {Roux},
  {Seabroke}, {Sordo}, {Th{\'e}venin}, {Gracia-Abril}, {Portell}, {Teyssier},
  {Altmann}, {Andrae}, {Audard}, {Bellas-Velidis}, {Benson}, {Berthier},
  {Blomme}, {Burgess}, {Busonero}, {Busso}, {C{\'a}novas}, {Carry}, {Cellino},
  {Cheek}, {Clementini}, {Damerdji}, {Davidson}, {de Teodoro}, {Nu{\~n}ez
  Campos}, {Delchambre}, {Dell'Oro}, {Esquej}, {Fern{\'a}ndez-Hern{\'a}ndez},
  {Fraile}, {Garabato}, {Garc{\'\i}a-Lario}, {Gosset}, {Haigron}, {Halbwachs},
  {Hambly}, {Harrison}, {Hern{\'a}ndez}, {Hestroffer}, {Hodgkin}, {Holl},
  {Jan{\ss}en}, {Jevardat de Fombelle}, {Jordan}, {Krone-Martins}, {Lanzafame},
  {L{\"o}ffler}, {Marchal}, {Marrese}, {Moitinho}, {Muinonen}, {Osborne},
  {Pancino}, {Pauwels}, {Recio-Blanco}, {Reyl{\'e}}, {Riello}, {Rimoldini},
  {Roegiers}, {Rybizki}, {Sarro}, {Siopis}, {Smith}, {Sozzetti}, {Utrilla},
  {van Leeuwen}, {Abbas}, {{\'A}brah{\'a}m}, {Abreu Aramburu}, {Aerts},
  {Aguado}, {Ajaj}, {Aldea-Montero}, {Altavilla}, {{\'A}lvarez}, {Alves},
  {Anders}, {Anderson}, {Anglada Varela}, {Antoja}, {Baines}, {Baker},
  {Balaguer-N{\'u}{\~n}ez}, {Balbinot}, {Balog}, {Barache}, {Barbato},
  {Barros}, {Barstow}, {Bartolom{\'e}}, {Bassilana}, {Bauchet}, {Becciani},
  {Bellazzini}, {Berihuete}, {Bernet}, {Bertone}, {Bianchi}, {Binnenfeld},
  {Blanco-Cuaresma}, {Blazere}, {Boch}, {Bombrun}, {Bossini}, {Bouquillon},
  {Bragaglia}, {Bramante}, {Breedt}, {Bressan}, {Brouillet}, {Brugaletta},
  {Bucciarelli}, {Burlacu}, {Butkevich}, {Buzzi}, {Caffau}, {Cancelliere},
  {Cantat-Gaudin}, {Carballo}, {Carlucci}, {Carnerero}, {Carrasco},
  {Casamiquela}, {Castellani}, {Castro-Ginard}, {Chaoul}, {Charlot}, {Chemin},
  {Chiaramida}, {Chiavassa}, {Chornay}, {Comoretto}, {Contursi}, {Cooper},
  {Cornez}, {Cowell}, {Crifo}, {Cropper}, {Crosta}, {Crowley}, {Dafonte},
  {Dapergolas}, {David}, {David}, {de Laverny}, {De Luise}, {De March}, {De
  Ridder}, {de Souza}, {de Torres}, {del Peloso}, {del Pozo}, {Delbo},
  {Delgado}, {Delisle}, {Demouchy}, {Dharmawardena}, {Di Matteo}, {Diakite},
  {Diener}, {Distefano}, {Dolding}, {Edvardsson}, {Enke}, {Fabre}, {Fabrizio},
  {Faigler}, {Fedorets}, {Fernique}, {Fienga}, {Figueras}, {Fournier},
  {Fouron}, {Fragkoudi}, {Gai}, {Garcia-Gutierrez}, {Garcia-Reinaldos},
  {Garc{\'\i}a-Torres}, {Garofalo}, {Gavel}, {Gavras}, {Gerlach}, {Geyer},
  {Giacobbe}, {Gilmore}, {Girona}, {Giuffrida}, {Gomel}, {Gomez},
  {Gonz{\'a}lez-N{\'u}{\~n}ez}, {Gonz{\'a}lez-Santamar{\'\i}a},
  {Gonz{\'a}lez-Vidal}, {Granvik}, {Guillout}, {Guiraud},
  {Guti{\'e}rrez-S{\'a}nchez}, {Guy}, {Hatzidimitriou}, {Hauser}, {Haywood},
  {Helmer}, {Helmi}, {Sarmiento}, {Hidalgo}, {Hilger}, {H{\l}adczuk}, {Hobbs},
  {Holland}, {Huckle}, {Jardine}, {Jasniewicz}, {Jean-Antoine Piccolo},
  {Jim{\'e}nez-Arranz}, {Jorissen}, {Juaristi Campillo}, {Julbe}, {Karbevska},
  {Kervella}, {Khanna}, {Kontizas}, {Kordopatis}, {Korn}, {K{\'o}sp{\'a}l},
  {Kostrzewa-Rutkowska}, {Kruszy{\'n}ska}, {Kun}, {Laizeau}, {Lambert},
  {Lanza}, {Lasne}, {Le Campion}, {Lebreton}, {Lebzelter}, {Leccia}, {Leclerc},
  {Lecoeur-Taibi}, {Liao}, {Licata}, {Lindstr{\o}m}, {Lister}, {Livanou},
  {Lobel}, {Lorca}, {Loup}, {Madrero Pardo}, {Magdaleno Romeo}, {Managau},
  {Mann}, {Manteiga}, {Marchant}, {Marconi}, {Marcos}, {Marcos Santos},
  {Mar{\'\i}n Pina}, {Marinoni}, {Marocco}, {Marshall}, {Martin Polo},
  {Mart{\'\i}n-Fleitas}, {Marton}, {Mary}, {Masip}, {Massari},
  {Mastrobuono-Battisti}, {Mazeh}, {McMillan}, {Messina}, {Michalik}, {Millar},
  {Mints}, {Molina}, {Molinaro}, {Moln{\'a}r}, {Monari}, {Mongui{\'o}},
  {Montegriffo}, {Montero}, {Mor}, {Mora}, {Morbidelli}, {Morel}, {Morris},
  {Muraveva}, {Murphy}, {Musella}, {Nagy}, {Noval}, {Oca{\~n}a}, {Ogden},
  {Ordenovic}, {Osinde}, {Pagani}, {Pagano}, {Palaversa}, {Palicio},
  {Pallas-Quintela}, {Panahi}, {Payne-Wardenaar}, {Pe{\~n}alosa Esteller},
  {Penttil{\"a}}, {Pichon}, {Piersimoni}, {Pineau}, {Plachy}, {Plum}, {Poggio},
  {Pr{\v{s}}a}, {Pulone}, {Racero}, {Ragaini}, {Rainer}, {Raiteri}, {Rambaux},
  {Ramos}, {Ramos-Lerate}, {Re Fiorentin}, {Regibo}, {Richards}, {Rios Diaz},
  {Ripepi}, {Riva}, {Rix}, {Rixon}, {Robichon}, {Robin}, {Robin}, {Roelens},
  {Rogues}, {Rohrbasser}, {Romero-G{\'o}mez}, {Rowell}, {Royer}, {Ruz Mieres},
  {Rybicki}, {Sadowski}, {S{\'a}ez N{\'u}{\~n}ez}, {Sagrist{\`a} Sell{\'e}s},
  {Sahlmann}, {Salguero}, {Samaras}, {Sanchez Gimenez}, {Sanna},
  {Santove{\~n}a}, {Sarasso}, {Schultheis}, {Sciacca}, {Segol}, {Segovia},
  {S{\'e}gransan}, {Semeux}, {Shahaf}, {Siddiqui}, {Siebert}, {Siltala},
  {Silvelo}, {Slezak}, {Slezak}, {Smart}, {Snaith}, {Solano}, {Solitro},
  {Souami}, {Souchay}, {Spagna}, {Spina}, {Spoto}, {Steele},
  {Steidelm{\"u}ller}, {Stephenson}, {S{\"u}veges}, {Surdej}, {Szabados},
  {Szegedi-Elek}, {Taris}, {Taylor}, {Teixeira}, {Tolomei}, {Tonello}, {Torra},
  {Torra}, {Torralba Elipe}, {Trabucchi}, {Tsounis}, {Turon}, {Ulla}, {Unger},
  {Vaillant}, {van Dillen}, {van Reeven}, {Vanel}, {Vecchiato}, {Viala},
  {Vicente}, {Voutsinas}, {Weiler}, {Wevers}, {Wyrzykowski}, {Yoldas}, {Yvard},
  {Zhao}, {Zorec}, {Zucker}, \& {Zwitter}}]{gaia_dr3}
{Gaia Collaboration}, {Vallenari}, A., {Brown}, A.~G.~A., {et~al.} 2023, \aap,
  674, A1, \dodoi{10.1051/0004-6361/202243940}

\bibitem[{{Gao} {et~al.}(2015){Gao}, {Hu}, {Robinson}, {Li}, \&
  {Yung}}]{Gao2015}
{Gao}, P., {Hu}, R., {Robinson}, T.~D., {Li}, C., \& {Yung}, Y.~L. 2015, \apj,
  806, 249, \dodoi{10.1088/0004-637X/806/2/249}

\bibitem[{{Gardner} {et~al.}(2023){Gardner}, {Mather}, {Abbott}, {Abell},
  {Abernathy}, {Abney}, {Abraham}, {Abraham}, {Abul-Huda}, {Acton}, {Adams},
  {Adams}, {Adler}, {Adriaensen}, {Aguilar}, {Ahmed}, {Ahmed}, {Ahmed},
  {Albat}, {Albert}, {Alberts}, {Aldridge}, {Allen}, {Allen}, {Altenburg},
  {Altunc}, {Alvarez}, {{\'A}lvarez-M{\'a}rquez}, {de Oliveira}, {Ambrose},
  {Anandakrishnan}, {Andersen}, {Anderson}, {Anderson}, {Anderson}, {Anderson},
  {Aprea}, {Archer}, {Arenberg}, {Argyriou}, {Arribas}, {Artigau}, {Arvai},
  {Atcheson}, {Atkinson}, {Averbukh}, {Aymergen}, {Bacinski}, {Baggett},
  {Bagnasco}, {Baker}, {Balzano}, {Banks}, {Baran}, {Barker}, {Barrett},
  {Barringer}, {Barto}, {Bast}, {Baudoz}, {Baum}, {Beatty}, {Beaulieu},
  {Bechtold}, {Beck}, {Beddard}, {Beichman}, {Bellagama}, {Bely}, {Berger},
  {Bergeron}, {Bernier}, {Bertch}, {Beskow}, {Betz}, {Biagetti}, {Birkmann},
  {Bjorklund}, {Blackwood}, {Blazek}, {Blossfeld}, {Bluth}, {Boccaletti},
  {Boegner}, {Bohlin}, {Boia}, {B{\"o}ker}, {Bonaventura}, {Bond}, {Bosley},
  {Boucarut}, {Bouchet}, {Bouwman}, {Bower}, {Bowers}, {Bowers}, {Boyce},
  {Boyer}, {Boyer}, {Boyer}, {Boyer}, {Bradley}, {Brady}, {Brandl}, {Brannen},
  {Breda}, {Bremmer}, {Brennan}, {Bresnahan}, {Bright}, {Broiles},
  {Bromenschenkel}, {Brooks}, {Brooks}, {Brown}, {Brown}, {Brown}, {Bruce},
  {Bryson}, {Bujanda}, {Bullock}, {Bunker}, {Bureo}, {Burt}, {Bush},
  {Bushouse}, {Bussman}, {Cabaud}, {Cale}, {Calhoon}, {Calvani}, {Canipe},
  {Caputo}, {Cara}, {Carey}, {Case}, {Cesari}, {Cetorelli}, {Chance},
  {Chandler}, {Chaney}, {Chapman}, {Charlot}, {Chayer}, {Cheezum}, {Chen},
  {Chen}, {Cherinka}, {Chichester}, {Chilton}, {Chittiraibalan}, {Clampin},
  {Clark}, {Clark}, {Clark}, {Claybrooks}, {Cleveland}, {Cohen}, {Cohen},
  {Col{\'o}n}, {Coleman}, {Colina}, {Comber}, {Comeau}, {Comer}, {Reis},
  {Connolly}, {Conroy}, {Contos}, {Contreras}, {Cook}, {Cooper}, {Cooper},
  {Correia}, {Correnti}, {Cossou}, {Costanza}, {Coulais}, {Cox}, {Coyle},
  {Cracraft}, {Crew}, {Curtis}, {Cusveller}, {Maciel}, {Dailey}, {Daugeron},
  {Davidson}, {Davies}, {Davis}, {Davis}, {Day}, {de Chambure}, {de Jong}, {De
  Marchi}, {Dean}, {Decker}, {Delisa}, {Dell}, {Dellagatta}, {Dembinska},
  {Demosthenes}, {Dencheva}, {Deneu}, {DePriest}, {Deschenes}, {Dethienne},
  {Detre}, {Diaz}, {Dicken}, {DiFelice}, {Dillman}, {Disharoon}, {Dixon},
  {Doggett}, {Dominguez}, {Donaldson}, {Doria-Warner}, {Santos}, {Doty},
  {Douglas}, {Doyon}, {Dressler}, {Driggers}, {Driggers}, {Dunn}, {DuPrie},
  {Dupuis}, {Durning}, {Dutta}, {Earl}, {Eccleston}, {Ecobichon}, {Egami},
  {Ehrenwinkler}, {Eisenhamer}, {Eisenhower}, {Eisenstein}, {El Hamel}, {Elie},
  {Elliott}, {Elliott}, {Engesser}, {Espinoza}, {Etienne}, {Etxaluze}, {Evans},
  {Fabreguettes}, {Falcolini}, {Falini}, {Fatig}, {Feeney}, {Feinberg}, {Fels},
  {Ferdous}, {Ferguson}, {Ferrarese}, {Ferreira}, {Ferruit}, {Ferry},
  {Filippazzo}, {Firre}, {Fix}, {Flagey}, {Flanagan}, {Fleming}, {Florian},
  {Flynn}, {Foiadelli}, {Fontaine}, {Fontanella}, {Forshay}, {Fortner}, {Fox},
  {Framarini}, {Francisco}, {Franck}, {Franx}, {Franz}, {Friedman}, {Friend},
  {Frost}, {Fu}, {Fullerton}, {Gaillard}, {Galkin}, {Gallagher}, {Galyer},
  {Garc{\'\i}a Mar{\'\i}n}, {Gardner}, {Garland}, {Garrett}, {Gasman},
  {G{\'a}sp{\'a}r}, {Gastaud}, {Gaudreau}, {Gauthier}, {Geers}, {Geithner},
  {Gennaro}, {Gerber}, {Gereau}, {Giampaoli}, {Giardino}, {Gibbons}, {Gilbert},
  {Gilman}, {Girard}, {Giuliano}, {Gkountis}, {Glasse}, {Glassmire}, {Glauser},
  {Glazer}, {Goldberg}, {Golimowski}, {Gonzaga}, {Gordon}, {Gordon},
  {Goudfrooij}, {Gough}, {Graham}, {Grau}, {Green}, {Greene}, {Greene},
  {Greenfield}, {Greenhouse}, {Greve}, {Greville}, {Grimaldi}, {Groe},
  {Groebner}, {Grumm}, {Grundy}, {G{\"u}del}, {Guillard}, {Guldalian}, {Gunn},
  {Gurule}, {Gutman}, {Guy}, {Guyot}, {Hack}, {Haderlein}, {Hagan}, {Hagedorn},
  {Hainline}, {Haley}, {Hami}, {Hamilton}, {Hammann}, {Hammel}, {Hanley},
  {Hansen}, {Hardy}, {Harnisch}, {Harr}, {Harris}, {Hart}, {Hartig}, {Hasan},
  {Hashim}, {Hashimoto}, {Haskins}, {Hawkins}, {Hayden}, {Hayden}, {Healy},
  {Hecht}, {Heeg}, {Hejal}, {Helm}, {Hengemihle}, {Henning}, {Henry}, {Henry},
  {Henshaw}, {Hernandez}, {Herrington}, {Heske}, {Hesman}, {Hickey}, {Hilbert},
  {Hines}, {Hinz}, {Hirsch}, {Hitcho}, {Hodapp}, {Hodge}, {Hoffman},
  {Holfeltz}, {Holler}, {Hoppa}, {Horner}, {Howard}, {Howard}, {Huber},
  {Hunkeler}, {Hunter}, {Hunter}, {Hurd}, {Hurst}, {Hutchings}, {Hylan},
  {Ignat}, {Illingworth}, {Irish}, {Isaacs}, {Jackson}, {Jaffe}, {Jahic},
  {Jahromi}, {Jakobsen}, {James}, {James}, {James}, {Jamieson}, {Jandra},
  {Jayawardhana}, {Jedrzejewski}, {Jeffers}, {Jensen}, {Joanne}, {Johns},
  {Johnson}, {Johnson}, {Johnson}, {Johnson}, {Johnson}, {Johnson},
  {Johnstone}, {Jollet}, {Jones}, {Jones}, {Jones}, {Jones}, {Jones}, {Jordan},
  {Jordan}, {Jue}, {Jurkowski}, {Justis}, {Justtanont}, {Kaleida}, {Kalirai},
  {Kalmanson}, {Kaltenegger}, {Kammerer}, {Kan}, {Kanarek}, {Kao}, {Karakla},
  {Karl}, {Kassin}, {Kauffman}, {Kavanagh}, {Kelley}, {Kelly}, {Kendrew},
  {Kennedy}, {Kenny}, {Keski-Kuha}, {Keyes}, {Khan}, {Kidwell}, {Kimble},
  {King}, {King}, {Kinzel}, {Kirk}, {Kirkpatrick}, {Klaassen}, {Klingemann},
  {Klintworth}, {Knapp}, {Knight}, {Knollenberg}, {Knutsen}, {Koehler},
  {Koekemoer}, {Kofler}, {Kontson}, {Kovacs}, {Kozhurina-Platais}, {Krause},
  {Kriss}, {Krist}, {Kristoffersen}, {Krogel}, {Krueger}, {Kulp}, {Kumari},
  {Kwan}, {Kyprianou}, {Labador}, {Labiano}, {Lafreni{\`e}re}, {Lagage},
  {Laidler}, {Laine}, {Laird}, {Lajoie}, {Lallo}, {Lam}, {LaMassa}, {Lambros},
  {Lampenfield}, {Lander}, {Langston}, {Larson}, {Larson}, {LaVerghetta},
  {Law}, {Lawrence}, {Lee}, {Lee}, {Lee}, {Leisenring}, {Leveille}, {Levenson},
  {Levi}, {Levine}, {Lewis}, {Lewis}, {Lewis}, {Libralato}, {Lidon},
  {Liebrecht}, {Lightsey}, {Lilly}, {Lim}, {Lim}, {Ling}, {Link}, {Link},
  {Lipinski}, {Liu}, {Lo}, {Lobmeyer}, {Logue}, {Long}, {Long}, {Long}, {Long},
  {L{\'o}pez-Caniego}, {Lotz}, {Love-Pruitt}, {Lubskiy}, {Luers}, {Luetgens},
  {Luevano}, {Lui}, {Lund}, {Lundquist}, {Lunine}, {L{\"u}tzgendorf}, {Lynch},
  {MacDonald}, {MacDonald}, {Macias}, {Macklis}, {Maghami}, {Maharaja},
  {Maiolino}, {Makrygiannis}, {Malla}, {Malumuth}, {Manjavacas}, {Marini},
  {Marrione}, {Marston}, {Martel}, {Martin}, {Martin}, {Martinez}, {Maschmann},
  {Masci}, {Masetti}, {Maszkiewicz}, {Matthews}, {Matuskey}, {McBrayer},
  {McCarthy}, {McCaughrean}, {McClare}, {McClare}, {McCloskey}, {McClurg},
  {McCoy}, {McElwain}, {McGregor}, {McGuffey}, {McKay}, {McKenzie}, {McLean},
  {McMaster}, {McNeil}, {De Meester}, {Mehalick}, {Meixner}, {Mel{\'e}ndez},
  {Menzel}, {Menzel}, {Merz}, {Mesterharm}, {Meyer}, {Meyett}, {Meza},
  {Midwinter}, {Milam}, {Miller}, {Miller}, {Miskey}, {Misselt}, {Mitchell},
  {Mohan}, {Montoya}, {Moran}, {Morishita}, {Moro-Mart{\'\i}n}, {Morrison},
  {Morrison}, {Morse}, {Moschos}, {Moseley}, {Mosier}, {Mosner}, {Mountain},
  {Muckenthaler}, {Mueller}, {Mueller}, {Muhiem}, {M{\"u}hlmann}, {Mullally},
  {Mullen}, {Munger}, {Murphy}, {Murray}, {Muzerolle}, {Mycroft}, {Myers},
  {Myers}, {Myers}, {Myers}, {Myrick}, {Nagle}, {Nayak}, {Naylor}, {Neff},
  {Nelan}, {Nella}, {Nguyen}, {Nguyen}, {Nickson}, {Nidhiry}, {Niedner},
  {Nieto-Santisteban}, {Nikolov}, {Nishisaka}, {Noriega-Crespo}, {Nota},
  {O'Mara}, {Oboryshko}, {O'Brien}, {Ochs}, {Offenberg}, {Ogle}, {Ohl},
  {Olmsted}, {Osborne}, {O'Shaughnessy}, {{\"O}stlin}, {O'Sullivan}, {Otor},
  {Ottens}, {Ouellette}, {Outlaw}, {Owens}, {Pacifici}, {Page}, {Paranilam},
  {Park}, {Parrish}, {Paschal}, {Patapis}, {Patel}, {Patrick}, {Pattishall},
  {Paul}, {Paul}, {Pauly}, {Pavlovsky}, {Pe{\~n}a-Guerrero}, {Pedder}, {Peek},
  {Pelham}, {Penanen}, {Perriello}, {Perrin}, {Perrine}, {Perrygo}, {Peslier},
  {Petach}, {Peterson}, {Pfarr}, {Pierson}, {Pietraszkiewicz}, {Pilchen},
  {Pipher}, {Pirzkal}, {Pitman}, {Player}, {Plesha}, {Plitzke}, {Pohner},
  {Poletis}, {Pollizzi}, {Polster}, {Pontius}, {Pontoppidan}, {Porges},
  {Potter}, {Prescott}, {Proffitt}, {Pueyo}, {Quispe Neira}, {Radich}, {Rager},
  {Rameau}, {Ramey}, {Alarcon}, {Rampini}, {Rapp}, {Rashford}, {Rauscher},
  {Ravindranath}, {Rawle}, {Rawlings}, {Ray}, {Regan}, {Rehm}, {Rehm}, {Reid},
  {Reis}, {Renk}, {Reoch}, {Ressler}, {Rest}, {Reynolds}, {Richon}, {Richon},
  {Ridgaway}, {Riedel}, {Rieke}, {Rieke}, {Rifelli}, {Rigby}, {Riggs},
  {Ringel}, {Ritchie}, {Rix}, {Robberto}, {Robinson}, {Robinson}, {Robinson},
  {Rock}, {Rodriguez}, {Rodr{\'\i}guez del Pino}, {Roellig}, {Rohrbach},
  {Roman}, {Romelfanger}, {Romo}, {Rosales}, {Rose}, {Roteliuk}, {Roth},
  {Rothwell}, {Rouzaud}, {Rowe}, {Rowlands}, {Roy}, {Royer}, {Rui}, {Rumler},
  {Rumpl}, {Russ}, {Ryan}, {Ryan}, {Saad}, {Sabata}, {Sabatino}, {Sabbi},
  {Sabelhaus}, {Sabia}, {Sahu}, {Saif}, {Salvignol}, {Samara-Ratna},
  {Samuelson}, {Sanders}, {Sappington}, {Sargent}, {Sauer}, {Savadkin},
  {Sawicki}, {Schappell}, {Scheffer}, {Scheithauer}, {Scherer}, {Schiff},
  {Schlawin}, {Schmeitzky}, {Schmitz}, {Schmude}, {Schneider}, {Schreiber},
  {Schroeven-Deceuninck}, {Schultz}, {Schwab}, {Schwartz}, {Scoccimarro},
  {Scott}, {Scott}, {Seaton}, {Seely}, {Seery}, {Seidleck}, {Sembach},
  {Shanahan}, {Shaughnessy}, {Shaw}, {Shay}, {Sheehan}, {Sheth}, {Shih},
  {Shivaei}, {Siegel}, {Sienkiewicz}, {Simmons}, {Simon}, {Sirianni},
  {Sivaramakrishnan}, {Slade}, {Sloan}, {Slocum}, {Slowinski}, {Smith},
  {Smith}, {Smith}, {Smith}, {Smith}, {Smith}, {Smolik}, {Soderblom}, {Sohn},
  {Sokol}, {Sonneborn}, {Sontag}, {Sooy}, {Soummer}, {Southwood}, {Spain},
  {Sparmo}, {Speer}, {Spencer}, {Sprofera}, {Stallcup}, {Stanley},
  {Stansberry}, {Stark}, {Starr}, {Stassi}, {Steck}, {Steeley}, {Stephens},
  {Stephenson}, {Stewart}, {Stiavelli}, {}, {Strada}, {Straughn}, {Streetman},
  {Strickland}, {Strobele}, {Stuhlinger}, {Stys}, {Such}, {Sukhatme},
  {Sullivan}, {Sullivan}, {Sumner}, {Sun}, {Sunnquist}, {Swade}, {Swam},
  {Swenton}, {Swoish}, {Tam Litten}, {Tamas}, {Tao}, {Taylor}, {Taylor},
  {Plate}, {Van Tea}, {Teague}, {Telfer}, {Temim}, {Texter}, {Thatte},
  {Thompson}, {Thompson}, {Thomson}, {Thronson}, {Tierney}, {Tikkanen},
  {Tinnin}, {Tippet}, {Todd}, {Tran}, {Trauger}, {Trejo}, {Vinh Truong},
  {Tsukamoto}, {Tufail}, {Tumlinson}, {Tustain}, {Tyra}, {Ubeda}, {Underwood},
  {Uzzo}, {Vaclavik}, {Valenduc}, {Valenti}, {Van Campen}, {van de Wetering},
  {Van Der Marel}, {van Haarlem}, {Vandenbussche}, {van Dishoeck},
  {Vanterpool}, {Vernoy}, {Vila Costas}, {Volk}, {Voorzaat}, {Voyton}, {Vydra},
  {Waddy}, {Waelkens}, {Wahlgren}, {Walker}, {Wander}, {Warfield}, {Warner},
  {Wasiak}, {Wasiak}, {Wehner}, {Weiler}, {Weilert}, {Weiss}, {Wells}, {Welty},
  {Wheate}, {Wheeler}, {White}, {Whitehouse}, {Whiteleather}, {Whitman},
  {Williams}, {Willmer}, {Willott}, {Willoughby}, {Wilson}, {Wilson}, {Wilson},
  {Windhorst}, {Wislowski}, {Wolfe}, {Wolfe}, {Wolff}, {Wondel}, {Woo},
  {Woods}, {Worden}, {Workman}, {Wright}, {Wu}, {Wu}, {Wun}, {Wymer},
  {Yadetie}, {Yan}, {Yang}, {Yates}, {Yeager}, {Yerger}, {Young}, {Young},
  {Yu}, {Yu}, {Zak}, {Zeidler}, {Zepp}, {Zhou}, {Zincke}, {Zonak}, \&
  {Zondag}}]{Gardner2023}
{Gardner}, J.~P., {Mather}, J.~C., {Abbott}, R., {et~al.} 2023, \pasp, 135,
  068001, \dodoi{10.1088/1538-3873/acd1b5}

\bibitem[{Grant \& Wakeford(2022)}]{david_grant_2022_7437681}
Grant, D., \& Wakeford, H.~R. 2022, Exo-TiC/ExoTiC-LD: ExoTiC-LD v3.0.0,
  v3.0.0,  Zenodo, \dodoi{10.5281/zenodo.7437681}

\bibitem[{{Greene} {et~al.}(2023){Greene}, {Bell}, {Ducrot}, {Dyrek}, {Lagage},
  \& {Fortney}}]{Greene2023}
{Greene}, T.~P., {Bell}, T.~J., {Ducrot}, E., {et~al.} 2023, \nat, 618, 39,
  \dodoi{10.1038/s41586-023-05951-7}

\bibitem[{{Guillot}(2010)}]{Guillot2010}
{Guillot}, T. 2010, \aap, 520, A27, \dodoi{10.1051/0004-6361/200913396}

\bibitem[{Harris {et~al.}(2020)Harris, Millman, van~der Walt, Gommers,
  Virtanen, Cournapeau, Wieser, Taylor, Berg, Smith, Kern, Picus, Hoyer, van
  Kerkwijk, Brett, Haldane, del R{'{\i}}o, Wiebe, Peterson,
  G{'{e}}rard-Marchant, Sheppard, Reddy, Weckesser, Abbasi, Gohlke, \&
  Oliphant}]{numpynew}
Harris, C.~R., Millman, K.~J., van~der Walt, S.~J., {et~al.} 2020, Nature, 585,
  357, \dodoi{10.1038/s41586-020-2649-2}

\bibitem[{{Horne}(1986)}]{Horne1986}
{Horne}, K. 1986, Publ. Astron. Soc. Pac., 98, 609, \dodoi{10.1086/131801}

\bibitem[{Hunter(2007)}]{matplotlib}
Hunter, J.~D. 2007, Computing in Science Engineering, 9, 90,
  \dodoi{10.1109/MCSE.2007.55}

\bibitem[{{Husser} {et~al.}(2013){Husser}, {Wende-von Berg}, {Dreizler},
  {Homeier}, {Reiners}, {Barman}, \& {Hauschildt}}]{Husser2013}
{Husser}, T.~O., {Wende-von Berg}, S., {Dreizler}, S., {et~al.} 2013, \aap,
  553, A6, \dodoi{10.1051/0004-6361/201219058}

\bibitem[{{Kirk} {et~al.}(2019){Kirk}, {L{\'o}pez-Morales}, {Wheatley},
  {Weaver}, {Skillen}, {Louden}, {McCormac}, \& {Espinoza}}]{Kirk2019}
{Kirk}, J., {L{\'o}pez-Morales}, M., {Wheatley}, P.~J., {et~al.} 2019, \aj,
  158, 144, \dodoi{10.3847/1538-3881/ab397d}

\bibitem[{{Kirk} {et~al.}(2021){Kirk}, {Rackham}, {MacDonald},
  {L{\'o}pez-Morales}, {Espinoza}, {Lendl}, {Wilson}, {Osip}, {Wheatley},
  {Skillen}, {Apai}, {Bixel}, {Gibson}, {Jord{\'a}n}, {Lewis}, {Louden},
  {McGruder}, {Nikolov}, {Rodler}, \& {Weaver}}]{Kirk2021}
{Kirk}, J., {Rackham}, B.~V., {MacDonald}, R.~J., {et~al.} 2021, \aj, 162, 34,
  \dodoi{10.3847/1538-3881/abfcd2}

\bibitem[{{Kite} \& {Schaefer}(2021)}]{Kite2021}
{Kite}, E.~S., \& {Schaefer}, L. 2021, \apjl, 909, L22,
  \dodoi{10.3847/2041-8213/abe7dc}

\bibitem[{{Kreidberg}(2015)}]{batman2015}
{Kreidberg}, L. 2015, \pasp, 127, 1161, \dodoi{10.1086/683602}

\bibitem[{{Kreidberg} {et~al.}(2014){Kreidberg}, {Bean}, {D{\'e}sert},
  {Benneke}, {Deming}, {Stevenson}, {Seager}, {Berta-Thompson}, {Seifahrt}, \&
  {Homeier}}]{Kreidberg2014}
{Kreidberg}, L., {Bean}, J.~L., {D{\'e}sert}, J.-M., {et~al.} 2014, \nat, 505,
  69, \dodoi{10.1038/nature12888}

\bibitem[{{Kreidberg} {et~al.}(2019){Kreidberg}, {Koll}, {Morley}, {Hu},
  {Schaefer}, {Deming}, {Stevenson}, {Dittmann}, {Vanderburg}, {Berardo},
  {Guo}, {Stassun}, {Crossfield}, {Charbonneau}, {Latham}, {Loeb}, {Ricker},
  {Seager}, \& {Vanderspek}}]{Kreidberg2019}
{Kreidberg}, L., {Koll}, D. D.~B., {Morley}, C., {et~al.} 2019, \nat, 573, 87,
  \dodoi{10.1038/s41586-019-1497-4}

\bibitem[{{Krissansen-Totton} \& {Fortney}(2022)}]{Krissansen-Totton2022}
{Krissansen-Totton}, J., \& {Fortney}, J.~J. 2022, \apj, 933, 115,
  \dodoi{10.3847/1538-4357/ac69cb}

\bibitem[{{Lammer} {et~al.}(2003){Lammer}, {Selsis}, {Ribas}, {Guinan},
  {Bauer}, \& {Weiss}}]{Lammer2003}
{Lammer}, H., {Selsis}, F., {Ribas}, I., {et~al.} 2003, \apjl, 598, L121,
  \dodoi{10.1086/380815}

\bibitem[{{Lim} {et~al.}(2023){Lim}, {Benneke}, {Doyon}, {MacDonald},
  {Piaulet}, {Artigau}, {Coulombe}, {Radica}, {L'Heureux}, {Albert}, {Rackham},
  {de Wit}, {Salhi}, {Roy}, {Flagg}, {Fournier-Tondreau}, {Taylor}, {Cook},
  {Lafreni{\`e}re}, {Cowan}, {Kaltenegger}, {Rowe}, {Espinoza}, {Dang}, \&
  {Darveau-Bernier}}]{Lim2023}
{Lim}, O., {Benneke}, B., {Doyon}, R., {et~al.} 2023, arXiv e-prints,
  arXiv:2309.07047, \dodoi{10.48550/arXiv.2309.07047}

\bibitem[{{Lincowski} {et~al.}(2018){Lincowski}, {Meadows}, {Crisp},
  {Robinson}, {Luger}, {Lustig-Yaeger}, \& {Arney}}]{Lincowski2018}
{Lincowski}, A.~P., {Meadows}, V.~S., {Crisp}, D., {et~al.} 2018, \apj, 867,
  76, \dodoi{10.3847/1538-4357/aae36a}

\bibitem[{{Line} {et~al.}(2014){Line}, {Knutson}, {Wolf}, \&
  {Yung}}]{Line2014-C/O}
{Line}, M.~R., {Knutson}, H., {Wolf}, A.~S., \& {Yung}, Y.~L. 2014, \apj, 783,
  70, \dodoi{10.1088/0004-637X/783/2/70}

\bibitem[{{Line} \& {Yung}(2013)}]{Line2013b}
{Line}, M.~R., \& {Yung}, Y.~L. 2013, \apj, 779, 3,
  \dodoi{10.1088/0004-637X/779/1/3}

\bibitem[{{Line} {et~al.}(2013){Line}, {Wolf}, {Zhang}, {Knutson}, {Kammer},
  {Ellison}, {Deroo}, {Crisp}, \& {Yung}}]{Line2013a}
{Line}, M.~R., {Wolf}, A.~S., {Zhang}, X., {et~al.} 2013, \apj, 775, 137,
  \dodoi{10.1088/0004-637X/775/2/137}

\bibitem[{{Luger} \& {Barnes}(2015)}]{Luger2015}
{Luger}, R., \& {Barnes}, R. 2015, Astrobiology, 15, 119,
  \dodoi{10.1089/ast.2014.1231}

\bibitem[{{Luque} \& {Pall{\'e}}(2022)}]{Luque2022}
{Luque}, R., \& {Pall{\'e}}, E. 2022, Science, 377, 1211,
  \dodoi{10.1126/science.abl7164}

\bibitem[{{Lustig-Yaeger} {et~al.}(2023{\natexlab{a}}){Lustig-Yaeger},
  {Meadows}, {Crisp}, {Line}, \& {Robinson}}]{Lustig-Yaeger2023earth}
{Lustig-Yaeger}, J., {Meadows}, V.~S., {Crisp}, D., {Line}, M.~R., \&
  {Robinson}, T.~D. 2023{\natexlab{a}}, arXiv e-prints, arXiv:2308.14804,
  \dodoi{10.48550/arXiv.2308.14804}

\bibitem[{{Lustig-Yaeger} {et~al.}(2022){Lustig-Yaeger}, {Sotzen}, {Stevenson},
  {Luger}, {May}, {Mayorga}, {Mandt}, \& {Izenberg}}]{Lustig-Yaeger2022}
{Lustig-Yaeger}, J., {Sotzen}, K.~S., {Stevenson}, K.~B., {et~al.} 2022, \aj,
  163, 140, \dodoi{10.3847/1538-3881/ac5034}

\bibitem[{{Lustig-Yaeger} {et~al.}(2023{\natexlab{b}}){Lustig-Yaeger}, {Fu},
  {May}, {Ortiz Ceballos}, {Moran}, {Peacock}, {Stevenson},
  {L{\'o}pez-Morales}, {MacDonald}, {Mayorga}, {Sing}, {Sotzen}, {Valenti},
  {Adams}, {Alam}, {Batalha}, {Bennett}, {Gonzalez-Quiles}, {Kirk}, {Kruse},
  {Lothringer}, {Rustamkulov}, \& {Wakeford}}]{Lustig-YaegerFu2023}
{Lustig-Yaeger}, J., {Fu}, G., {May}, E.~M., {et~al.} 2023{\natexlab{b}}, arXiv
  e-prints, arXiv:2301.04191, \dodoi{10.48550/arXiv.2301.04191}

\bibitem[{{Magic} {et~al.}(2015){Magic}, {Chiavassa}, {Collet}, \&
  {Asplund}}]{Magic2015}
{Magic}, Z., {Chiavassa}, A., {Collet}, R., \& {Asplund}, M. 2015, \aap, 573,
  A90, \dodoi{10.1051/0004-6361/201423804}

\bibitem[{{May} {et~al.}(2023){May}, {MacDonald}, {Bennett}, {Moran},
  {Wakeford}, \& {Peacock}}]{MayMacDonald2023}
{May}, E.~M., {MacDonald}, R.~J., {Bennett}, K.~A., {et~al.} 2023, ApJL,
  submitted

\bibitem[{{McElwain} {et~al.}(2023){McElwain}, {Feinberg}, {Perrin}, {Clampin},
  {Mountain}, {Lallo}, {Lajoie}, {Kimble}, {Bowers}, {Stark}, {Acton},
  {Atkinson}, {Barinek}, {Barto}, {Basinger}, {Beck}, {Bergkoetter}, {Bluth},
  {Boucarut}, {Brady}, {Brooks}, {Brown}, {Byard}, {Carey}, {Carrasquilla},
  {Chae}, {Chaney}, {Chayer}, {Chonis}, {Cohen}, {Cole}, {Comeau}, {Coon},
  {Coppock}, {Coyle}, {Dean}, {Dziak}, {Eisenhower}, {Flagey}, {Franck},
  {Gallagher}, {Gilman}, {Glassman}, {Green}, {Grieco}, {Haase},
  {Hadjimichael}, {Hagopian}, {Hahn}, {Hartig}, {Havey}, {Hayden}, {Hellekson},
  {Hicks}, {Holfeltz}, {Howard}, {Huguet}, {Jahne}, {Johnson}, {Johnston},
  {Jurling}, {Kegley}, {Kennard}, {Keski-Kuha}, {Knight}, {Kulp}, {Levi},
  {Levine}, {Lightsey}, {Luetgens}, {Mather}, {Matthews}, {McKay}, {Mehalick},
  {Mel{\'e}ndez}, {Mosier}, {Murphy}, {Nelan}, {Niedner}, {Nol}, {Ohara},
  {Ohl}, {Olczak}, {Osborne}, {Park}, {Perrygo}, {Pueyo}, {Redding}, {Regan},
  {Reynolds}, {Rifelli}, {Rigby}, {Sabatke}, {Saif}, {Scorse}, {Seo}, {Shi},
  {Sigrist}, {Smith}, {Smith}, {Smith}, {Sohn}, {Stahl}, {Telfer}, {Terlecki},
  {Texter}, {Van Buren}, {Van Campen}, {Vila}, {Voyton}, {Waldman}, {Walker},
  {Weiser}, {Wells}, {West}, {Whitman}, {Wolf}, \& {Zielinski}}]{McElwain2023}
{McElwain}, M.~W., {Feinberg}, L.~D., {Perrin}, M.~D., {et~al.} 2023, \pasp,
  135, 058001, \dodoi{10.1088/1538-3873/acada0}

\bibitem[{{Ment} {et~al.}(2023){Ment}, {Charbonneau}, {Irwin}, {Winters},
  {Pass}, {Shporer}, {Essack}, {Kostov}, {Kunimoto}, {Levine}, {Seager},
  {Vanderspek}, \& {Winn}}]{Ment2023}
{Ment}, K., {Charbonneau}, D., {Irwin}, J., {et~al.} 2023, arXiv e-prints,
  arXiv:2304.01920, \dodoi{10.48550/arXiv.2304.01920}

\bibitem[{{Misra} {et~al.}(2014){Misra}, {Meadows}, \& {Crisp}}]{Misra2014}
{Misra}, A., {Meadows}, V., \& {Crisp}, D. 2014, \apj, 792, 61,
  \dodoi{10.1088/0004-637X/792/1/61}

\bibitem[{{Moran} {et~al.}(2023){Moran}, {Stevenson}, {Sing}, {MacDonald},
  {Kirk}, {Lustig-Yaeger}, {Peacock}, {Mayorga}, {Bennett},
  {L{\'o}pez-Morales}, {May}, {Rustamkulov}, {Valenti}, {Adams Redai}, {Alam},
  {Batalha}, {Fu}, {Gonzalez-Quiles}, {Highland}, {Kruse}, {Lothringer}, {Ortiz
  Ceballos}, {Sotzen}, \& {Wakeford}}]{MoranStevenson2023}
{Moran}, S.~E., {Stevenson}, K.~B., {Sing}, D.~K., {et~al.} 2023, \apjl, 948,
  L11, \dodoi{10.3847/2041-8213/accb9c}

\bibitem[{{Owen} \& {Jackson}(2012)}]{Owen2012}
{Owen}, J.~E., \& {Jackson}, A.~P. 2012, \mnras, 425, 2931,
  \dodoi{10.1111/j.1365-2966.2012.21481.x}

\bibitem[{{Owen} \& {Mohanty}(2016)}]{Owen2016}
{Owen}, J.~E., \& {Mohanty}, S. 2016, \mnras, 459, 4088,
  \dodoi{10.1093/mnras/stw959}

\bibitem[{{Peacock} {et~al.}(2019){Peacock}, {Barman}, {Shkolnik},
  {Hauschildt}, \& {Baron}}]{Peacock2019b}
{Peacock}, S., {Barman}, T., {Shkolnik}, E.~L., {Hauschildt}, P.~H., \&
  {Baron}, E. 2019, \apj, 871, 235, \dodoi{10.3847/1538-4357/aaf891}

\bibitem[{P\'erez \& Granger(2007)}]{ipython}
P\'erez, F., \& Granger, B.~E. 2007, Computing in Science and Engineering, 9,
  21, \dodoi{10.1109/MCSE.2007.53}

\bibitem[{{Pizzolato} {et~al.}(2003){Pizzolato}, {Maggio}, {Micela},
  {Sciortino}, \& {Ventura}}]{Pizzolato2003}
{Pizzolato}, N., {Maggio}, A., {Micela}, G., {Sciortino}, S., \& {Ventura}, P.
  2003, \aap, 397, 147, \dodoi{10.1051/0004-6361:20021560}

\bibitem[{{Preibisch} \& {Feigelson}(2005)}]{Preibisch2005}
{Preibisch}, T., \& {Feigelson}, E.~D. 2005, \apjs, 160, 390,
  \dodoi{10.1086/432094}

\bibitem[{{Rieke} {et~al.}(2023){Rieke}, {Kelly}, {Misselt}, {Stansberry},
  {Boyer}, {Beatty}, {Egami}, {Florian}, {Greene}, {Hainline}, {Leisenring},
  {Roellig}, {Schlawin}, {Sun}, {Tinnin}, {Williams}, {Willmer}, {Wilson},
  {Clark}, {Rohrbach}, {Brooks}, {Canipe}, {Correnti}, {DiFelice}, {Gennaro},
  {Girard}, {Hartig}, {Hilbert}, {Koekemoer}, {Nikolov}, {Pirzkal}, {Rest},
  {Robberto}, {Sunnquist}, {Telfer}, {Wu}, {Ferry}, {Lewis}, {Baum},
  {Beichman}, {Doyon}, {Dressler}, {Eisenstein}, {Ferrarese}, {Hodapp},
  {Horner}, {Jaffe}, {Johnstone}, {Krist}, {Martin}, {McCarthy}, {Meyer},
  {Rieke}, {Trauger}, \& {Young}}]{Rieke2023}
{Rieke}, M.~J., {Kelly}, D.~M., {Misselt}, K., {et~al.} 2023, \pasp, 135,
  028001, \dodoi{10.1088/1538-3873/acac53}

\bibitem[{{Rigby} {et~al.}(2023){Rigby}, {Perrin}, {McElwain}, {Kimble},
  {Friedman}, {Lallo}, {Doyon}, {Feinberg}, {Ferruit}, {Glasse}, {Rieke},
  {Rieke}, {Wright}, {Willott}, {Colon}, {Milam}, {Neff}, {Stark}, {Valenti},
  {Abell}, {Abney}, {Abul-Huda}, {Acton}, {Adams}, {Adler}, {Aguilar}, {Ahmed},
  {Albert}, {Alberts}, {Aldridge}, {Allen}, {Altenburg},
  {{\'A}lvarez-M{\'a}rquez}, {Alves de Oliveira}, {Andersen}, {Anderson},
  {Anderson}, {Argyriou}, {Armstrong}, {Arribas}, {Artigau}, {Arvai},
  {Atkinson}, {Bacon}, {Bair}, {Banks}, {Barrientes}, {Barringer}, {Bartosik},
  {Bast}, {Baudoz}, {Beatty}, {Bechtold}, {Beck}, {Bergeron}, {Bergkoetter},
  {Bhatawdekar}, {Birkmann}, {Blazek}, {Blome}, {Boccaletti}, {B{\"o}ker},
  {Boia}, {Bonaventura}, {Bond}, {Bosley}, {Boucarut}, {Bourque}, {Bouwman},
  {Bower}, {Bowers}, {Boyer}, {Bradley}, {Brady}, {Braun}, {Breda},
  {Bresnahan}, {Bright}, {Britt}, {Bromenschenkel}, {Brooks}, {Brooks},
  {Brown}, {Brown}, {Brown}, {Bunker}, {Burger}, {Bushouse}, {Cale}, {Cameron},
  {Cameron}, {Canipe}, {Caplinger}, {Caputo}, {Cara}, {Carey}, {Carniani},
  {Carrasquilla}, {Carruthers}, {Case}, {Catherine}, {Chance}, {Chapman},
  {Charlot}, {Charlow}, {Chayer}, {Chen}, {Cherinka}, {Chichester}, {Chilton},
  {Chonis}, {Clampin}, {Clark}, {Clark}, {Coe}, {Coleman}, {Comber}, {Comeau},
  {Connolly}, {Cooper}, {Cooper}, {Coppock}, {Correnti}, {Cossou}, {Coulais},
  {Coyle}, {Cracraft}, {Curti}, {Cuturic}, {Davis}, {Davis}, {Dean}, {DeLisa},
  {deMeester}, {Dencheva}, {Dencheva}, {DePasquale}, {Deschenes}, {Hunor
  Detre}, {Diaz}, {Dicken}, {DiFelice}, {Dillman}, {Dixon}, {Doggett},
  {Donaldson}, {Douglas}, {DuPrie}, {Dupuis}, {Durning}, {Easmin}, {Eck},
  {Edeani}, {Egami}, {Ehrenwinkler}, {Eisenhamer}, {Eisenhower}, {Elie},
  {Elliott}, {Elliott}, {Ellis}, {Engesser}, {Espinoza}, {Etienne}, {Etxaluze},
  {Falini}, {Feeney}, {Ferry}, {Filippazzo}, {Fincham}, {Fix}, {Flagey},
  {Florian}, {Flynn}, {Fontanella}, {Ford}, {Forshay}, {Fox}, {Franz}, {Fu},
  {Fullerton}, {Galkin}, {Galyer}, {Garc{\'\i}a Mar{\'\i}n}, {Gardner},
  {Gardner}, {Garland}, {Garrett}, {Gasman}, {Gaspar}, {Gaudreau}, {Gauthier},
  {Geers}, {Geithner}, {Gennaro}, {Giardino}, {Girard}, {Giuliano},
  {Glassmire}, {Glauser}, {Glazer}, {Godfrey}, {Golimowski}, {Gollnitz},
  {Gong}, {Gonzaga}, {Gordon}, {Gordon}, {Goudfrooij}, {Greene}, {Greenhouse},
  {Grimaldi}, {Groebner}, {Grundy}, {Guillard}, {Gutman}, {Ha}, {Haderlein},
  {Hagedorn}, {Hainline}, {Haley}, {Hami}, {Hamilton}, {Hammel}, {Hansen},
  {Harkins}, {Harr}, {Hart}, {Hart}, {Hartig}, {Hashimoto}, {Haskins},
  {Hathaway}, {Havey}, {Hayden}, {Hecht}, {Heller-Boyer}, {Henriques}, {Henry},
  {Hermann}, {Hernandez}, {Hesman}, {Hicks}, {Hilbert}, {Hines}, {Hoffman},
  {Holfeltz}, {Holler}, {Hoppa}, {Hott}, {Howard}, {Howard}, {Hunter},
  {Hunter}, {Hurst}, {Husemann}, {Hustak}, {Ilinca Ignat}, {Illingworth},
  {Irish}, {Jackson}, {Jahromi}, {Jakobsen}, {James}, {James}, {Januszewski},
  {Jenkins}, {Jirdeh}, {Johnson}, {Johnson}, {Jones}, {Jones}, {Jones},
  {Jones}, {Jordan}, {Jordan}, {Jurczyk}, {Jurling}, {Kaleida}, {Kalmanson},
  {Kammerer}, {Kang}, {Kao}, {Karakla}, {Kavanagh}, {Kelly}, {Kendrew},
  {Kennedy}, {Kenny}, {Keski-kuha}, {Keyes}, {Kidwell}, {Kinzel}, {Kirk},
  {Kirkpatrick}, {Kirshenblat}, {Klaassen}, {Knapp}, {Knight}, {Knollenberg},
  {Koehler}, {Koekemoer}, {Kovacs}, {Kulp}, {Kumari}, {Kyprianou}, {La Massa},
  {Labador}, {Labiano}, {Lagage}, {Lajoie}, {Lallo}, {Lam}, {Lamb}, {Lambros},
  {Lampenfield}, {Langston}, {Larson}, {Law}, {Lawrence}, {Lee}, {Leisenring},
  {Lepo}, {Leveille}, {Levenson}, {Levine}, {Levy}, {Lewis}, {Lewis},
  {Libralato}, {Lightsey}, {Link}, {Liu}, {Lo}, {Lockwood}, {Logue}, {Long},
  {Long}, {Loomis}, {Lopez-Caniego}, {Lorenzo Alvarez}, {Love-Pruitt}, {Lucy},
  {Luetzgendorf}, {Maghami}, {Maiolino}, {Major}, {Malla}, {Malumuth},
  {Manjavacas}, {Mannfolk}, {Marrione}, {Marston}, {Martel}, {Maschmann},
  {Masci}, {Masciarelli}, {Maszkiewicz}, {Mather}, {McKenzie}, {McLean},
  {McMaster}, {Melbourne}, {Mel{\'e}ndez}, {Menzel}, {Merz}, {Meyett}, {Meza},
  {Miskey}, {Misselt}, {Moller}, {Morrison}, {Morse}, {Moseley}, {Mosier},
  {Mountain}, {Mueckay}, {Mueller}, {Mullally}, {Murphy}, {Murray}, {Murray},
  {Mustelier}, {Muzerolle}, {Mycroft}, {Myers}, {Myrick}, {Nanavati}, {Nance},
  {Nayak}, {Naylor}, {Nelan}, {Nickson}, {Nielson}, {Nieto-Santisteban},
  {Nikolov}, {Noriega-Crespo}, {O'Shaughnessy}, {O'Sullivan}, {Ochs}, {Ogle},
  {Oleszczuk}, {Olmsted}, {Osborne}, {Ottens}, {Owens}, {Pacifici}, {Pagan},
  {Page}, {Park}, {Parrish}, {Patapis}, {Paul}, {Pauly}, {Pavlovsky}, {Pedder},
  {Peek}, {Pena-Guerrero}, {Penanen}, {Perez}, {Perna}, {Perriello},
  {Phillips}, {Pietraszkiewicz}, {Pinaud}, {Pirzkal}, {Pitman}, {Piwowar},
  {Platais}, {Player}, {Plesha}, {Pollizi}, {Polster}, {Pontoppidan},
  {Porterfield}, {Proffitt}, {Pueyo}, {Pulliam}, {Quirt}, {Quispe Neira},
  {Ramos Alarcon}, {Ramsay}, {Rapp}, {Rapp}, {Rauscher}, {Ravindranath},
  {Rawle}, {Regan}, {Reichard}, {Reis}, {Ressler}, {Rest}, {Reynolds}, {Rhue},
  {Richon}, {Rickman}, {Ridgaway}, {Ritchie}, {Rix}, {Robberto}, {Robinson},
  {Robinson}, {Robinson}, {Rock}, {Rodriguez}, {Rodriguez Del Pino}, {Roellig},
  {Rohrbach}, {Roman}, {Romelfanger}, {Rose}, {Roteliuk}, {Roth}, {Rothwell},
  {Rowlands}, {Roy}, {Royer}, {Royle}, {Rui}, {Rumler}, {Runnels}, {Russ},
  {Rustamkulov}, {Ryden}, {Ryer}, {Sabata}, {Sabatke}, {Sabbi}, {Samuelson},
  {Sapp}, {Sappington}, {Sargent}, {Sauer}, {Scheithauer}, {Schlawin},
  {Schlitz}, {Schmitz}, {Schneider}, {Schreiber}, {Schulze}, {Schwab}, {Scott},
  {Sembach}, {Shanahan}, {Shaughnessy}, {Shaw}, {Shawger}, {Shay}, {Sheehan},
  {Shen}, {Sherman}, {Shiao}, {Shih}, {Shivaei}, {Sienkiewicz}, {Sing},
  {Sirianni}, {Sivaramakrishnan}, {Skipper}, {Sloan}, {Slocum}, {Slowinski},
  {Smith}, {Smith}, {Smith}, {Smith}, {Snyder}, {Soh}, {Sohn}, {Soto},
  {Spencer}, {Stallcup}, {Stansberry}, {Starr}, {Starr}, {Stewart},
  {Stiavelli}, {Straughn}, {Strickland}, {Stys}, {Summers}, {Sun}, {Sunnquist},
  {Swade}, {Swam}, {Swaters}, {Swoish}, {Taylor}, {Taylor}, {Te Plate}, {Tea},
  {Teague}, {Telfer}, {Temim}, {Thatte}, {Thompson}, {Thompson}, {Thomson},
  {Tikkanen}, {Tippet}, {Todd}, {Toolan}, {Tran}, {Trejo}, {Truong},
  {Tsukamoto}, {Tustain}, {Tyra}, {Ubeda}, {Underwood}, {Uzzo}, {Van Campen},
  {Vandal}, {Vandenbussche}, {Vila}, {Volk}, {Wahlgren}, {Waldman}, {Walker},
  {Wander}, {Warfield}, {Warner}, {Wasiak}, {Watkins}, {Weaver}, {Weilert},
  {Weiser}, {Weiss}, {Weissman}, {Welty}, {West}, {Wheate}, {Wheatley},
  {Wheeler}, {White}, {Whiteaker}, {Whitehouse}, {Whiteleather}, {Whitman},
  {Williams}, {Willmer}, {Willoughby}, {Wilson}, {Wirth}, {Wislowski}, {Wolf},
  {Wolfe}, {Wolff}, {Workman}, {Wright}, {Wu}, {Wu}, {Wymer}, {Yates},
  {Yeager}, {Yeates}, {Yerger}, {Yoon}, {Young}, {Yu}, {Zak}, {Zeidler},
  {Zhou}, {Zielinski}, {Zincke}, \& {Zonak}}]{Rigby2023}
{Rigby}, J., {Perrin}, M., {McElwain}, M., {et~al.} 2023, \pasp, 135, 048001,
  \dodoi{10.1088/1538-3873/acb293}

\bibitem[{{Rustamkulov} {et~al.}(2022){Rustamkulov}, {Sing}, {Liu}, \&
  {Wang}}]{Rustamkulov2022}
{Rustamkulov}, Z., {Sing}, D.~K., {Liu}, R., \& {Wang}, A. 2022, \apjl, 928,
  L7, \dodoi{10.3847/2041-8213/ac5b6f}

\bibitem[{{Rustamkulov} {et~al.}(2023){Rustamkulov}, {Sing}, {Mukherjee},
  {May}, {Kirk}, {Schlawin}, {Line}, {Piaulet}, {Carter}, {Batalha}, {Goyal},
  {L{\'o}pez-Morales}, {Lothringer}, {MacDonald}, {Moran}, {Stevenson},
  {Wakeford}, {Espinoza}, {Bean}, {Batalha}, {Benneke}, {Berta-Thompson},
  {Crossfield}, {Gao}, {Kreidberg}, {Powell}, {Cubillos}, {Gibson}, {Leconte},
  {Molaverdikhani}, {Nikolov}, {Parmentier}, {Roy}, {Taylor}, {Turner},
  {Wheatley}, {Aggarwal}, {Ahrer}, {Alam}, {Alderson}, {Allen}, {Banerjee},
  {Barat}, {Barrado}, {Barstow}, {Bell}, {Blecic}, {Brande}, {Casewell},
  {Changeat}, {Chubb}, {Crouzet}, {Daylan}, {Decin}, {D{\'e}sert},
  {Mikal-Evans}, {Feinstein}, {Flagg}, {Fortney}, {Harrington}, {Heng}, {Hong},
  {Hu}, {Iro}, {Kataria}, {Kempton}, {Krick}, {Lendl}, {Lillo-Box}, {Louca},
  {Lustig-Yaeger}, {Mancini}, {Mansfield}, {Mayne}, {Miguel}, {Morello},
  {Ohno}, {Palle}, {Petit dit de la Roche}, {Rackham}, {Radica},
  {Ramos-Rosado}, {Redfield}, {Rogers}, {Shkolnik}, {Southworth}, {Teske},
  {Tremblin}, {Tucker}, {Venot}, {Waalkes}, {Welbanks}, {Zhang}, \&
  {Zieba}}]{Rustamkulov2023}
{Rustamkulov}, Z., {Sing}, D.~K., {Mukherjee}, S., {et~al.} 2023, \nat, 614,
  659, \dodoi{10.1038/s41586-022-05677-y}

\bibitem[{{Salvatier} {et~al.}(2016){Salvatier}, {Wiecki{\^a}}, \&
  {Fonnesbeck}}]{Salvatier2016}
{Salvatier}, J., {Wiecki{\^a}}, T.~V., \& {Fonnesbeck}, C. 2016, {PyMC3: Python
  probabilistic programming framework}, Astrophysics Source Code Library,
  record ascl:1610.016.
\newblock \doeprint{1610.016}

\bibitem[{{Schaefer} {et~al.}(2016){Schaefer}, {Wordsworth}, {Berta-Thompson},
  \& {Sasselov}}]{Schaefer2016}
{Schaefer}, L., {Wordsworth}, R.~D., {Berta-Thompson}, Z., \& {Sasselov}, D.
  2016, \apj, 829, 63, \dodoi{10.3847/0004-637X/829/2/63}

\bibitem[{{Schlawin} {et~al.}(2020){Schlawin}, {Leisenring}, {Misselt},
  {Greene}, {McElwain}, {Beatty}, \& {Rieke}}]{schlawin_jwst_2020}
{Schlawin}, E., {Leisenring}, J., {Misselt}, K., {et~al.} 2020, \aj, 160, 231,
  \dodoi{10.3847/1538-3881/abb811}

\bibitem[{{Shkolnik} \& {Barman}(2014)}]{Shkolnik2014}
{Shkolnik}, E.~L., \& {Barman}, T.~S. 2014, \aj, 148, 64,
  \dodoi{10.1088/0004-6256/148/4/64}

\bibitem[{{Skilling}(2004)}]{Skilling2004}
{Skilling}, J. 2004, in American Institute of Physics Conference Series, Vol.
  735, Bayesian Inference and Maximum Entropy Methods in Science and
  Engineering: 24th International Workshop on Bayesian Inference and Maximum
  Entropy Methods in Science and Engineering, ed. R.~{Fischer}, R.~{Preuss}, \&
  U.~V. {Toussaint}, 395--405, \dodoi{10.1063/1.1835238}

\bibitem[{{Speagle}(2020)}]{Speagle2020}
{Speagle}, J.~S. 2020, \mnras, 493, 3132, \dodoi{10.1093/mnras/staa278}

\bibitem[{{STScI Development Team}(2013)}]{STScIDevelopmentTeam2013}
{STScI Development Team}. 2013, {pysynphot: Synthetic photometry software
  package}, Astrophysics Source Code Library, record ascl:1303.023.
\newblock \doeprint{1303.023}

\bibitem[{{Swain} {et~al.}(2021){Swain}, {Estrela}, {Roudier}, {Sotin},
  {Rimmer}, {Valio}, {West}, {Pearson}, {Huber-Feely}, \& {Zellem}}]{Swain2021}
{Swain}, M.~R., {Estrela}, R., {Roudier}, G.~M., {et~al.} 2021, \aj, 161, 213,
  \dodoi{10.3847/1538-3881/abe879}

\bibitem[{{The JWST Transiting Exoplanet Community Early Release Science Team}
  {et~al.}(2022){The JWST Transiting Exoplanet Community Early Release Science
  Team}, {Ahrer}, {Alderson}, {Batalha}, {Batalha}, {Bean}, {Beatty}, {Bell},
  {Benneke}, {Berta-Thompson}, {Carter}, {Crossfield}, {Espinoza}, {Feinstein},
  {Fortney}, {Gibson}, {Goyal}, {Kempton}, {Kirk}, {Kreidberg},
  {L{\'o}pez-Morales}, {Line}, {Lothringer}, {Moran}, {Mukherjee}, {Ohno},
  {Parmentier}, {Piaulet}, {Rustamkulov}, {Schlawin}, {Sing}, {Stevenson},
  {Wakeford}, {Allen}, {Birkmann}, {Brande}, {Crouzet}, {Cubillos}, {Damiano},
  {D{\'e}sert}, {Gao}, {Harrington}, {Hu}, {Kendrew}, {Knutson}, {Lagage},
  {Leconte}, {Lendl}, {MacDonald}, {May}, {Miguel}, {Molaverdikhani}, {Moses},
  {Murray}, {Nehring}, {Nikolov}, {Petit dit de la Roche}, {Radica}, {Roy},
  {Stassun}, {Taylor}, {Waalkes}, {Wachiraphan}, {Welbanks}, {Wheatley},
  {Aggarwal}, {Alam}, {Banerjee}, {Barstow}, {Blecic}, {Casewell}, {Changeat},
  {Chubb}, {Col{\'o}n}, {Coulombe}, {Daylan}, {de Val-Borro}, {Decin}, {Dos
  Santos}, {Flagg}, {France}, {Fu}, {Garc{\'\i}a Mu{\~n}oz}, {Gizis},
  {Glidden}, {Grant}, {Heng}, {Henning}, {Hong}, {Inglis}, {Iro}, {Kataria},
  {Komacek}, {Krick}, {Lee}, {Lewis}, {Lillo-Box}, {Lustig-Yaeger}, {Mancini},
  {Mandell}, {Mansfield}, {Marley}, {Mikal-Evans}, {Morello}, {Nixon}, {Ortiz
  Ceballos}, {Piette}, {Powell}, {Rackham}, {Ramos-Rosado}, {Rauscher},
  {Redfield}, {Rogers}, {Roman}, {Roudier}, {Scarsdale}, {Shkolnik},
  {Southworth}, {Spake}, {E Steinrueck}, {Tan}, {Teske}, {Tremblin}, {Tsai},
  {Tucker}, {Turner}, {Valenti}, {Venot}, {Waldmann}, {Wallack}, {Zhang}, \&
  {Zieba}}]{ERSFirstLook}
{The JWST Transiting Exoplanet Community Early Release Science Team}, {Ahrer},
  E.-M., {Alderson}, L., {et~al.} 2022, arXiv e-prints, arXiv:2208.11692.
\newblock \doarXiv{2208.11692}

\bibitem[{van~der Walt {et~al.}(2011)van~der Walt, Colbert, \&
  Varoquaux}]{numpy}
van~der Walt, S., Colbert, S.~C., \& Varoquaux, G. 2011, Computing in Science
  Engineering, 13, 22, \dodoi{10.1109/MCSE.2011.37}

\bibitem[{{Virtanen} {et~al.}(2020){Virtanen}, {Gommers}, {Oliphant},
  {Haberland}, {Reddy}, {Cournapeau}, {Burovski}, {Peterson}, {Weckesser},
  {Bright}, {van der Walt}, {Brett}, {Wilson}, {Jarrod Millman}, {Mayorov},
  {Nelson}, {Jones}, {Kern}, {Larson}, {Carey}, {Polat}, {Feng}, {Moore}, {Vand
  erPlas}, {Laxalde}, {Perktold}, {Cimrman}, {Henriksen}, {Quintero}, {Harris},
  {Archibald}, {Ribeiro}, {Pedregosa}, {van Mulbregt}, \&
  {Contributors}}]{scipy}
{Virtanen}, P., {Gommers}, R., {Oliphant}, T.~E., {et~al.} 2020, Nature
  Methods, 17, 261, \dodoi{https://doi.org/10.1038/s41592-019-0686-2}

\bibitem[{{Wakeford} {et~al.}(2019){Wakeford}, {Lewis}, {Fowler}, {Bruno},
  {Wilson}, {Moran}, {Valenti}, {Batalha}, {Filippazzo}, {Bourrier},
  {H{\"o}rst}, {Lederer}, \& {de Wit}}]{Wakeford2019}
{Wakeford}, H.~R., {Lewis}, N.~K., {Fowler}, J., {et~al.} 2019, \aj, 157, 11,
  \dodoi{10.3847/1538-3881/aaf04d}

\bibitem[{{Wheatley} {et~al.}(2017){Wheatley}, {Louden}, {Bourrier},
  {Ehrenreich}, \& {Gillon}}]{Wheatley2017}
{Wheatley}, P.~J., {Louden}, T., {Bourrier}, V., {Ehrenreich}, D., \& {Gillon},
  M. 2017, \mnras, 465, L74, \dodoi{10.1093/mnrasl/slw192}

\bibitem[{Zahnle \& Catling(2017)}]{Zahnle2017}
Zahnle, K.~J., \& Catling, D.~C. 2017, The Astrophysical Journal, 843, 122,
  \dodoi{10.3847/1538-4357/aa7846}

\bibitem[{{Zieba} {et~al.}(2023){Zieba}, {Kreidberg}, {Ducrot}, {Gillon},
  {Morley}, {Schaefer}, {Tamburo}, {Koll}, {Lyu}, {Acu{\~n}a}, {Agol}, {Iyer},
  {Hu}, {Lincowski}, {Meadows}, {Selsis}, {Bolmont}, {Mandell}, \&
  {Suissa}}]{Zieba2023}
{Zieba}, S., {Kreidberg}, L., {Ducrot}, E., {et~al.} 2023, \nat, 620, 746,
  \dodoi{10.1038/s41586-023-06232-z}

\end{thebibliography}
\bibliographystyle{aasjournal}



\end{document}